\documentclass[fleqn,a4paper]{article}

\usepackage{german}
\usepackage{graphicx}
\usepackage{setspace}
\usepackage{longtable}
\usepackage{amssymb}
\usepackage{hyperref}

\def\R2Lurl#1#2{\mbox{\href{#1}{\tt #2}}}

\newcommand{\tab}{\hspace{5mm}}

\begin{document}

\newcommand{\keywords}{electromagnetic theory, projective geometries }

\newcommand{\PACS}{02.40.Dr, 03.50.De, 92.40.Cy}

\title{Zum kleinschen Modell des Elektromagnetismus}
\author{S.\ L.\ Vesely$^{1}$, A.\ A.\ Vesely} 

\newcommand{\address}
  {$^{1}$I.T.B., C.N.R., via Fratelli Cervi 93, I-20090 Segrate(MI)
   \\ \hspace*{0.5mm} Italy \\ 
   }

\newcommand{\email}{\tt sara.vesely@itb.cnr.it, vesely@tana.it}

\maketitle

{\small
\noindent \address
\par
\par
\noindent email: \email
}
\par

{\small
 \par
\par
\noindent{\bf PACS:} \PACS 
}

\begin{abstract}

Weil das kleinsche Modell des Elektromagnetismus den meisten 
kein Begriff sein d\"{u}rfte, sagen wir sofort aus welchem Anlass 
wir gerade diesen Titel gew\"{a}hlt haben. Seit J. Maxwells Zeit 
hat eine ziemliche Schwierigkeit bei dem von ihm aufgestellten 
Gleichungssystem in der Deutung dessen L\"{o}sungen bestanden.

Sie wurden haupts\"{a}chlich auf das mechanische Denkmodell gest\"{u}tzt. 
1905 hat dann der junge A. Einstein gegen das damalige Establishment 
die Meinung vertreten, dass rein formale Gesichtspunkte ausgedient 
h\"{a}tten. Er hat bahnbrecherisch statuiert, dass die Elektrodynamik 
lauter physikalische Aussagen \"{u}ber Vermessungen betrifft. Kein 
Wunder dass er, als er 1917 mit vier freundlichen, und interessanten, 
und inzwischen wohl verschollenen Brieflein von F. Klein konfrontiert 
wurde, es ablehnte die darin erw\"{a}hnten Arbeiten mathematischen 
Inhalts zur Kenntnis zu nehmen, oder gar Hand an deren physikalischen 
Interpretation zu legen. Damals war Klein ein furchterregender 
Kontrahent.

Er war sich dessen auch bewusst. Heute d\"{u}rften 
dagegen all die mit der Relativit\"{a}tstheorie verbundenen Priorit\"{a}tsfragen 
genug abgeflaut sein, um gleichfalls einen Blick auf seinen Ansatz 
zu erlauben. Denn er war bestimmt ein Vertreter des deutschen 
Establishments, aber er war zugleich selber ein origineller Denker. 
Wir heben hier einige Punkte hervor, die mit Bezug auf Relativit\"{a}t 
und Elektromagnetismus heute noch von einigem Interesse sein 
k\"{o}nnten.

\end{abstract}

\section{
\"{U}bersicht}

Im ersten Kapitel wird die Anlehnung der Dynamik an die analytische 
Geometrie an den Beispielen der elastischen Anziehung und der 
Gravitation klar gemacht. Im Anschluss daran wird der mechanische 
Zeitbegriff diskutiert.
Die t-parametrisierte Bahn wird als station\"{a}re Bewegung gedeutet, 
und es wird dargetan, dass sich der Ansatz zur r\"{a}umlichen Bewegung 
ausgedehnter K\"{o}rper aus geometrischen Gr\"{u}nden nicht eindeutig 
parametrisieren l\"{a}sst.

Die Deutung der Ausbreitung auf wahrscheinlichkeitstheoretischer 
Grundlage scheint keinen Ersatz f\"{u}r die geometrische Auffassung 
zu bieten, weil sie sich auf v\"{o}llig andere Merkmale st\"{u}tzt.
Im zweiten Kapitel wird das Problem angesprochen, ob \"{u}berhaupt 
Felder und Bewegungen zusammenh\"{a}ngend behandelt werden sollen. 
Die Schwierigkeit besteht darin, Massenpunkte im Feld aufzul\"{o}sen. 
Es wird nahegelegt, dass elektromagnetische Felder zu ihrer physikalischen 
Kennzeichnung allerdings keinerlei Vorstellung von Flugbahnen 
bed\"{u}rfen, so dass t-parametrisierte Ausdr\"{u}cke vermutlich 
nur zur L\"{o}sung der Gleichungen herangezogen worden sind. Werden 
ferner freie elektromagnetische Felder als Signale aufgefasst, 
was sich angesichts der Relativit\"{a}tstheorie Einsteins empfiehlt, 
dann k\"{o}nnen station\"{a}re Wellenfronten phasendifferenztreu 
in der gau{\ss}schen Zahlenebene bzw. auf der riemannschen Kugel 
abgebildet werden.

Dabei gestaltet sich die geometrische Modellierung, 
trotzdem sie ihre veranschaulichende Funktion beibeh\"{a}lt, anders 
als eine r\"{a}umliche Versinnbildlichung der \"{a}u{\ss}eren Welt.
Im dritten Kapitel greifen wir zwecks der geometrischen Modellierung 
die kleinschen Arbeiten \"{u}ber die geometrischen Grundlagen der 
Lorentzgruppe und \"{u}ber die mit dem Nullsystem verbundenen Raumverwandtschaften 
wieder auf.

Sie enthalten Beitr\"{a}ge aus verschiedenen Gebieten 
der mathematischen Physik, und handeln laut dem Erlanger Programm 
von zwei verschiedenen Transformationsgruppen. Als 1910 Klein, 
gelegentlich seiner Besch\"{a}ftigung mit der speziellen Relativit\"{a}tstheorie, 
die elektromagnetischen Gleichungen im Vakuum an die letztgenannte 
Gruppe anlehnte, zog er nur deshalb den lorentzschen Hyperkegel 
zus\"{a}tzlich heran, weil er der Au{\ss}enwelt eine pseudoeuklidische 
raumzeitliche Struktur verleihen zu sollen meinte.

Wenn man aus 
seiner geometrischen Schilderung der speziellen Relativit\"{a}tstheorie 
die Ma{\ss}bestimmung wieder abstreift, bleibt also ein selbstst\"{a}ndiges 
geometrisches Modell des Elektromagnetismus \"{u}brig. Weil die 
geometrische Modellierung relational aufgefasst werden kann, 
wirft das nicht nur die Frage nach dem eventuellen Tr\"{a}ger der 
raumzeitlichen Struktur der Au{\ss}enwelt, sondern auch diejenige 
nach der Interpretation der maxwellschen Gleichungen von neuem 
auf.

\section{
Einleitung}

Um 1900 ist die weitl\"{a}ufige Frage gestellt worden, ob denn 
unsere Welt anhand der elektromagnetischen als der vereinheitlichenden 
Naturgesetze gefasst werden k\"{o}nne. Doch suchte man zu jener 
Zeit zun\"{a}chst einmal nach einer rationalen Erkl\"{a}rung f\"{u}r 
eine Menge r\"{a}tselhaft anmutender Erscheinungen. Man hoffte 
sich ein besseres Verst\"{a}ndnis der elektrisch ausgel\"{o}sten 
Ph\"{a}nomene dadurch zu verschaffen, dass man deren Verkn\"{u}pfung 
mit der mechanischen Lehre klarstellte. Gesetzt man k\"{o}nnte 
die mechanischen Gesetze mit dem Wahrheitswert ,,wahr`` kennzeichnen, 
trotzdem w\"{u}rde die mechanische Begr\"{u}ndung aller Befunde h\"{o}chstens 
einen logischen Zusammenhang aufdecken, und sie verm\"{o}chte keinen 
Einblick in die zugrundeliegenden Prozesse zu geben. Da sich 
zudem die gewohnte bildliche Darstellung nicht zwanglos auf die 
Elektrodynamik \"{u}bertragen lie{\ss}, diente jene Formalisierung 
keineswegs der Anschaulichkeit. Deshalb trat damals neben den 
Bedarf an einheitlichem Verst\"{a}ndnis noch die Frage nach Rolle 
und Grundlage der graphisch-anschaulichen Schilderung hinzu. 
Aus jener Auffassung heraus hat sich 1905 Albert Einstein der 
elektromagnetischen Raum-Zeitmessung bahnbrecherisch zugewandt 
und in ihrer Namen s\"{a}mtliche bisher gewohnte Raumvorstellungen 
gest\"{u}rzt.

Heute, wo man zur Versinnbildlichung der Daten ohnehin zunehmend 
rein elektrische und optische Mittel einsetzt, und man sich kurzum 
der baren elektrischen Antwort materieller Sachverhalte annimmt, 
verlagert sich das um 1900 gestellte Deutungsproblem um einen 
Deut von dem Verst\"{a}ndnis der elektrischen Erregung als solcher 
auf die Interpretation von elektrischen Messergebnissen. Es entspricht 
aber die spezielle Relativit\"{a}tstheorie dieser Wendung auch 
nicht ganz, denn das empfangene Signal veranschaulicht raumzeitliche 
Abmessungen zumindest nicht unmittelbar. Auch gelten Raum und 
Zeit nicht von Haus aus als typisch elektrische Gr\"{o}{\ss}en. Ihre 
Ma{\ss}einheiten werden heute zwar elektrisch festgelegt. Aber 
bis keine fundamentalen Uhren und Ma{\ss}st\"{a}be in die Relativit\"{a}tstheorie 
aufgenommen werden k\"{o}nnen, bleibt Kants Deutung der raumzeitlichen 
Ereignisse, wie Einstein wusste, trotzdem eine nicht zu verwerfende 
Alternative. Endlich scheinen die weiterf\"{u}hrenden Theorien 
gegen\"{u}ber dem seit 1900 erreichten enormen Fortschritt in der 
Nachrichtentechnik nur einen wahrscheinlichkeitstheoretischen 
Zugang zur Interpretation des empfangenes Signals hervorgebracht 
zu haben.\\
Das kleinsche Modell des Elektromagnetismus bietet insofern eine 
m\"{o}gliche Antwort auf dieses Problem, als es ein Bindeglied 
zwischen Erscheinungen und deren Formalisierung darlegt. Historisch 
hat es F. Klein dazu getrieben, sich zur Geometrisierung der 
Physik zu \"{a}u{\ss}ern, um die Zeit als H. Minkowski seine vierdimensionale 
Geometrie der Welt aufstellte, und D. Hilbert mit Einstein um 
eine einheitliche physikalische Theorie der mechanischen und 
elektromagnetischen Erscheinungen zu wetteifern begann. Zu dieser 
Zeit machte Klein keinen Hehl\footnote{Es lohnt sich zu betonen, dass 
Klein von Anfang an an die restlose Durchf\"{u}hrbarkeit des hilbertsschen 
Programms nicht glaubte. Vorz\"{u}glich aus diesem Grund h\"{a}tte 
er es sehr begr\"{u}{\ss}t, wenn seine eigenen Entwicklungen auf 
irgendein Gebiet der Naturwissenschaften Anwendung gefunden h\"{a}tten.} 
daraus, dass er die Aufmerksamkeit der Physiker auf seine \"{u}ber 
lange Jahre hinweg geleistete Arbeit \"{u}ber nichteuklidische 
Geometrien zu lenken hoffte und wom\"{o}glich damit eine physikalische 
Theorie zu verbinden w\"{u}nschte. Er hat entsprechend im Artikel: 
``\"{U}ber die geometrischen Grundlagen der Lorentzgruppe,, das 
bestehende Problem der grundlegendsten physikalischen Gesetze 
etwa folgenderma{\ss}en eingegrenzt. Man habe anfangs geglaubt, 
dass es einen absoluten mit \"{A}ther gef\"{u}llten Raum g\"{a}be, 
weil die elektromagnetischen Gesetze auf eine charakteristische 
Weise von den Koordinatentransformationen abh\"{a}ngen. Laut seinem 
Erlanger Programm h\"{a}tte man demnach, wenn man gleich die Zeit 
als Raumver\"{a}nderliche dazunimmt, die galileische Gruppe der 
mechanischen Bewegungen als G$_{10}$, diejenige des Elektromagnetismus 
aber als G$_{7}$ bezeichnen sollen. Die Angeh\"{o}rigkeit zur Lorentzgruppe 
verleiht dem Elektromagnetismus unter Zugrundelegung einer nichteuklidischen 
Ma{\ss}bestimmung dann wieder die volle Bewegungsfreiheit. Diese 
Aussage kn\"{u}pft die Lorentz-Quadrik an eine vierdimensionale 
Raumstruktur. Andererseits kn\"{u}pfen die geometrischen Betrachtungen 
\"{u}ber das elektromagnetische Gleichungssystem bei Klein direkt 
an die maxwellschen Entwicklungen, und sind noch vor 1905 angestellt 
worden. Indem er nun die Lorentzinvarianz als projektive Ma{\ss}bestimmung 
einordnet, und zugleich den maxwellschen Gleichungen in der hertzschen 
analytischen Bezeichnung\cite{klein1973},
unabh\"{a}ngig 
von jeder Metrik auch ein lineares geometrisches Gebilde zuschreibt, 
kommt er zum Ergebnis, dass die zwei Transformationsgruppen nicht 
\"{u}bereinstimmen. Das veranlasst ihn 1921 zur testamentarischen 
Frage nach der Interpretation des Elektromagnetismus.

Wir denken, dass man beim Vorhaben, den Sinn graphisch-anschaulicher 
Darstellungen zu diskutieren, man gleichfalls klarstellen sollte, 
ob die Versinnbildlichung vorz\"{u}glich der Beschreibung oder 
der Modellierung des gemeinten Umstandes dienen soll.

Was die reine \textit{Beschreibung} betrifft, gibt die restlose Identifizierung 
einer Beobachtung mit der Gestalt des zu beschreibenden Objekts 
im Nachhinein keinen Aufschluss mehr \"{u}ber sein Verhalten unter 
bestimmten Versuchsbedingungen an die Hand. Man m\"{o}ge deshalb 
viererlei bedenken.\\
Zum Ersten. Gesetzt es g\"{a}be eine wirklichere Beschaffenheit 
als diejenige, die man durch Sinneswahrnehmung erf\"{a}hrt, beinhalten 
physikalische Gesetze, sofern sie sich nach den Befunden richten, 
schwerlich mehr Wahrheit als Beobachtungen.\\
Zum Zweiten. Obwohl die Physiker davon ausgehen, dass es eine 
\"{a}u{\ss}ere Welt gibt, mag das Vorhaben ihre wirkliche Beschaffenheit 
mit lauter optischen Mitteln zu erschlie{\ss}en etwas naiv sein.\\
Zum Dritten hat auch die umgekehrte Ansicht, dass uns beim Auseinandernehmen 
eines Mechanismus unmissverst\"{a}ndlich seine Funktionen, einschlie{\ss}lich 
der elektrischen und optischen Eigenschaften, bekannt werden, 
etwas verh\"{a}ngnisvolles mit sich.

Viertens k\"{o}nnen wir voraussetzen, dass ein Informationsgehalt 
bereits in den optischen bzw. elektrischen Messungen enthalten 
sei. Aber die Vorstellung, wonach jedem Signal eindeutig ein 
Musterbaustein zukommt, mag sich einem einheitlichen Naturverst\"{a}ndnis 
sowohl fordernd wie hinderlich zeigen.

Was die \textit{Modellierung} der Erscheinungen betrifft, wird sie 
vorz\"{u}glich argumentierend oder vergleichend gebaut.\\
Im ersten Fall ist das Verlangen, dass die nachtr\"{a}gliche Interpretation 
einer mathematischen Formulierung kein allzu befremdendes Bild 
der Naturereignisse liefern soll freilich eine Geschmacksfrage. 
Wichtiger erscheint uns, dass zwei verschiedene Theorien nicht 
unbedingt dasselbe leisten, und auch nicht ineinander aufgehen. 
Weiter kann eine mathematische Struktur rein logisch-formalen 
Prinzipien gehorchen oder, in Anlehnung an F. Klein, zus\"{a}tzlich 
an die Plastik appellieren. Dar\"{u}ber hinaus gibt es \"{u}blichere 
Weisen bestimmte mathematische Aufgaben zu l\"{o}sen, an die manchmal 
neue Erkenntnisse mit Vorteil gekn\"{u}pft werden k\"{o}nnen.

Im zweiten Fall zweifeln wir nicht daran, dass sich jedes mal 
wegweisende \"{A}hnlichkeiten mit den herk\"{o}mmlichen Vorstellungen 
der Physiklehre herausarbeiten lassen. Es gibt aber sehr viele 
Kriterien, wonach sich Gegenst\"{a}nde vergleichen lassen. Eine 
Analogie k\"{o}nnte, nachdem sie einmal in Erw\"{a}gung gezogen wurde, 
sich auf die Grundlage purer Beobachtung immer wieder aufdr\"{a}ngen, 
oder sich erst bei der Beschreibung bieten. Bestimmte modellierende 
Verh\"{a}ltnisse k\"{o}nnten die Aufmerksamkeit auf sich ziehen, 
oder man k\"{o}nnte die M\"{o}glichkeit dazu erst nach der mathematischen 
Einkleidung erkennen. Es ist nicht so, dass man sich unbedingt 
an ein bestimmtes Kriterium zu halten habe, aber es ist manchmal 
von Interesse die Richtlinien deutlich herauszukehren.\\
Das kleinsche Modell scheint uns interessant, weil F. Klein den 
Unterschied zwischen Geometrie, graphischer Beschreibung und 
versinnlichtem geometrischem Modell zuletzt scharf erkannt hat. 
Auch m\"{u}ssen wir hervorheben, dass sich der Geometer die Frage, 
woran es bei der physikalischen Deutung einer Geometrie ankommt, 
nicht weniger als Einstein \"{u}berlegt hat. Im Wesentlichen haben 
beide angenommen, dass physikalische Aussagen experimentell auf 
ihren Wahrheitsgehalt getestet zu werden brauchen. Dazu war aber 
der Physiker offensichtlich immer bestrebt den physikalischen 
Teil seiner Theorie damit zu begr\"{u}nden, dass es nur eine einzige 
von der Physiklehre zu fassende Wirklichkeit geben kann. Er hat 
sich bem\"{u}ht zu zeigen, wie sich in erster Ann\"{a}herung seine
allgemeine 
Relativit\"{a}tstheorie auf das newtonsche als ein anerkanntes 
und gut gepr\"{u}ftes Beobachtungsgesetz zur\"{u}ckf\"{u}hren l\"{a}sst. 
Daher tut es leid, dass sich Newton selber, als er die Gravitationss\"{a}tze 
erforschte, mit der transzendentalen Raumanschauung nicht hat 
vertraut machen k\"{o}nnen. Trotzdem wir n\"{a}mlich auch von der 
Einheit der Welt fest \"{u}berzeugt sind, sehen wir nicht ein, 
wie wir sie eindeutig und wom\"{o}glich anschaulich auf mathematische 
Lehrs\"{a}tze reduzieren k\"{o}nnten.

\section{
Drehungen l\"{a}ngs geometrischer \"{O}rter}

\subsection{
Lineare physikalische Theorien}

Wie weit stellen Theorien die experimentell vorherrschenden Verh\"{a}ltnisse 
sinnvoll dar?\\
Diese Frage vom Grund auf zu beantworten suchen hie{\ss}e zu vermuten, 
dass sich hinter den Erscheinungen eine dem menschlichen Verstand 
fassbare Logik berge, die es zu erraten gilt. Weil uns nur unsere 
Vernunft zwecks der Deutung von Experimenten zur Verf\"{u}gung 
steht, d\"{u}rfte das Beklopfen der Natur nach ihrer Funktionsweise 
au{\ss}erhalb des von uns erkennbaren Bereiches zu liegen kommen.

Die alten Griechen haben aber eine etwas anders formulierte Frage 
positiv beantwortet. N\"{a}mlich die, ob man den Naturereignissen 
auch etwa vern\"{u}nftig begegnen k\"{o}nne\footnote{Das Argument von 
der vern\"{u}nftigen Begegnung der Fakten hat etwas f\"{u}r sich. 
Wenn wir von umstrittenen sowie v\"{o}llig subjektiven Erfahrungen 
wie etwa Poltergeistern und Telepathie, oder Visionen und Tr\"{a}umen 
absehen, k\"{o}nnen wir annehmen, dass unser Verst\"{a}ndnis der 
Erscheinungen zwar nicht die Naturereignisse selber, wohl aber 
unsere Lage beeinflussen kann.}. Im Folgenden versuchen wir, 
uns rein pragmatisch an diese Art des Naturverst\"{a}ndnisses zu 
halten.

Die zu linearen Theorien f\"{u}hrenden Erkl\"{a}rungen sind in den 
oben erw\"{a}hnten Sinn allgemeiner und am leichtesten durchschaubar 
zugleich.

\subsection{
Gesetzliche Bahnen und deren analytische Darstellung}

In der klassischen Mechanik nimmt man bekanntlich an, dass die 
Bewegungsgesetze der K\"{o}rper weder vom Aufenthaltsort des Beobachters 
-- z.B. Kos bzw. Lule{\aa} -, noch von dessen konstanten Geschwindigkeit 
abh\"{a}ngen. Das l\"{a}uft auf Homogenit\"{a}t und Isotropie unserer 
Welt hinaus. Die letzte als Tr\"{a}gheitsprinzip bekannte Aussage 
wird am Vergleich der vom Gestade aus beobachteten Vorf\"{a}lle 
mit den sich auf einem einen Fluss hinabgleitenden Schiff abspielenden 
Ereignissen illustriert. Obwohl sich an Bord und am Boden (neulich 
kommt die Eisenbahn als Betriebsmittel auch in Betracht), keine 
\"{u}bereinstimmende Ortung angeben l\"{a}sst, nimmt man an, die \textit{Lage} 
eines Gegenstandes relativ zum Beschauer, innerhalb einer gewissen 
Klasse von Bezugssystemen, immer berechnen zu k\"{o}nnen. Die Verkn\"{u}pfung 
der Flugbahn zum Gesetz wird sogar ganz wichtig, wenn man den 
Meter und die Sekunde als fundamentale Ma{\ss}einheiten w\"{a}hlt.

Auch in Betreff der Kraftvorstellung tritt bei Newton eine Neuerung 
auf. In der Impetustheorie verband man mit der Kraft die Erfahrung 
eines tastsinnlichen, muskelbewegungsempfindlichen Reizes und 
den entsprechenden Eindruck einer strikt lokalen, \"{u}ber Ber\"{u}hrung 
vermittelten Wirkung. Newton hat zur Herleitung der Beobachtungss\"{a}tze 
Keplers die Kraft zur rein logischen Notwendigkeit erhoben\footnote{Newton 
hat das Wesen der Gravitationskraft am Magnet triftig gezeigt.}. 
Dabei hat er die auf geometrischen Betrachtungen fu{\ss}enden Gesetze 
noch in Worte gekleidet. Weil aber inzwischen alle Geometrie 
auf die analytische Geometrie zur\"{u}ckgef\"{u}hrt worden ist, dient 
heute der mathematische Ausdruck $F(r, \rm t) = \rm mb$ unmittelbar 
als allgemeine Kraftdefinition.
 Wenn \textit{b} die zweite Ableitung \textit{b} 
= d$^{2}$\textit{r}/dt$^{2}$ eines mit dem Zeitparameter t ver\"{a}nderlichen 
Standorts \textit{r} bedeutet, erh\"{a}lt man f\"{u}r endliche \textit{b}-Werte 
den sich unter den gegebenen Bedingungen einstellenden Zeitablauf 
einer K\"{o}rpermarkierung \textit{r} = \textit{r}(t) mittels Integration. 
Hier misst \textit{r} nicht die von der Anfangsstelle P\ensuremath{^\circ} 
an zur\"{u}ckgelegte Strecke als Bogenl\"{a}nge
\footnote{Die heute f\"{u}r 
die Bogenl\"{a}nge einer nach dem Parameter u zweimal ableitbaren 
Kurve $f$ \"{u}blichen Ausdr\"{u}cke
 $\int{\sqrt{\sum{a_{\kappa\lambda}f'_{\kappa}f'_{\lambda}}}}du$ 
bzw.
 $\int{\sqrt{E({df \over du})^2 + 2F({df \over du}) + G}du}$ setzen 
bereits die Abwickelbarkeit der Fl\"{a}che mit den Koordinaten 
$u = {\rm Konst}$., $v = f(u) = {\rm Konst}$.
 auf der Ebene voraus. Die geod\"{a}tischen 
Linien auf der u,v-Fl\"{a}che ergeben sich aus dem obigen Ausdruck 
durch einen Variationsansatz.}
$\int^{P^{'}}_{P^\circ}\sqrt{{({d{\bf r} \over dt})}^2}dt$
auf der Bahnkurve, sondern es stellt ein unbestimmtes 
Zeitintegral dar. Wird dieses jedoch mit einer zur\"{u}ckgelegten 
Strecke \textit{s} verkn\"{u}pft, dann wird das Problem die Ein-K\"{o}rper 
Bewegung zu bestimmen linearisiert. Es ist also
$s = \int_0^w\int_0^u du dv = \int_0^w du \int_0^u dv = \frac12 w^2$
 laut Descartes ein zur zweiten Potenz erhobener Ausdruck 
f\"{u}r das Ma{\ss} einer Strecke. Das war nicht immer so. In der 
klassischen Antike hat man Strecken und Fl\"{a}chen auseinandergehalten 
und nicht zahlenm\"{a}{\ss}ig verstanden.

Die Griechen spalteten anscheinend von der Geometrielehre \"{u}berhaupt 
alles ab, was sich nicht in ihre eigene analytische Methode f\"{u}gte, 
und gingen geometrische Aufgaben mit Zirkel und Lineal nur an, 
um anhand der Konstruktionen zu beweisen, dass es die entsprechenden 
Figuren gibt.

Zur Bew\"{a}ltigung der mit Zirkel und Lineal nicht zu l\"{o}senden, 
und also philosophisch nicht zu begr\"{u}ndenden Probleme haben 
freilich die meisten unter ihnen besondere Kurven gezeichnet. 
Schon der Kreis darf in dieser Hinsicht als besondere Kurve gelten. 
Zur praktischen Durchf\"{u}hrung einer weiteren Klasse von Rechnungen 
wurden Kegelschnitte verwertet. Schlie{\ss}lich gab es noch die 
Klasse der ``linearen Probleme``, zu deren Bearbeitung man komplizierterer 
Kurven -- z.B. der Quadratrix -- bedurfte. Weil die Altgriechen 
die zur Sch\"{a}tzung von Fl\"{a}chen und Volumen ben\"{o}tigten Kurven 
meist nur mit Hilfe mechanischer Vorrichtungen zeichnen konnten, 
und dabei vermutlich nicht \"{u}ber mit Fehlern behaftete Zahlenwerte 
kamen, nannten sie die entsprechenden Probleme absch\"{a}tzend 
auch ,,mechanisch``.
 Das waren sie auch
\footnote{Die mechanische Art 
B\"{o}gen zu strecken hat sp\"{a}ter wieder Beachtung erlangt. Besonders 
zur Bef\"{o}rderung zu Lande kommen mechanische Getriebe, die der 
Umsetzung von Kreisbewegungen in lineare Bewegungen und umgekehrt 
dienen, h\"{a}ufig vor. Es auch nichtsdestoweniger mathematisch 
interessante Gelenke, wie der Inversor von Peaucellier.}
.

Die Integralrechnung diente in erster Linie der Bew\"{a}ltigung 
dieser mechanischen Probleme, weil sie sie endlich in mathematische 
Obhut f\"{u}hrte. Man gibt in der klassischen Mechanik an verwickelte 
in einfachere Bewegungen irgendwie zerlegen zu k\"{o}nnen -- das 
lehnt an die descartesschen Entwicklungen -, so dass sich Kr\"{a}fte \textit{F} 
spalten lassen. Kann man umgekehrt bei Zugrundelegung eines kartesischen 
Koordinatensystems mit Einheitsvektoren
 \textbf{\^{\i}} und \textbf{\^{\j}} \textit{F} 
in jedem Augenblick
 $t_0$ laut
$F \equiv {\bf F}_D({\bf r},t_0) \equiv F_{D_x}${\bf\^{\i}}$ + F_{D_y}$ {\bf\^ {\j}}
 zerlegen, so sollen 
auch ihre orthogonalen Komponenten unmittelbar und unabh\"{a}ngig 
voneinander den Bewegungszustand in der jeweiligen Richtung bestimmen. 
Somit nimmt das Grundgesetz die Form ordentlicher Differentialgleichungen 
mit einem einzigen gemeinsamen Parameter t an. Die Ortkurve f\"{u}r 
eine Masse $m$ kann in der kartesischen $(x, y)$-Ebene, sobald die 
Hilfsvariable $t$ aus dem Zeitablauf eliminiert wird veranschaulicht 
werden. Das bedeutet, dass bei festgehaltenem Anfangspunkt der 
Vektor ${\bf r = r}(t)$ analytisch eine Funktion der Koordinaten 
des Endpunktes beschreibt, die geometrisch als Ortkurve gedeutet 
werden darf.

\textbf{Elastische Anziehung. -} Nehmen wir das Paradebeispiel einer \textit{andauernd} 
von einem federnden Instrument veranlassten hookeschen elastischen 
Anziehung auf einen Punkt der Masse m. Das Grundgesetz f\"{u}hrt 
f\"{u}r den relevanten Ortpunkt (x, y) zu parametrischen Gleichungen 
$md^{2}x/dt^{2} = - mk^{2}x$
und
$md^{2}y/dt^{2} = - mj^{2}y$,
mit den einfach periodischen Integralen
$x = a \sin 2{\pi}/T(t - t_{0}) = a \sin k(t- t_{0}) = a[ \sin kt \cos kt_{0} - \cos kt \sin kt_{0}]$,
$y = b \sin 2{\pi}/T'(t - t_{0}') = b \sin j(t - t_{0}') = b[ \sin jt \cos j t_{0}' - \cos jt \sin j t_{0}']$,
wobei $T = 2{\pi}/k$
und
$T' = 2{\pi}/j$
als pure 
Tautologien hinzunehmen sind.

Bisher waren es lauter mathematische Spekulationen. Tats\"{a}chlich 
lassen sich aber leicht elektrische, und mit geringf\"{u}giger 
M\"{u}he auch mechanische Ger\"{a}te bauen, an die man nahezu periodische, 
auf die hookeschen Gesetze zur\"{u}ckf\"{u}hrbare Bewegungen feststellen 
kann. Man kann geradezu bewirken, dass bei der Bewegung eine 
Spur, die sogenannte Lissajous-Figur, hinterlassen wird. Gar 
manche Bahnen sind auch anhand eines Spirographs, eines Kinderspielzeugs, 
nachzuahmen. Wir hypotisieren, dass hier unweigerlich das hookesche 
Gesetz zur Geltung kommt, und fragen nach der analytischen Kurve.

Wir \"{u}berschauen die Methode die Ortkurve analytisch festzulegen 
am besten, wenn wir weitere Spezialisierungen einf\"{u}hren. Setzen 
wir synchrone Bewegung l\"{a}ngs der zwei orthogonalen \textbf{\^{\i}} 
und \textbf{\^{\j}} Richtungen voraus, ergo $j = k$, so folgt mit $a = b = 
R$: $R^{2} \sin k(t_{0}' - t_{0}) \sin kt = R[x \sin k t_{0}' - y \sin kt_{0}]$, 
$R^{2} \sin k(t_{0}' - t_{0}) \cos kt = R[x \cos kt_{0}' - y \cos kt_{0}]$,
woraus man die Zeitverl\"{a}ufe $R \sin kt = Ax - By$ und $R \cos kt = Cx - Dy$
 als lineare Funktionen von $x, y$ bestimmt\footnote{$A = \sin 
k t_{0}'/[\sin k(t_{0}' - t_{0})]$ u.s.w.}
 und die Ortkurven
 $R^{2} = (Ax - By)^{2} + (Cx - Dy)^{2} = (A^{2} + C^{2})x^{2} - 2(AB + CD)xy
+ (B^{2} + D^{2})y^{2}$
 erh\"{a}lt. Diese geometrischen \"{O}rter sind 
konzentrische Ellipsen mit gegen \textbf{\^{\i}} um $45^\circ$ verdrehter 
Hauptachse. Ihre Exzentrizit\"{a}t h\"{a}ngt von $(t_{0}' - t_{0})$ ab, 
weshalb die spezielle Wahl $kt_{0}' = 0, kt_{0} = - \pi/2$ zu dem 
mit gleichm\"{a}{\ss}iger Geschwindigkeit vom Betrag
 $|{\bf v}| = Rk$
umgelaufenen Kreis
 $R^{2} = x^{2} + y^{2}$ f\"{u}hrt.

Die mathematisch zum kreisf\"{o}rmigen Locus f\"{u}hrende Methode 
kann umgekehrt werden. Stellt also ein Kreis oder irgendeine 
andere endlich lange Kurve eine m\"{o}gliche Bahn analytisch dar, 
dann kann sie in t-abh\"{a}ngige Bewegungszust\"{a}nde \"{u}bersetzt 
werden.\\
 Weil die Masse m dabei aus der Vektor-Differentialgleichung 
ausf\"{a}llt, \"{u}berholt die Integration die Exhaustionsmethode \textit{kinematisch}
\footnote{Eine 
Bewegungsvorstellung liegt bereits der Konstruktion der Quadratrix 
zugrunde. Man stellte sie sich als geometrischer Ort der Schnittpunkte 
eines mit konstanter Winkelgeschwindigkeit im Uhrzeigersinn drehenden 
Strahls mit einer zu sich selbst parallel und ebenfalls mit konstanter 
Geschwindigkeit fallenden Linie vor. Weil diese Kurve im Altertum 
obendrein zum Quadrieren von Kreisen, d.h. zur Bestimmung der 
Seite eines mit einem Kreis inhaltsgleichen Vierecks, diente 
lohnt es sich zu erw\"{a}hnen, dass sich damit in erster Linie 
B\"{o}gen $u$ rektifizieren lie{\ss}en. Dazu wurde die Quadratrix 
innerhalb eines Quadrats der Seite $\ell$ eingezeichnet. 
Dann teilte man die Basisseite ${\rm OL} = \ell$ in genauso viele 
kongruente Streifen ein, wie der Winkel in O gleiche Zonen von 
$0^\circ$ bis $90^\circ$ hatte. Schlie{\ss}lich ordnete man 
dem n-ten Winkel den n-ten Abschnitt zu. Damit wurden Winkel 
als Abschnitte auf die Ordinatenachse lesbar. Wenn A der Durchschnitt 
des Winkels $0^\circ$ mit dem Basisabschnitt OL bezeichnet, 
also ${\rm OA} = a <\ell$, ist es nach Pappus
 $u : \ell = \ell : a$,
 womit sich der Bogen als $u = \ell^{2}/a$ 
ergibt. Wegen $u = \frac14 2\pi\ell$ (in moderner 
Schreibweise), muss man zur Bestimmung des mit dem Kreis inhaltsgleichen 
Quadrat das Rechteck $\ell\times2u$ anschlie{\ss}end 
,,quadrieren``.}
. Bezeichne n\"{a}mlich $\frac12r\varphi$, 
mit $\varphi$ im \textit{Bogenma{\ss}}, den Fl\"{a}cheninhalt eines halben 
Kreissektors,
 $\frac12 rv_t = \frac12 r^2(d\varphi/dt)= \frac12r^2\omega$
 aber den infinitesimalen Fl\"{a}cheninhalt 
(die Fl\"{a}chengeschwindigkeit). Wenn die Tangentialgeschwindigkeit 
$v_t$ anstelle von \ensuremath{\varphi} tritt, wird der Ausdruck
$\frac12(d\varphi/dt)r^2$
 im \textit{linearen Ma{\ss}} messbar.
 L\"{a}sst man folglich die zu 
den Fluxionen f\"{u}hrende \textit{mathematische Entwicklung} zu, so 
l\"{a}sst sich im Prinzip jeder gekr\"{u}mmte Weg linear messen. 
Denkt man dar\"{u}ber nach, dass seit Newton ein L\"{a}ngenma{\ss} 
die irrationale Zahl $\pi$ als Faktor enthalten kann, so hat man 
den Beitrag Newtons zur Rektifizierung des Kreisumfangs richtig 
eingesch\"{a}tzt. Beil\"{a}ufig kann der Kreis derart parametrisiert 
werden, dass unendlich viele rationale Punkte auf dem Umfang 
zu liegen kommen. Die parametrische Darstellung
 $x = R 2t / (1+ t^2)$, $y = R (1 - t^2)/(1 + t^2)$
 leistet das, wenn man f\"{u}r 
$t \in[0, 1]$ rationale Zahlen w\"{a}hlt. Die diophantischen 
Gleichungen stellen lauter ebene Kurven dar, welche solche Darstellungen 
nicht gestatten.

\textbf{Gravitationskraft. -} Nun wollen wir die k\"{u}hne durch Integration 
erlangte L\"{o}sung des Quadraturproblems mit der dynamischen Vorstellung 
der gesetzm\"{a}{\ss}igen Drehungen um einen festen anziehenden Punkt 
vergleichen.\\
Der Gegen\"{u}berstellung schicken wir folgende Bemerkung voraus. 
W\"{a}re man mit den einem Massenpunkt zugeschriebenen Bewegungszust\"{a}nden 
genauso k\"{u}hn verfahren, dann h\"{a}tte es gehei{\ss}en das Tr\"{a}gheitsgesetz 
mit dem Grundsatz der Gravitation zu vereinigen und, sofern sich 
die Gleichungen integrieren lie{\ss}en, die Integrale als Geod\"{a}ten 
zu vermerken. Das wurde sp\"{a}ter Einsteins Ansatz. Newton wollte 
jedoch die Meinung durchsetzen, dass sich nat\"{u}rlich ergebende 
Flugbahnen gesetzlich fassen lie{\ss}en, und hat entsprechend einen 
rational erkl\"{a}rbaren Unterschied zwischen Kurventypen aufgestellt. 
Diese Festlegung der krummen Bahnen als kraftbedingte Drehungen 
scheint quasi auftischen zu wollen, dass es f\"{u}r einen K\"{o}rper 
eine in jedem Zeitabschnitt eindeutig bestimmte Bahnkurve gibt, 
die als Funktion eines in alle Ewigkeit andauernden \textit{Zeitflusses} $t$ 
analytisch bestimmt werden kann.

Der Zeitablauf $x = a t + b$ auf unbeschr\"{a}nkter gerader 
Bahn hebt sich durch die besondere Hypothese 
$md^2x/dt^2 \equiv 0, \forall t$
 vor allen anderen Bewegungen hervor.

Newton hat bekanntlich die drei von Kepler auf experimenteller 
Grundlage aufgestellten Gesetze seinem Kraftgesetz nach interpretiert. 
Der Bewegungsablauf gehorcht dann folgendem, den eben besprochenen 
beobachtbaren Lissajous Schwingungen \"{a}hnlich aussehendem Gravitationsgesetz 
$md^2{\bf r}/dt^2 = - mk^2{\bf r}$.
 Mit \textbf{r} = $x${\bf \^{\i}}$ + y${\bf \^{\j}} 
dem Fahrstrahl Brennpunkt -- Ortlinie schreibt man
 $d^2x/dt^2 = -k^2x$ und
 $d^2y/dt^2 = -k^2y$,
 wobei
 $k^2 = GM/\sqrt{[x^2 + y^2]^3}$,
 M die Sonnenmasse und G die Gravitationskonstante 
bedeuten. Experimentell uml\"{a}uft nun die Erde auf einer nach 
Keplers 1. Gesetz elliptischen Kurve die Sonne einmal j\"{a}hrlich, 
was ja einer fr\"{u}her festgelegten \textit{Zeitmessung} entspricht. 
Und es ist, wenn $a$ die gro{\ss}e Halbachse der Ellipse, und $T$ die 
Periode eines vollen Umlaufs bezeichnen, laut dem 3. keplerschen 
Gesetz $v\sqrt{a^3} \div T$
\footnote{Daraus hat Newton die besondere 
im Faktor $k$ enthaltene $1/r^2$ Abh\"{a}ngigkeit der Anziehungskraft 
abgeleitet.}
. Zum Schluss gilt, laut 2. Gesetz, f\"{u}r jeden \textit{endlichen} 
vom Perihel an erstreckten Ellipsensektor $\Sigma$ das Doppeltverh\"{a}ltnis
$\tau : \Sigma   = T : \pi ab$.
 Die seit Newton m\"{o}gliche alternative \textit{infinitesimale} 
Schreibweise f\"{u}r dieses Doppeltverh\"{a}ltnis hei{\ss}t:
$\pi ab : d \varphi r^2/2 = T : dt$.
 In dieser Form bezieht 
sich, weil die Winkelgeschwindigkeit
$\omega = d\varphi/dt$ 
nicht konstant bleibt, sondern von der auf der Kurve mit \ensuremath{\varphi} 
zu kennzeichnenden Lage abh\"{a}ngt, die Umlaufsfrequenz vornehmlich 
auf einen Ber\"{u}hrungskreis vom Halbmesser $r$, und der Satz besagt, 
dass die Bewegung auf dieser Kreisbahn gleichf\"{o}rmig mit
$\omega= 2\pi/T$
 erfolgt
\footnote{Das \"{u}bertrifft in gewisser Hinsicht 
die einfache galileische Aussage, da es in diesem Fall hinsichtlich 
des Ausdrucks des Gesetzes sicher ein mit der Sonne ruhendes 
bevorzugtes Bezugssystem gibt. Diesen von der relativen Raumlage 
abh\"{a}ngigen Ausdruck erh\"{a}lt man f\"{u}r den Fl\"{a}chensatz in 
Polarkoordinaten. Es ist allerdings m\"{o}glich die Punktlage als 
$z = r e^{i\theta}$
 niederzuschreiben. Auf jeden Fall lautet, mit
$\omega = d\phi/dt$,
 die radiale Beschleunigung
$b_r= d^2r/dr^2 - r\omega^2$,
 und die tangentiale Beschleunigung 
$b_t = r d\omega/dt + 2 (dr/dt)\omega = (1/r) d(r^2\omega)/dt$. 
Ist
$b_t = 0$,
 dann
$r^2\omega = r^2d\phi/dt = \rm konst$.}. 
Da sich der Bewegungszustand auf den Kreis nicht \"{a}ndert, wenn 
man auf die kreisrunde Scheibe aus einem allm\"{a}hlich schieferen 
Blickwinkeln hinschaut, obwohl man den Kreismittelpunkt graduell 
in einen Ellipsenbrennpunkt r\"{u}cken sieht, gilt der f\"{u}r Kreise 
\"{u}ber das endliche Doppeltverh\"{a}ltnis ausgedr\"{u}ckte Fl\"{a}chensatz 
auch f\"{u}r Ellipsen, ohne dass man eigens dazu ein Differentialgesetz 
aufzustellen braucht.\\
Hooke hatte nun, wie vorweggenommen, seinerseits die M\"{o}glichkeit 
in Aussicht gestellt Ellipsenbahnen aus einem etwas anders aussehenden 
Gesetz zu erhalten. Die dort f\"{u}r
$ k = j$
 benutzten Gleichungen 
$d^2x/dt^2 = - k^2x$
 und
$d^2y/dt^2 = - k^2y$
 f\"{u}hren zum Ausdruck: 
$x(d^2y/dt^2) - y(d^2x/dt^2) = 0$,
 woraus sich der Fl\"{a}chensatz 
in der Form:
$ \frac12 {d \over dt}[x(dy/dt) - y(dx/dt)] = 0$
 schreiben 
l\"{a}sst. Mit
$ a = b = R$
 lautet das Differentialgesetz
$ x(dy/dt)- y(dx/dt) = {\bf r} \times d{\bf r} = R^2k \sin k(t'_0 - t_0)$, 
wobei die Differenz
$ (t'_0 - t_0)$
 f\"{u}r die Exzentrizit\"{a}t bestimmend 
ist, und die Winkelgeschwindigkeit synchroner harmonischer Schwingungen 
nach Definition
$ \omega = 2\pi/T \propto  k = j$
 lautet. 
F\"{u}r den Kreis schreibt man
$ (t'_0 - t_0) = \frac12\pi$. 
Die Unabh\"{a}ngigkeit des Fl\"{a}cheninhalts einer allgemeinen ebenen 
Figur vom Innenpunkt in den man den Polarm eines Polarplanimeters 
(von Jakob Amsler \cite{kleinElem}) einsetzt, um mit dem Fahrarm 
um die Kurve umzufahren kann nachgepr\"{u}ft werden. Sind aber 
gesetzliche Bewegungen von Belang, so ergeben elastische Anziehung 
und Gravitationskraft trotzdem verschiedene periodische Bewegungen. 
Wir m\"{o}chten hier zeigen, dass sich das keplersche Problem nicht 
auf den isotropen Oszillator zur\"{u}ckf\"{u}hren l\"{a}sst, obwohl 
beides mal die schwer integrierbaren Ellipsen als Ortkurven in 
Betracht kommen. Sollte der Fl\"{a}chensatz beides mal denselben 
Sinn haben, dann w\"{u}rden harmonische Rundschwingungen um den 
Mittelpunkt einer Ellipse und Sonnenuml\"{a}ufe nach dem 1. keplerschen 
Gesetz zu derselben Zeitmessung f\"{u}hren. Leider haben wir uns 
doch das erste mal auf den Zentriwinkel \ensuremath{\psi}, das zweite 
auf die Anomalie \ensuremath{\varphi} bezogen. Wollte man auch die Planetenbahnen 
auf den Zentriwinkel beziehen, so w\"{u}rde dasselbe Doppeltverh\"{a}ltnis
 $\tau  : (\Delta+\Sigma) = T : \pi ab$
 wie bei den elliptischen Lissajous-Figuren 
gelten. Der Dreieck
$\Sigma$ mit Basis
$ ae = \sqrt{[a^2 - b^2]}$,
 H\"{o}he 
$b \sin \psi$
  und dem Fl\"{a}cheninhalt
$ ae/2 b\sin\psi$
  dient 
dazu, den Ellipsensektor auf den Ellipsenmittelpunkt zu beziehen. 
Der Fl\"{a}cheninhalt des entsprechenden Kreissektors vom Radius 
a gilt
 $\Sigma' = a^2 \psi /2$.
 Wegen
$ (\Delta + \Sigma) : \Sigma' = b : a$,
 hat man
$ \Sigma + ae/2 b\sin\psi  = ab \psi /2$, 
und somit
$ \Sigma = ab/2 (\psi  - e\sin \psi)$.
 Auf den Mittelpunkt 
bezogen, h\"{a}tte der keplersche Satz vom periodischen Umlauf 
f\"{u}r jeden Zeitabschnitt $\tau$  also
 $\tau  = T/(2\pi)[\psi  - e\sin \psi ]$
 gehei{\ss}en, was sogleich nach Ansetzen 
der Bewegung eine zeitmodulierte Winkelabh\"{a}ngigkeit f\"{u}r das 
Gravitationsgesetz liefert. Zur Zeit von Hooke und Newton stand 
die Frage nach der nat\"{u}rlicheren Gesetzm\"{a}{\ss}igkeit zur Diskussion. 
Weil wir immerhin nur geometrische Betrachtungen zu beobachtbaren 
Bewegungen angestellt haben, kann die Modulation unm\"{o}glich 
auf einen Messfehler zur\"{u}ckgef\"{u}hrt werden. Es folgt, dass 
entweder die Planetenbahnen oder die Lissajous-Ellipsen nicht 
mit konstanter Fl\"{a}chengeschwindigkeit umgelaufen werden, und 
dass dasselbe geometrische Ort (nach Wahl einer Uhr
\footnote{Als Uhr 
meint man eine Pendeluhr. Oberhalb der Erde wirkt auf sie laut 
Newton eine $1/r^2$ proportionale Kraft. Huygens hat bewiesen, 
dass erst das Zykloidenpendel eine von der Schwingungsamplitude 
unabh\"{a}ngige Zeit misst. Ptolemaios hatte aber die Zykloide 
auf die Beschreibung der Planetenbewegung angewendet.}) mit zwei 
verschiedenen Kraftgesetzen verbunden werden kann.

\subsection{
Gesetze zur Drehung der K\"{o}rper}

Die Bewegungen der Himmelsk\"{o}rper sind reversibel und k\"{o}nnen 
rechnerisch bis in vorhistorische Zeitalter zur\"{u}ckverfolgt 
und mit historischen Angaben kontrolliert werden. Es trifft sich 
jedoch, dass die linearen Gleichungen der Mechanik nicht gen\"{u}gen, 
um die Endlage von K\"{o}rperbewegungen unter irdischen Einfl\"{u}ssen 
zu berechnen. Es fragt sich dann, ob nicht unsere kausale Vorstellungen 
der Gravitation hierzu einer feineren analytischen Anpassung 
bedarf.\\
Jedermann wei{\ss}, dass ein ehrlicher Spielw\"{u}rfel, welcher von derselben 
Anfangsstellung gezogen wird, praktisch auf jede seiner Fl\"{a}chen enden 
kann, ohne dass es m\"{o}glich w\"{a}re, daraus seine Anfangslage 
eindeutig wieder zu erschlie{\ss}en. Das beruht einer gel\"{a}ufigen 
Erkl\"{a}rung nach auf den Umstand, dass obschon sich Anfangs- 
und Endstellung jedes Mal experimentell genau beschreiben lassen, 
das angegebene Kraftgesetz f\"{u}r s\"{a}mtliche Punkte in etwa dieselbe 
gerade oder parabolische Bahn vorschreibt. Die gravitative Kraft 
vorbesteht sozusagen, und wirkt augenblicklich nur aufgrund ihrer 
H\"{o}he \"{u}ber dem Boden, wogegen sich w\"{a}hrend des Flugs weitere 
Wirkungen entwickeln k\"{o}nnen. F\"{u}r den jetzt zur Diskussion 
gestellten Fall scheint das Problem darin zu liegen, dass der 
Spieler dem W\"{u}rfel ballistisch auch ein von der Schwungmasse 
abh\"{a}ngiges Drehmoment erteilt. Der Impetus kann aber hart wie 
eine kleine St\"{o}rung behandelt werden. Der W\"{u}rfel rotiert 
n\"{a}mlich unbek\"{u}mmert des stracks ausfallenden Handkontakts 
augenscheinlich weiter, so dass der Effekt einer infinitesimal 
kurzen St\"{o}rung mit der Zeit anw\"{a}chst, bis er betr\"{a}chtlich 
wird. Es steht \textit{uns} frei eine von der Fallbewegung unabh\"{a}ngige 
Ursache f\"{u}r die Kreiselbewegung auszumachen. Man k\"{o}nnte beispielsweise 
behaupten, dass man nun keine genaue geometrische Flugbahn markierter 
Punkte angeben kann, weil elastische Verzerrungen eine wesentliche 
Rolle auf die Bahnen der einzelnen Massenpunkte spielen\footnote{Es 
bleibe dahingestellt, ob die Vorstellung von Unterschieden zwischen 
schwerer und tr\"{a}ger Masse in dieses Bild passt.}. Zieht man 
aber die Koh\"{a}sion heran, so l\"{a}sst sich dem W\"{u}rfel als Ganzes, 
bis er sich wie ein starrer System von Massenpunkten verh\"{a}lt, 
trotzdem kein Drall aus linearen dynamischen Gesetzen zuschreiben. 
Nimmt man statt dessen,  von vornherein \textit{einen} Erhaltungssatz 
an, so l\"{a}sst sich freilich das Torkeln nach dem Loslassen qualitativ 
damit erkl\"{a}ren. Das bedeutet immerhin, dass beim Loslassen 
des W\"{u}rfels gleich eine Kreiselbewegung station\"{a}r, also gesetzm\"{a}{\ss}ig, 
verl\"{a}uft. Soweit, so gut. Der Knobel kommt tats\"{a}chlich jedes 
Mal auf einer unberechenbaren Position zu liegen, wenn markierte 
Punkte auf ihm eben keine im Voraus absch\"{a}tzbare Flugbahn zur\"{u}cklegen. 
Wozu das Erhaltungsgesetz? Es ist in dieser Hinsicht genauso 
stichhaltig zu sagen, dass die W\"{u}rfelbewegung infolge des Schusses 
erst gar keinen station\"{a}ren Zustand erreicht, und dass deshalb 
keine Bahn im Voraus festgelegt werden kann. Wenn man gedenkt, 
dass sowohl das Kraftgesetz wie die Bahnangabe mit demselben 
Gegenstand -- namentlich dem t-parametrisierten Ausdruck der Bewegung 
- befasst sind, und dass der Unterschied nur in der Beurteilung 
seiner Zwangsl\"{a}ufigkeit liegt, muss man zugeben, dass die Mechanik 
nicht immer mit der t-Parametrisierung endlicher Kurvenstrecken 
auskommt. Durch die Unm\"{o}glichkeit eine \textit{feste Bahn} mit t 
zu parametrisieren wird allerdings die Reversibilit\"{a}t der Bewegung 
im Sinn ihrer Umkehrung auf demselben Gleis verhindert, ohne 
die Eindeutigkeit einer bereits durchgelaufenen Flugbahn anzutasten. 
Newton hat anscheinend zwischen t-parametrisierbaren und nicht 
parametrisierbaren Kurven unterschieden, und hat seine Dynamik 
auf die Ersteren gegr\"{u}ndet. Diese Unterscheidung l\"{a}sst keinen 
Schluss auf das Bestehen zweierlei Arten von Bewegungen in unserer 
Welt zu. Vorausgesetzt dem Chaos sei eine physikalische Bedeutung 
erteilt, kann man nicht behaupten, die nicht integrierbaren Bahnen 
entspr\"{a}chen \textit{chaotischeren} Bewegungen.

\textbf{Dynamische Kr\"{a}fte als Vektoren. --} Um diesen \"{U}belstand 
zu beheben, kann man f\"{u}r eine gegebene Massenverteilung ein 
einziges Drehmoment aus der Statik \"{u}bernehmen. Unter den Bedingungen,\\
{\textbullet}\tab 
dass \textit{dynamische Kr\"{a}fte} mittels des Dynamometers, also im 
Zweik\"{o}rpersystem K\"{o}rper + Messger\"{a}t, messbar sind - wozu 
das 3. Kraftgesetz seinen Dienst erweist -, dass also
${\bf F}_D \sim {\bf F}_S$
 dyn gilt,\\
{\textbullet}\tab 
dass man die graphische Statik zur Erkl\"{a}rung einer K\"{o}rperdrehung 
heranziehen kann,\\
{\textbullet}\tab und dass die Folge der Ruhelagen eine stetige Bewegung 
bildet,

l\"{a}sst sich das auf einen W\"{u}rfel wirkende Kr\"{a}ftesystem in 
eine am Schwerpunkt angreifende Resultante und ein Paar bez\"{u}glich 
eines beliebigen weiteren Referenzpunktes zerlegen. Fasst man 
innerhalb der Dynamik die Kraft also als Vektor\footnote{Bis jetzt 
haben wir angenommen, dass dynamische Kr\"{a}fte am mitgef\"{u}hrten 
Massenpunkt angreifen. Statische Kr\"{a}fte sind dagegen linienfl\"{u}chtige 
Vektoren, und k\"{o}nnen deshalb addiert werden.} der Statik auf, 
dann sollte das am torkelnden W\"{u}rfel angreifende Kr\"{a}ftesystem 
momentan durch \textit{eine} Resultante und \textit{ein} Paar zu ersetzen 
sein. Um zur dynamischen Auffassung zu \"{u}bergehen, gibt man 
jeweils die Unterlage an, worauf liegend der Schwerpunkt zur 
Ruhe k\"{a}me. So ist in jedem Augenblick der statische Druck der 
Beschleunigung, welche der haltlose K\"{o}rper erfahren w\"{u}rde, 
\"{a}quivalent. Man bestimmt ferner das Kr\"{a}ftepaar in jedem Augenblick, 
indem man \"{a}quivalente Gegenkr\"{a}fte so ansetzt, dass der bewegte 
W\"{u}rfel in die Gleichgewichtslage versetzt wird. Damit ist das 
Problem auf die Statik zur\"{u}ckgef\"{u}hrt. Hier ist es geometrisch \textit{f\"{u}r 
den Raum} nicht l\"{o}sbar. W\"{a}hrend es ein Leichtes ist Kr\"{a}fte 
graphisch nach der Parallelogrammregel in der Ebene zu addieren, 
weil man Polygonkanten aus einem Zug schlie{\ss}en kann, bestimmt 
man mit Polyederkanten die Gleichgewichtslage nicht. Deshalb 
definieren dynamische Kr\"{a}fte, sofern sie nicht von Anfang an 
zum Zweck der Addition von Vektoren nach einem Dreibein gespaltet 
worden sind, der graphischen Statik v\"{o}llig fremde Gr\"{o}{\ss}en. 
Da keine vektorielle Resultante aus der Statik heraufbeschw\"{o}rt 
wird, hilft unserer Meinung nach auch keine Kraftmixtur: Gravitation 
+ Luftreibung + von der Erdkugelumdrehung verursachten Flieh- 
und zusammengesetzter Kraft.

\textbf{Statistische \"{U}berlegungen zu den Endlagen. -} Da jede wissenschaftliche 
Theorie, als Mittel zum Austausch von Erfahrungen, auf Abmachungen 
seitens der Beteiligten beruht, kann man wohl die oben angef\"{u}hrte 
beschreibende Behandlung durch die \textit{nicht auf naturbedingten 
Zufall}, sondern auf logischen Schl\"{u}ssen fu{\ss}ende (lineare) 
Wahrscheinlichkeitstheorie ersetzen\footnote{Wir Menschen sind eigentlich 
auch Naturerscheinungen, doch wollen wir, weil wir zugleich Objekt 
und Subjekt der angestellten Untersuchung w\"{a}ren, hier die Erforschung 
unserer geistigen F\"{a}higkeiten unterlassen.}. Mit diesem Abkommen 
verf\"{u}gt man sicher \"{u}ber ein neues Mittel zur Verst\"{a}ndigung. 
Inwieweit das auch zu einem bestimmten Ziel \"{u}ber die Mechanik 
hinaus weiterf\"{u}hrt w\"{u}ssten wir nicht zu sagen. Indessen erzielt 
die Statistik keine Linearisierung der beobachteten W\"{u}rfe, 
sondern sie kennzeichnet denselben Gegenstand durch ein \textit{anders 
geartetes Merkmal}, die Augenzahl. Wir versuchen den neuen Aufschluss 
besser zu schildern.

Was das Informationsgehalt angeht, lehrt uns im Rahmen dieser 
Darstellung ein einzelner Zug nichts Neues \"{u}ber die \"{a}u{\ss}eren 
Umst\"{a}nde die es bedingt haben, da das Spielresultat nicht vom 
gesamten Verlauf der Bewegung, sondern nur von der schlie{\ss}lich 
nach oben zu gewandten W\"{u}rfelfl\"{a}che abh\"{a}ngt, was laut newtonscher 
Dynamik mit der Versetzung der beliebig zu w\"{a}hlenden Anfangsbedingungen 
in die station\"{a}re Endlage gleichkommt. Die f\"{u}r einen Spielw\"{u}rfel 
charakteristische Endlage wird zum Aufbau der Statistik mit einer 
von 6 (4, 8, 12 oder 20) Augenzahlen im Voraus identifiziert. 
Die Statistik \"{u}ber eine gen\"{u}gende Anzahl von Z\"{u}gen lehrt 
uns \"{u}ber die Bewegung des achteckigen ehrlichen W\"{u}rfels nichts. 
Die Wahrscheinlichkeit daf\"{u}r, dass er eine Augenzahl von 1 
bis 6 zeigt haben wir n\"{a}mlich aus Symmetriegr\"{u}nden durch 
logisches Schluss\footnote{Das Laplacesche Prinzip des unzureichenden 
Grundes.} auf jeweils 1/6 festgelegt. Also geh\"{o}rt einerseits 
die a priori Wahrscheinlichkeit eine Anzahl Augen zu ziehen nicht 
zu den W\"{u}rfeleigenschaften. Andererseits kann ein Gl\"{u}cksspieler 
durch geschicktes Mogeln\footnote{Dadurch, dass er sich in eine bestimmte 
Weise zu ziehen ein\"{u}bt, und sich die dazugeh\"{o}rige Tabelle 
merkt.} jedes willk\"{u}rliche Resultat erreichen, was nachtr\"{a}glich 
(a posteriori) daran hindert eine statistisch festgestellte mittlere 
Augenzahl ausschlie{\ss}lich mit den dem W\"{u}rfel zugeordneten 
Eigenschaften zu erkl\"{a}ren. Dessen unbeachtet setzt man meistens 
beim Ersetzen der Dynamik mit einer statistischen Theorie voraus, 
dass das wahrscheinlichkeitstheoretisch als Einzelfall betrachtete 
Ereignis sich mit einer aus der Dynamik resultierenden (station\"{a}ren) 
Endlage deckt. \"{U}ber die wahrscheinlichkeitstheoretische Deutung 
des einzelnen Zuges ist viel diskutiert worden. Wie oben verdeutlicht, 
kann man n\"{a}mlich \textit{einen} (bereits gezogenen) Zug eines bestimmten 
W\"{u}rfels beliebig genau beschreiben. Aber alle m\"{o}glichen W\"{u}rfe 
kann man weder aufz\"{a}hlen, noch prinzipiell im Voraus bestimmen. 
Deshalb kann kein Einzelzug als mittlerer Zug in die Wahrscheinlichkeitstheorie 
eingehen. Und umgekehrt kommt kein statistisch berechneter Mittelwert 
als m\"{o}gliche Augenzahl bei einem Wurf vor. Man kann h\"{o}chstens 
ein Kompromiss zwischen Beschreibung des einzelnen Falls (logisches \textit{es 
gibt}) und Wiedergabe des allgemeinen Verhaltens (logisches \textit{f\"{u}r 
alle}) suchen. Das gelingt zum Zweck des W\"{u}rfelns, wenn man 
die auf der nach oben zu gewandten Fl\"{a}che erscheinenden Punktezahl 
als Einzellfall zu betrachten vereinbart. Aus dem Merkmal ,,Punktezahl 
auf der oberen Fl\"{a}che`` l\"{a}sst sich jedoch nicht erschlie{\ss}en, 
wie sich ein ,,Las Vegas`` W\"{u}rfel als Kreisel verhalten mag, 
weil es beim Drehen gar nicht auf die von einem W\"{u}rfel schlie{\ss}lich 
gezeigten Punktezahl ankommt. Dazu merkt man, sobald es darum 
geht aus dem Spielmodell durch logische Schl\"{u}sse auf kinematische 
Befunde n\"{a}herungsweise zu schlie{\ss}en, dass der
$ 0.1\bar 6$-W\"{u}rfel 
weder experimentell existiert noch \textit{logisch verwertbar ist}.

\textbf{Rechnerische Hilfsmittel zur N\"{a}herung der L\"{o}sungen. -} 
Wie bereits erw\"{a}hnt, ist die Beschreibung der Drehungen von 
ausgedehnten K\"{o}rpern um endliche Winkel kaum linear zu bew\"{a}ltigen. 
Vorausgesetzt dem Einzelfall sei irgendwie durch Optimierung 
oder numerische Simulation beizukommen\cite{yao}, 
k\"{o}nnte man auf die Idee kommen, die bei einem Wurf vorherrschenden 
Verh\"{a}ltnisse rein numerisch mit beliebiger Genauigkeit zu ermitteln. 
Heutzutage erlauben in der Tat die rechnerischen Hilfsmittel 
numerische L\"{o}sungen von komplizierten Gleichungen umgehend 
zu bestimmen. Eine mathematische Beziehung zwischen physikalischen 
Gr\"{o}{\ss}en lie{\ss}e sich aber nur best\"{a}tigen, wenn sich das 
Ger\"{a}t, das wir von der angewendeten Software verschieden annehmen 
wollen, einem Algorithmus entsprechend verhielte.

 Die reine \"{U}bereinstimmung 
von berechnetem und gemessenem Wert ist kein Beweis hierf\"{u}r, 
sondern sie ist die notwendige Bedingung zur Durchf\"{u}hrung der 
numerischen Kontrolle. Sie unbedingt voraussetzen zu wollen hie{\ss}e, 
dass Naturgesetze vor ihrer ,,Entdeckung`` schon als mathematische 
Formeln parat daliegen. Indessen st\"{o}{\ss}t die Vorstellung, dass 
z.B. der Teilchenimpuls in Wirklichkeit einen bestimmten Wert 
hat, auch gegen ein Prinzip der Quantenmechanik.
Der Frage, ob sich irgendwelche Zahlensysteme auf Erscheinungen 
zur\"{u}ckf\"{u}hren lassen, so dass die Naturerscheinungen unmittelbar 
mit numerischen Gleichungen zu belegen sind, wollen wir nicht 
nachgehen. Vorgreifend wollen wir aber schreiben, dass sobald 
der Zahlenbereich nicht mit in die Formulierung des Gesetzes 
eingehen muss, die Wahl des Zahlenbereichs frei zur Verf\"{u}gung 
steht, wenn man zur mathematischen Modellierung \"{u}bergeht. Hinsichtlich 
dieser Wahl kann man es dann versuchen, lineare physikalische 
Theorien aufzustellen.

\section{
Algebraische Vektorfelder und Bahnen}

Weil man heute mit nunmehr algebraischen Vektorfeldern vielfach 
sehr abstrakte Wirklichkeiten in Verbindung bringt, wollen wir 
von den moderneren Fragestellungen absehen, und uns zu den Betrachtungen 
aus der Zeit wenden, wo man noch die Wirklichkeit wie sie nun 
einmal ist physikalisch durchdringen zu k\"{o}nnen glaubte. Insbesondere 
m\"{o}chten wir zeigen, dass sich die Frage nach der wirklichen 
Raumstruktur nicht beantworten l\"{a}sst. Dazu versuchen wir ein 
paar \"{a}ltere Feldbegriffe ganz knapp zur\"{u}ckzuverfolgen.
W\"{a}hrend unserer Analyse des Kraftbegriffs, haben wir beil\"{a}ufig 
hinzugef\"{u}gt, dass graphische Darstellungen unter Umst\"{a}nden 
von den geometrischen Ausf\"{u}hrungen abweichen k\"{o}nnen. Dieses 
Problem taucht schon bei Euklid auf, weil er graphische \textit{Existenzbeweise} 
(logisches \ensuremath{\exists}) neben formelleren Begr\"{u}ndungen bringt, 
die Graphik aber, im Gegenteil zur Logik, keinen Anspruch auf 
Allgemeinheit (logisches \ensuremath{\forall}) hat. Infolgedessen st\"{o}{\ss}t 
man, je nachdem ob man von der Statik oder von der Dynamik ausgeht, 
auf verschiedene Ausf\"{u}hrungen.

\subsection{
Wanderungseigenschaften der Punktmassen in statischen Kraftfeldern}

Die klassische Statiklehre ist nicht nur graphisch begr\"{u}ndet, 
sondern immerhin \"{a}lter als die Differentialrechnung selber. 
Sie liefert nach Varignon eine spezielle zeichnerische Abbildung 
der Kraftverh\"{a}ltnisse an gew\"{a}hlten Punkten eines starren 
im Gleichgewicht befindlichen K\"{o}rpers. Darum geht es innerhalb 
dieser Lehre nur um die Wahl der Angriffspunkte \textit{endlicher} 
Kr\"{a}fte auf bereits vorliegenden K\"{o}rpern, und haupts\"{a}chlich 
noch nicht um die Auskundschaftung einer Anregungs\textit{dichte}, 
vorausgesetzt der Aufpunkt sei mit einer \textit{Einheits}masse belegt. 
Man sucht innerhalb dieser Lehre Konzepte f\"{u}r volumenverteilte 
Massendichten und Vakuumfelder umsonst.

Die Kr\"{a}fte werden jeweils als kalibrierte Vektoren auf ein 
Zeichenblatt eingetragen, wobei sich des Gleichgewichts halber 
Wirkung und Gegenwirkung im Angriffspunkt unabwendbar aufheben 
sollen.

\textbf{Zur Erweiterung der Statik auf den Raum.}
 - Zur Sch\"{a}tzung 
des mechanischen Vorteils bei der Anwendung einfacher Maschinen, 
z.B. eines Hebels, berechnet man das Gleichgewicht der ausge\"{u}bten 
Kraft $K_1$  mit der Last $K_2$  als umgekehrtes Verh\"{a}ltnis zu den 
Hebelarmen, $\ell_2  : \ell_1$ . Ein Kr\"{a}ftepaar wird 
anschlie{\ss}end als Verh\"{a}ltnis entgegengesetzt gleicher durch 
denselben Hebelarm $\ell$ verbundener Kr\"{a}fte $K_1$  und
$ K_2= - K_1$  dargestellt. Es ist dann
$ L_1  + L_2  = \ell K_1 + \ell K_2  = \ell (K_1  + K_2 ) = 0$.
 Nennt man alsdann 
$K_i \ell_i  = L_i$
  Moment der Kraft $K_i$ , so gilt f\"{u}r 
den Hebel $L_2  = - L_1$ . Die Summe
$\Sigma _i  L_i$
  aller mit 
den richtigen Vorzeichen genommenen Momente ist im Gleichgewicht 
immer gleich Null, wenn man der Zwangskr\"{a}fte Rechnung tr\"{a}gt. 
Im allgemeinen definiert man aber noch in der Statik das von 
Null verschiedene Moment der an einem Punkt $P$ angreifenden \"{a}u{\ss}eren 
Kr\"{a}fte $K_i$  bez\"{u}glich einem willk\"{u}rlichen doch festen Punkt 
$O$ als
$ L = \Sigma_i  L_i  = \Sigma_i  \ell_i K_i  = \ell K$, 
wobei $K$ die Summe i eingepr\"{a}gter Kr\"{a}fte ist. In dem eben 
geschriebenen Ausdruck wird $\ell K$ als Plangr\"{o}{\ss}e verstanden, 
und man zeigt, dass Plangr\"{o}{\ss}en als \textit{orientierte Fl\"{a}chen} 
Kraft mal Hebelarm summierbar sind. Dasselbe statuiert man f\"{u}r 
deren \textit{Erg\"{a}nzungen}, d.h. die freien (axialen) Vektoren
${\bf M}_i = {\bf \ell}_i \times {\bf K}_i$.
 Folglich unterscheiden 
sich Kr\"{a}fte und Momente auch wenn beide graphisch als Vektoren 
dargestellt sind. Sollten Kr\"{a}fte\cite{hertz1884} einerlei sein, dann hinge das 
Problem der Polyederkanten innerhalb der graphischen Statik mit 
der Summierbarkeit von axialen und polaren Vektoren zusammen. 
Daran ankn\"{u}pfend diskutierte A. M\"{o}bius graphische Konstruktionen 
in Verbindung mit der Dualit\"{a}t\cite{mobius1886}.
 Es gelang ihm die statische Kraftdarstellung 
\"{u}ber die Liniengeometrie auf den Raum zu erweitern, was wir 
auf sp\"{a}ter aufschieben. Die Vektoranalyse nimmt Notiz davon, 
zumal diese Wendung direkt zum stokesschen Theorem f\"{u}hrt.

\textbf{Die Entwicklung der Verpflanzung aus der Statik. --} Euklid 
hat niemals kongruente Bewegungen in die Geometrie eingef\"{u}hrt. 
In der Statik sind \textit{virtuelle infinitesimale Verschiebungen} 
blo{\ss} zur Behandlung der Gleichgewichtslage von Ketten, d.h. 
von teilbaren Systemen, eingef\"{u}hrt worden. Der Stetigkeitsbegriff, 
wie er zwecks der Infinitesimalrechnung ben\"{o}tigt wird, hat 
Newton erdacht.\\
Die \textit{Vorstellung laut dem Prinzip der virtuellen Geschwindigkeiten} 
l\"{a}sst sich aus dem Begriff von stabiler Gleichgewichtslage 
herausarbeiten, und dient der mathematischen Charakterisierung 
eines Feldes zur linearen Ordnung in der \ensuremath{\delta}-Umgebung 
eines Punktes\textit{.} Virtuelle infinitesimale Verschiebungen aus 
der Gleichgewichtslage sind, weil die Reaktionskr\"{a}fte unbekannt 
bleiben, zur Kennzeichnung der stabilen Gleichgewichtszust\"{a}nde 
eingef\"{u}hrt worden. Gelegentlich bezieht man die newtonsche 
zur lebendigen Kraft folgenderweise:\\
 ${\bf F}(\ell) \delta \ell = m d^2\ell /dt^2 {\partial \ell\over\partial t} \delta t =
 {\bf K}\delta t$,
 wobei ${\bf K}$ die lebendige Kraft,
$ {\bf F}(\ell) \delta \ell$ 
die vom \textit{linienfl\"{u}chtigen} Kraftvektors
${\bf F}(\ell)$ 
virtuell zu leistende Arbeit\footnote{Die Arbeit ist f\"{u}r die Zwecke 
der Thermodynamik mit einer angeblich experimentell begr\"{u}ndeter 
Erhaltung der Energie, wie z.B. die Unm\"{o}glichkeit des Perpetuum 
Mobile, in Zusammenhang gebracht worden.}, $\delta \ell$ \"{u}berdies 
eine beliebige von jedem Bahnbegriff gel\"{o}ste und ausschlie{\ss}lich 
mit den Bindungen vertr\"{a}gliche kleine Verr\"{u}ckung meint. Noch 
M\"{o}bius und Pl\"{u}cker waren der Meinung, dass sich die f\"{u}r 
die Statik geltende Beziehung zwischen Lage eines Punktes und 
daran angreifende Resultierende ohne Weiteres an die sogenannte 
materielle Bewegungsvorstellung anschlie{\ss}en lie{\ss}e. Sie glaubten, 
dass die Kennzeichnung eines Punktes $P$ durch seine Lage
 $x(t), y(t), z(t) \forall t = t_0$
 eine duale, d.h. zweifache Deutung 
in demselben Bezugssystem zulie{\ss}e: einmal galten die Koordinaten 
den auf ihn wirkenden Kraft und Moment, das andere mal dienten 
dieselben dessen infinitesimalen Drehung und Verschiebung respektive. 
Soll aber etwa
 ${\bf F}_D \equiv {\bf F}(\ell)$
 gesetzt werden? 
Bei Newton f\"{u}hrt die Bedingung, dass das Kraftgesetz einer 
gew\"{o}hnlichen Differentialgleichung
 ${\bf F}_D dt = m d{\bf v}$,
 mit
 ${\bf v} = d{\bf r}/dt$
 gen\"{u}gen soll, zu folgenden Ans\"{a}tzen. Es ist
$ {\bf F}_D = {\bf F}_D ({\bf r})$
 f\"{u}r eine aus einem Potential abgeleitete rein \textit{lageabh\"{a}ngige} 
Kraft, oder
$ {\bf F}_D  = {\bf F}_D (t)$
 f\"{u}r eine Kraft deren Zeitabh\"{a}ngigkeit 
nach Art der Maschinendynamik \textit{vorgeschrieben} ist\footnote{Von 
einer beliebigen Zeitabh\"{a}ngigkeit kann man nicht annehmen, 
dass sie gesetzlich sei}. Weil
$ {\bf F}_D  = {\bf F}_D ({\bf v})$
 f\"{u}r 
eine sich gegen das Beharrungsverm\"{o}gen auswirkende Reibungskraft 
g\"{a}lte, ist es um die Vertr\"{a}glichkeit dieses Ausdrucks mit 
der Galilei-Invarianz nicht gut bestellt.

\textbf{Zum dynamischen Fundament der Ausdehnung.} - Da die Differentialrechnung 
mathematisch begr\"{u}ndet ist, und die Mathematik ihrerseits allgemein 
gehalten wird, kann man fragen, inwiefern die dynamischen Kraftgesetze 
nach Integration die Raumform aus der Bewegung zu bestimmen erlauben. 
Die Frage passt zu dem Gedanke, dass es haupts\"{a}chlich auf die 
Abstrahierung aus unserer Welt einer logisch-mathematischen Struktur 
der Wirklichkeit ankomme, welche im Nachhinein direkt best\"{a}tigt 
oder wiederlegt werden k\"{o}nne. In diesem Sinn entspricht ein 
algebraisches Feldgebilde als ausgedehnte Mannigfaltigkeit der 
Punkte $P = (x, y, z)$ einer Isomorphie zu \textit{gemessenen} Gr\"{o}{\ss}en.

\textbf{Transport. --} Unter Transport verstehen wir eine gerichtete 
Bef\"{o}rderung von Massenpunkten. Der Lokalisierbarkeit eines 
bestimmten Massenpunktes, sagen wir mal auf einer Kettenlinie, 
liegt freilich die Voraussetzung zugrunde, dass sich die Kette 
in einer Gleichgewichtslage befindet. Man k\"{o}nnte nun geneigt 
sein zu glauben, dass ein von festsitzenden anziehenden Massen 
bestimmtes algebraisches Vektorkraftfeld die Bewegung eines jeglichen 
in ihm befindlichen Massenpunktes auf \"{a}hnliche Weise zu bestimmen 
erlaube. Obschon die Bewegung kausal erfolgt, unterscheidet sich 
die \textit{newtonsche Vorstellung der gesetzm\"{a}{\ss}igen Bahn} dennoch 
von einer Kette, weil die genaue Ortkurve erst mit den Anfangsbedingungen, 
und eigentlich nur mit ihnen festgelegt wird.\\
Sollte sich das Problem auf die anschauliche Ermittlung einer 
mit der Beziehung
 ${\bf F} = md^2{\bf r}/dt^2$
 in einem Feld erkl\"{a}rten 
Ortkurve allein belaufen, dann w\"{u}rde das Potentialfeld die 
Aufgabe l\"{o}sen. Es f\"{u}hrt der Ausdruck:
$ f(x, y) dx + g(x, y) dy = dV$
 zwar zu einem von der partikul\"{a}ren Bahn zwischen $a$
und $b$ unabh\"{a}ngigen Wert des bestimmten Integrals $V_{ab}$. Aber 
Felder d\"{u}rfen, wenn f\"{u}r sie etwa
 ${\bf Fv} dt = {\bf F} d{\bf r}$,
 ${\bf v} = d{\bf r}/dt$
 gilt, durch eine Potentialfunktion ersetzt werden. 
Es wird dabei vorausgesetzt, dass die \"{A}nderung des Potentials 
$V$ beim Fortschreiten der Probe von einem zum benachbarten Aufpunkt 
um
 $dV = d{\bf r} \nabla V$,
 mit
$ d{\bf r} = \pm {\bf v} dt$,
 zu- 
oder abnimmt. M\"{o}bius hat diese Gleichungen im Paragraph ,,Analogie 
zwischen dem Gleichgewichte an einem Faden und der Bewegung eines 
Punktes`` erstellt.\\
Dazu soll man noch den Punkt auf der Kurve fortlaufend orten. 
Weil Potentiale immerhin Punktfunktionen r\"{a}umlicher Argumente 
sind, sieht es so aus, als ob man das k\"{o}nnte.

\subsection{
Ergr\"{u}ndung des Weltraumes aus der Bewegung}

Wir haben im vorigen Kapitel zu schildern versucht, wie Newton 
den Bewegungen zur prinzipiellen Messung von Strecken auf Kreisumf\"{a}ngen, 
begegnet ist. Man k\"{o}nnte zugespitzt sagen, dass er, nachdem 
der altgriechische Begriff der Bewegung als einer eventuell unendlich 
fein zerlegbaren Folge von z\"{a}hlbaren gegenw\"{a}rtigen Stellungen 
an dem ber\"{u}hmten Paradox von Achilles und der Schildkr\"{o}te 
gescheitert war, f\"{u}r Massenpunkte einen stetigen Zeitverlauf 
aus der Differentialrechnung heraus \textit{neu} interpretiert hat. 
Aber das neue Verfahren l\"{o}st nur zum Schein das alte Paradoxon 
\"{u}ber Raum und Bewegung. Eine Ortkurve kann erst dann als Flugbahn 
angesehen werden, wenn sie als partikul\"{a}re L\"{o}sung vorliegt. 
Sofern mit L\"{o}sung eine einmal iterierte Quadratur des Fundamentalsatzes 
gemeint ist\footnote{Die Eindeutigkeit der Bestimmung h\"{a}ngt mit 
dem Ausschluss singul\"{a}rer Integrale zusammen.}, stellt sie 
immerhin ein K\"{u}rzel f\"{u}r den Inbegriff aller zweiparametrigen 
Kurven $r(t)$ dar, wovon nur eine der beobachteten Flugbahn 
entspricht. Die Anfangsbedingungen dienen dazu, diese einzige 
Bahn aus der Klasse der L\"{o}sungen auszusondern. Ansonsten sind 
sie nur deshalb f\"{u}r einen bestimmten Massenpunkt charakteristisch, 
weil sie \textit{beliebig w\"{a}hlbare} Zahlen sind. Als solche geh\"{o}ren 
sie aber dem Gesetz nicht an \footnote{Deshalb sind Punktmassen Fremdlinge 
in jeder Feldtheorie.}. Sind sie etwa experimentell festzulegen?

\textbf{Die Beziehung algebraischer Felder zum Kosmos. -} Fassen wir 
das Problem der Bahnbestimmung gravitierender Massenpunkte wieder 
ins Auge und lassen wir die f\"{u}r praktische Berechnungen unumg\"{a}ngliche 
St\"{o}rungsrechnung beiseite. In den Kraftausdr\"{u}cken meinen 
die Ableitungen, d.h. die Grenzwerte der Differentialquotienten 
exakte endliche Werte. So sind die Differentiale selber in diesem 
Limes genau gleich Null, und man kann das L\"{a}ngenma{\ss} einer 
Strecke nicht mit dem Stetigkeitsaxiom beglaubigen. Werden aber 
die Differentialgleichungen f\"{u}r Massendichten aufgeschrieben, 
dann werden trotzdem infinitesimale Ausdr\"{u}cke bis zur ersten 
Ordnung keine archimedeischen Gr\"{o}{\ss}en. Mithin wird die von 
Newton zur L\"{o}sung des Paradoxes von Achilles und der Schildkr\"{o}te 
erfundene Monotonie der Bewegung, sobald sie mit der Raumausdehnung 
verbunden wird, gest\"{u}rzt.

Die Erkundung eines algebraischen Feldes \"{u}ber eine \textit{mitgef\"{u}hrte} 
Sonde l\"{a}sst sich auch aus einem anderen Grund nicht mit den 
newtonschen Vorstellungen identifizieren. Gesetzt die Sonde vertrete 
jeden beliebigen K\"{o}rper, mindestens was dessen Bewegungsmerkmale 
betrifft, kann man das Potential dahin verstehen, dass sein Vorhandensein 
den ,,physikalischen Raum`` verzerrt. Auf die Verzerrung kann 
man, wenn man laut Riemann eine lineare, lokal dem Raum isomorphe 
Mannigfaltigkeit definiert, dank der Differentialgeometrie R\"{u}cksicht 
nehmen\footnote{Fraglich bleibt freilich dabei, ob sich die Raumstruktur 
auch im Gro{\ss}en ausmachen l\"{a}sst. Allerdings haben Helmholtz 
und S. Lie bewiesen, dass kongruente Bewegungen im mechanischen 
Sinn nur in R\"{a}umen von konstanter Kr\"{u}mmung stattfinden k\"{o}nnen.}. 
Die Verzerrung zu beweisen kann unm\"{o}glich das Anliegen Newtons 
gewesen sein, denn er hielt den euklidischen geometrischen Raum 
just f\"{u}r das mathematische Abbild unserer Welt.

W\"{a}re die Verkn\"{u}pfung zwischen \textit{Raumvorstellung} und \textit{Bewegung} 
leicht \"{u}berschaubar, dann lie{\ss}e sich letzten Endes der Grund 
nicht entwirren, aus welchem die Altgriechen, von derselben Raumvorstellung 
ausgehend, -- Euklid war schlie{\ss}lich Grieche -- zu einer aus 
ihrer eigenen Sicht irrigen Auffassung der Bewegung gekommen 
w\"{a}ren.

\subsection{
Mathematische Physik}

Unter dem Stichwort ,,mathematische Physik`` verstand Newton die 
Behandlung mathematischer Systeme von \textit{Einzelk\"{o}rpern}, die 
den jeweils in Natur anzutreffenden Systemen \"{a}hnlich sind. 
Dazu meinte er, dass sich der mathematisch definierte Raum mit 
der beobachteten Au{\ss}enwelt decken soll. Wie er wusste, l\"{a}sst 
sich trotzdem die Existenz der von ihm vorausgesetzten zentripetalen 
Kraft nicht pragmatisch \"{u}ber die geometrische Konstruktion 
der Bahnen beweisen. Nun hat Maxwell, im Gegenteil zu ihm, der 
zur Deutung der \textit{beobachteten} Bewegung - d.h. der partikul\"{a}ren 
Integrale - das Kraftgesetz aus heiterem Himmel vorgeschrieben 
hat, seine \textit{allgemeinen} elektromagnetischen Gesetze aus den \textit{Beobachtungen} 
Faradays zusammengestellt, w\"{a}hrend seine Interpretation \textit{s\"{a}mtlicher} 
L\"{o}sungen in der Luft h\"{a}ngen geblieben ist. Welches Wirklichkeitsgehalt 
will man den L\"{o}sungen seiner Gleichungen zuschreiben? Inwiefern 
gibt es in Natur \textit{Einzelsignale}? Soll man f\"{u}r sie denselben 
Raum wie in der Mechanik zugrundelegen? Das l\"{a}sst sich in diesem 
File nicht ersch\"{o}pfen. Allerdings besch\"{a}ftigen wir uns mit 
diesen Problemen, weil wir hoffen, dass eine allgemeine lineare 
Theorie partikul\"{a}rer Signale aufgestellt werden kann. Erweist 
sich etwas in dieser Richtung als m\"{o}glich, so beruht es darauf, 
dass man den Signalen zum Gegenteil von K\"{o}rpern keine volle 
Individualit\"{a}t zumutet. Es folgen vorerst einige uns bekannte 
Deutungsans\"{a}tze.

\textbf{Zu einigen \"{u}blichen Deutungen der elektromagnetischen Felder.} 
- Wird der stilisierte K\"{o}rper, d.h. der Massenpunkt der Mechanik 
einfach durch die Ladung in demselben Raum ersetzt, dann redet 
man von klassischer Elektrodynamik, einer f\"{u}r \textit{\ensuremath{\pm} 
elektrisch geladene Punkte} geltenden rationalen Dynamik. Sobald 
man aber den elektrischen Strom der Bewegung von mehreren elektrischen 
Stromtr\"{a}gern gleichsetzt, werden sowohl das galileische als 
auch das newtonsche Prinzip verletzt. Das kommt weil die maxwellschen 
Gleichungen, wenn sie nach Analogie mit denjenigen der Dynamik 
interpretiert werden, \textit{zeitabh\"{a}ngige} Kr\"{a}ftedichten aufweisen 
und, speziell was die Induktion betrifft, von der Geschwindigkeit 
der Ladungstr\"{a}ger im Leiter abh\"{a}ngen. Man kann nur sagen, 
dass sie sich entkoppeln lassen, wenn die Zeitver\"{a}nderung des 
Systems gegen Null strebt, d.h. wenn man der Induktion mittels 
der St\"{o}rungsrechnung gerecht werden kann.

F\"{u}r die elektrodynamischen Integrale weist man nach Helmholtz 
nach, dass sich die allgemeinsten Felder aus der Summe eines 
skalaren Potentials (${\bf Fv}$) und eines Vektorpotentials
($\bf v \times F$) 
berechnen lassen. Die Feldgleichungen bestimmen somit, wenn ${\bf v}$ 
klein ist, die elektrodynamische Bewegung als geschwindigkeitsabh\"{a}ngige 
Verallgemeinerung der dynamischen Bewegungsgleichungen. Von einem \textit{allgemeinen 
Feldintegral} kann indes schon deshalb nicht die Rede sein, weil 
nach Voraussetzung die vorgegebenen Ladungen und Str\"{o}me das 
Feld bestimmen. Zu ihm gesellen sich die von den Ladungen unabh\"{a}ngigen 
sich fortpflanzenden elektromagnetischen Schwingungen. Sie scheinen 
das allgemeine Integral zu ergeben. Werden nun die Schwingungen 
mechanisch gedeutet, dann muss man aber notd\"{u}rftig annehmen, 
dass die Felder auch auf pauschal \textit{neutrale Materie} wirken. 
Die dann von physikalischer Sicht notwendig erscheinende Erweiterung 
des $\bf E$-$\bf H$-Feldes um ein rein gravitatives Feld erschwert 
sowohl die St\"{o}rungsrechnung als ihre dynamische Deutung erheblich. 
Rein mathematisch bleibt ungekl\"{a}rt, ob die Summe von quellenbedingtem 
und freiem Feld das \textit{allgemeine} Integral bildet.\\
Die Deutung des freien elektromagnetischen Feldes ist kein einfaches 
Unterfangen. Lagrange, dem wir die energetische Felddeutung verdanken, 
war sich bewusst, dass die r\"{a}umliche Konfiguration eines physikalischen 
Systems nur m\"{u}hsam aus den verallgemeinerten Bewegungsgleichungen 
zu entflechten ist, so hat er sich urspr\"{u}nglich ausschlie{\ss}lich 
mit solchen Potentialfunktionen befasst, die Punkt f\"{u}r Punkt 
eine Kinetik abzuleiten gestatteten. Wenn man dieses Bild verallgemeinert, 
und zum \"{a}thererf\"{u}llten Raum voranschreitet, erteilt man den 
einzelnen Raumpunkten sowohl elastische Energie, als auch Geschwindigkeitskoordinaten 
und man tut, als ob man ohne weiteres so viele Gleichungen zwischen 
den Kr\"{a}ften niederschreiben k\"{o}nnte, wie es Verbindungslinien 
zwischen je zwei elastisch gekoppelten Punkten gibt. Obwohl das 
Feld anschlie{\ss}end energetisch als Oszillatorfeld gedeutet wird, 
besteht neben dem Kollektivum die Idee weiter, dass die einzelnen 
Punkte kleine Schwingungen um Gleichgewichtslagen durchf\"{u}hren. 
Sollte jedenfalls ein mit elektromagnetischer Energie erf\"{u}lltes 
Gebiet wirklich ins Schwingen geraten, w\"{u}rde es trotzdem laut 
jeder mechanischen Auffassung eine unheimliche Energiemenge speichern. 
Ob den normalen Schwingungsmoden, wie sie aus der Fourierentwicklung 
hervorgehen, immer ein mechanisches Gegenst\"{u}ck nach demselben 
Kriterium zukommt, ist schon deshalb angeblich hart zu entscheiden.

Neben der Formulierung \"{u}ber freie Koordinaten mit energetischer 
Interpretation verdanken wir Lagrange ein weiteres, noch abstrakteres 
L\"{o}sungsverfahren, das darin besteht als formelle L\"{o}sung einer 
Gleichung je Freiheitsgrad eine Potenzreihe
$ f(z) = \Sigma_k c_k z^k$,
 $c_k \in \mathbb{C}$
 anzusetzen\footnote{Dieses Verfahren 
findet auf die ,,Wellengleichungen`` Anwendung. Es k\"{o}nnte sich 
genauso gut f\"{u}r die maxwellschen Gleichungen eignen.}. Es ist 
$c_k  = 1/(2\pi i) \oint_R f(\zeta)/\zeta^{n+1}$ 
d\ensuremath{\zeta}, wobei $R$ etwa ein ringf\"{o}rmiges Bereich im Innern 
eines Kreisringes sein mag. Auf diese Weise erh\"{a}lt man, sofern 
die Konvergenz gesichert ist, \textit{partikul\"{a}re} L\"{o}sungen $f(z)$ 
mit dem Vorteil gegen\"{u}ber Helmholtz, dass Feld- und Wellenanteil 
nicht weiter unterschieden zu werden brauchen. Man beruft sich 
bei der sich hieran anschlie{\ss}enden Interpretation der Terme 
der formalen Potenzreihenentwicklung meistens stracks auf Intuition. 
D.h. man setzt 
$k \in \mathbb{N} \cup \{0\}$,
 und man deutet 
die Entwicklung kurzum als Summe aller m\"{o}glichen Schwingungsfrequenzen 
des dargestellten Systems, multipliziert mit ihren jeweiligen 
Amplituden. Wie es in der Fourieranalyse Brauch ist, schreibt 
man folglich:
$ f(z) = \Sigma_{k=0} c_k z^k = u + iv = c_0 
+ c_1 r(\cos \vartheta  + i \sin \vartheta ) + c_2 r^2(\cos 2\vartheta  
+ i \sin 2\vartheta ) + ... = \frac12 a_0 + \Sigma_{k=1} 
r^k (a_k  \cos k\vartheta  + b_k  \sin k\vartheta ) - \frac12 i b_0 
+ i\{\Sigma_{k=1} r^k (a_k  \sin k\vartheta  - b_k  \cos k\vartheta )
\rightarrow$
 \footnote{$r=1$ 
soll innerhalb des Konvergenzkreises fallen.}
$ u(\vartheta )_{r=1} = a_0 + \Sigma_{k=1} a_k  \cos k\vartheta  + \Sigma_{k=1} 
b_k  \sin k\vartheta  \rightarrow \Sigma_{k=0} a_k  \cos k\vartheta$. 
Das ist offensichtlich kein partikul\"{a}res Integral nach dem \textit{allgemeinen 
Verfahren}, wovon die Tatsache, dass der urspr\"{u}ngliche lineare 
Zeitverlauf aus dem sogenannten Frequenzspektrum \textit{nicht mehr} 
erlangt wird, Zeugnis ablegt. Wiewohl also die eingef\"{u}hrte 
Einschr\"{a}nkung des $k$-Zeigers physikalisch gerechtfertigt erscheint, 
so l\"{a}sst sich trotzdem fragen, \textit{welche} Menge man f\"{u}r die 
Gesamtheit aller m\"{o}glichen Systeme der Zahlenwerte $\{c_k \}$ 
in der Reihenentwicklung zulassen will. Das bernouillische Prinzip 
sagt n\"{a}mlich aus, dass wenn man \"{u}ber eine \textit{vollst\"{a}ndige 
Gruppe} von Fundamentall\"{o}sungen verf\"{u}gt, man \textit{jede} partikul\"{a}re 
L\"{o}sung erhalten kann. Da jeder mit einer Frequenz zu kennzeichnende 
partikul\"{a}re Zustand \textit{station\"{a}r} hei{\ss}t, weil er erst nach 
einem Anlauf eintreten muss und davon abh\"{a}ngt, m\"{u}ssen wir 
allerdings feststellen, dass das Energiespektrum i.allg. mit \textit{keiner} 
partikul\"{a}ren L\"{o}sung der gegebenen Differentialgleichung \"{u}bereinstimmt. 
Ob und inwiefern die analytische L\"{o}sung, wenn wir das betrachtete 
,,Frequenzintervall`` von - \ensuremath{\infty} bis + \ensuremath{\infty} erstrecken, 
einer partikul\"{a}ren \textit{Transiente} gerecht wird, lassen wir 
vorerst noch offen.

Fazit. Die Schreibweise der sogenannten allgemeinen L\"{o}sung 
der laplaceschen oder poissonschen Gleichung als \"{U}berlagerung 
trigonometrischer Funktionen setzt - bis man jedes Reihenglied 
als physikalisch begr\"{u}ndet h\"{a}lt -- voraus, dass beobachtete 
fl\"{u}chtige Prozesse \"{u}berhaupt nicht mit elementaren materiellen 
Zerfallsprozessen, falls es diese gibt, verbunden sind. Also 
erfolgt jede beobachtete Signalabklingung eventuell trotz der 
Stabilit\"{a}t der Materiebauteile, oder aber das Signal ist von 
sich aus periodisch, nur wiederholt es sich nach einem unmessbar 
langen Zeitabschnitt.\\
Wir erw\"{a}hnen noch, dass man mit einer hydrodynamischen Umdeutung 
der von Maxwell aufgestellten Gleichungen unter der Einwendung, 
dass sich die Elektrizit\"{a}t wie eine stetige Str\"{o}mung verhalte 
zum Ergebnis kommt, dass die beiden die stofflich gedachten Ladungen 
und Str\"{o}me enthaltenden Gleichungen die Felder bestimmen, w\"{a}hrend 
die beiden anderen, darunter die elektromagnetische Induktion, 
zur mathematischen Festlegung der Potentiale dienen. Wozu ist 
das gut? Maxwell hat letzten Endes unumwunden elektrische Versuchsergebnisse 
interpretiert. Die Hydrodynamik bezieht zwar ihre Vorstellungen 
reichlich aus dem Verhalten str\"{o}mender Fl\"{u}ssigkeiten, aber 
sie hat sich weitgehend selbstst\"{a}ndig als rationale energetisch 
gepr\"{a}gte Verallgemeinerung der Lehre der substantiellen Bewegung 
entwickelt. Auch die sich an diese Entwicklung anschlie{\ss}ende 
hydrodynamische Interpretation der elektrischen Erscheinungen 
gr\"{u}ndet nicht sowohl auf Einblicke aus der Hydrologie, als 
vielmehr auf den Umstand, dass der Satz von Gleichungen ein und 
derselbe ist, wobei die Gau{\ss}- und Stokes-Theoreme es verhindern, 
dass mit den Integrationsfl\"{a}chen Randbedingungen f\"{u}r die 
elektromagnetischen L\"{o}sungen im Vakuum gekn\"{u}pft werden\footnote{Bisweilen 
werden anschlie{\ss}end die Randbedingungen des nicht entkoppelten 
elektromagnetischen Systems mit den fresnelschen Beziehungen 
f\"{u}r statische Felder in Zusammengang gebracht.}. Sollte sich 
eines Tages herausstellen, dass es mit lauter Antennen als Quellen 
des elektromagnetischen Feldes\cite{jackson}
schon klappt, w\"{u}rde man vermutlich versuchen die Grenzfl\"{a}chen 
auf lineare nicht-isotrope Strahler zu beziehen.\\
Als nach H. Hertz die Existenz sich fortpflanzender und abklingender 
elektromagnetischer Wellen weitgehend akzeptiert wurde, wurde 
man einer zus\"{a}tzlichen M\"{o}glichkeit gewahr. Man nahm wieder 
seine Zuflucht zur Potentialgleichung\cite{riemann}.
 Aber die partikul\"{a}re L\"{o}sung wurde 
in einer gewissen Hinsicht verallgemeinert, da man sie nunmehr 
als durchschnittliches Verhalten vieler Systeme deutete. Wie 
tr\"{u}gerisch diese Annahme sein kann, zeigt schon die einfache 
\"{U}berlegung, dass Mittelwerte von Schwingungsfrequenzen das 
Verhalten keines partikul\"{a}ren Oszillators mehr beschreiben.

\textbf{Mathematisches zu den Integralen des elektromagnetischen 
Gleichungssystems. -} Bei dem von Maxwell aufgestellten gekoppelten 
Gleichungssystem, das in Vektorrechnungsschreibweise:
$ \nabla  \times \textbf{E} = - 1/c \partial \textbf{B}/\partial t$,
$ \nabla  \times \textbf{H} = (4\pi/c) \textbf{J} + 1/c \partial \textbf{D}/\partial t$,
$ \nabla \textbf{D} = 4\pi\rho , \nabla \textbf{B} = 0$
 lautet, treten vom 
mechanischen Standpunkt aus, \textit{der expliziten Zeitabh\"{a}ngigkeit 
wegen}, allgemeinere Bahnen auf, darunter Schraubungen und nicht 
als St\"{o}rungen zu verdr\"{a}ngende abklingende Raum-Zeitverlaufe. 
Das Gleichungssystem l\"{a}sst sich dennoch im Vakuum auf zwei 
einfache Weisen entkoppeln:\\
{\textbullet}\tab 
im zeitunabh\"{a}ngigen Fall erh\"{a}lt man:
$ \nabla  \times \textbf{E} = 0, \nabla \textbf{E} = 4\pi\rho$,
$ \nabla  \times \textbf{B} = (4\pi/c) \textbf{J}$,
$ \nabla \textbf{B} = 0$
\footnote{Es ist dabei zu 
beachten, dass Richtungsdifferentialquotienten keine Differentiale 
sind. Nablasymbole sind in dieser Hinsicht keine Kurzformen, 
sondern dr\"{u}cken bereits das Vorhandensein einer Potentialfunktion 
aus.}.\\
{\textbullet}\tab 
f\"{u}r quell- und wirbelfreie Bereiche erh\"{a}lt man, wenn
$ \rho = \textbf{J} = 0$
 sind:
 $\Delta \textbf{E} - 1/c^2 \partial ^2\textbf{E}/\partial t^2 = 0$,
$ \Delta \textbf{B} - 1/c^2 \partial ^2\textbf{B}/\partial t^2= 0$.

Wenn das System mathematisch entkoppelt wird, verf\"{u}gt die Physiklehre 
im ersten Fall \"{u}ber (inhomogene) poissonsche und zugeh\"{o}rige 
(homogene) laplacesche\footnote{Wenn man kein vom Leiter berandetes 
magnetisches Blatt zuzieht, um gleichsam auch bildhaft die Umwicklung 
des Integrationsweges um den Leiter zu verhindern, existiert 
das magnetische Potential im \"{u}blichen Sinn zun\"{a}chst nicht.} 
skalare Gleichungen. Legt man komponentenweise laplacesche bzw. 
poissonsche Gleichungen auch f\"{u}r vektorielle Potentiale bzw. 
f\"{u}r Felder zugrunde, dann kann die L\"{o}sung m\"{u}helos um ein 
der Zeit proportionales Argument erweitert werden. Trifft man 
die richtige Wahl f\"{u}r den Proportionalit\"{a}tsfaktor, dann kann 
man die beiden F\"{a}lle gleich vereinigen, indem man mit dem d'alembertschen 
Operator
$\square = \partial^\mu\partial_\mu$
 ($\mu = 1, 2, 3, 4$) f\"{u}r die homogenen Feldgleichungen respektive
$\square\textbf{E} = 0$
 und
$\square\textbf{B} = 0$
 setzt. Da die mathematische Klassifizierung 
der Integrale als stehende bzw. fortlaufende Wellen nur von dem 
Vorzeichen der Diskriminante abh\"{a}ngt, folgert man aus dieser 
Schreibweise, dass auch fortlaufende Wellen station\"{a}r zu sein 
haben. Ob man folglich zwischen komponentenweise Potential\"{u}bertragung 
und Fortpflanzung von Wellenfronten unterscheiden will oder nicht, 
ergeben sich, wie Maxwell selber gezeigt hat, in ladungs- und 
stromlosen Bereiche formell in etwa dieselbe Gleichungen.

Unterschiede zwischen den beiden F\"{a}llen k\"{o}nnen nur dann eintreten, 
wenn sich zeitabh\"{a}ngige, ged\"{a}mpfte bzw. einschwingende Verhalten 
entwickeln. Mit deren dynamischer Interpretation hat es aber 
seine eigene Bewandtnis: Es scheint uns indessen ungekl\"{a}rt, 
ob vorauseilende und \textit{verz\"{o}gerte} Wirkungen begrifflich noch 
mit direkten kausalen Kraftvorstellungen in Einklang zu bringen 
sind.

\textbf{1. Fall der Entkopplung von E und B.} -- Wir verabreden jetzt 
diesen Fall einfach durch Tilgung der t-Variable zu erhalten. 
Die elektromagnetischen Feldgleichungen sehen dann, abgesehen 
davon dass sie vektoriell sind, ungef\"{a}hr wie die Potentialgleichung 
der Gravitationstheorie aus. Mathematisch enthalten aber die \textit{allgemeinen 
unbestimmten} Integrale partieller Differentialgleichungen erster 
Ordnung, im Gegensatz zu den gew\"{o}hnlichen Differentialgleichungen, 
eine unbestimmte Funktion \ensuremath{\phi}. Grob gesprochen steigt man 
von einem partikul\"{a}ren Feld $\varphi(x, y, z)$ zum allgemeinen 
Integral \ensuremath{\Phi} einer homogenen partiellen Differentialgleichung, 
indem man $\Phi = \phi[\varphi]$
 setzt, wobei man 
mit \ensuremath{\phi} eine beliebige stetig abbildende Funktion meint. 
Da man \ensuremath{\varphi} meistens keine Bild\textit{kurve} zuordnet, stellt \ensuremath{\Phi} 
tats\"{a}chlich eine unendliche Klasse von L\"{o}sungen dar. Mit 
den inhomogenen partiellen Differentialgleichungen verf\"{a}hrt 
man im Prinzip auf \"{a}hnliche Weise, nur dass die Anzahl der 
voneinander unabh\"{a}ngigen partikul\"{a}ren L\"{o}sungen zunimmt.

Die Schwierigkeit besteht allemal darin, das mehrfache partikul\"{a}re 
Integral \ensuremath{\varphi} niederzuschreiben. F\"{u}r \textit{begrenzte} Bereiche 
stehen dazu Iterationsverfahren zur Verf\"{u}gung. Weil die Integrationsgrenzen 
Funktionen sind, d.h.:\\
$\varphi (x, y, z) = \int_{x^0}^x \int_{y^0(u)}^{y(u)} \int_{z^0(u,v)}^{z(u,v)} \chi (u, 
v, w) du dv dw$
versucht man sich im Raum, wie angedeutet, auf Normalbereiche 
zur\"{u}ckzuf\"{u}hren.

Dem einfacheren Parametrisierungsverfahren stehen, da man sich 
auf gew\"{o}hnliche Differentialgleichungen erster Ordnung zur\"{u}ckf\"{u}hrt, 
bereits station\"{a}re Kurven zugrunde. Ein Feld aus Bahnen kann 
nun irgendwie geschichtet werden, wogegen es r\"{a}umliche Geometrien 
gibt, wie das kleinsche geometrische Modell der maxwellschen 
Gleichungen, die sich nicht in Scheiben schneiden lassen. Aber 
wir befassen uns vorerst mit Aufschneiden. Liegt f\"{u}r
$ t \in [a,b]$
 ein ebenes Feld, dessen Argumente $x$ und $y$ sind, als L\"{o}sung 
einer linearen Differentialgleichung erster Ordnung vor, dann 
lautet das allgemeine Integral:
$\textbf{F}\{x(t), y(t)\} = \bf C$.
 Zieht 
man durch ein beliebiges Punktepaar
$ x_0 = x(t_0)$,
$ y_0 = y(t_0)$ 
eine Kurve, dann wird $C$ auf $C_0$ spezialisiert. W\"{a}hlt man eine 
Folge $t_i$,  ( $t_i  \in [a, b], i = 0, 1,..., N$) von Werten 
des Parameters auf einer der Feldebene $(x, y)$ senkrechten t-Hilfsachse, 
so erh\"{a}lt man auf der jeweils entsprechenden H\"{o}he $C_i$  je 
eine ebene Kurve durch die $(x_i, y_i)$-Werte als partikul\"{a}res 
Integral. Wegen der eindeutigen Bestimmung der Gleichungen f\"{u}r 
$t \in [a, b]$, darf man den Inbegriff der Kurven, d.h. das 
allgemeine Integral, orthogonal auf eine einzige $(x, y)$-Ebene 
projizieren, was H\"{o}henkurven ergibt. Diese orthogonale Projektion 
wollen wir als algebraisch geometrische Felddarstellung verstehen. 
Der Fall einer stetigen \"{A}nderung des Parameters $t$ wird dabei 
zeichnerisch mittels unterschiedlich schattierter Fl\"{a}chenscharen 
wiedergegeben.

Dieselbe Untersuchung l\"{a}sst sich unter Ausfall der Graphik 
f\"{u}r gew\"{o}hnliche Differentialgleichungen der voneinander unabh\"{a}ngigen 
r\"{a}umlichen Argumente $x, y, z, u, v, w,...$ wiederholen. Allerdings 
gilt diese Darstellung f\"{u}r freie Massenpunkte, oder wenn holonome 
Differentialbedingungen vorgeschrieben sind.

\textbf{2. Fall der Entkopplung von E und B. --} Aus den elektromagnetischen 
Gleichungen im Vakuum leitet man unmittelbar zwei homogene unged\"{a}mpfte 
Schwingungsgleichungen f\"{u}r die Vektorfelder ab. Die L\"{o}sungen 
der in \textit{kartesischen Koordinaten} ausgedr\"{u}ckten Gleichungen 
stellen ebene Wellenfronten mit zur Front transversaler Ausbreitungsrichtung 
dar. Aber die Wellenform h\"{a}ngt mit der reduzierten Schwingungsgleichung, 
und also mit der Trennung der Argumente zusammen. \textit{Nach Trennung} 
der r\"{a}umlichen Variablen ergibt sich nichtsdestoweniger in 
einer Mannigfaltigkeit von drei Dimensionen f\"{u}r ebene transversale 
Wellen einen analytisch auf der kartesischen Ebene darzustellenden 
Zeitverlauf.

Die allgemeinste analytisch auf der Ebene darzustellende reduzierte 
Schwingungsgleichung elektromagnetischer Systeme ist immerhin 
die lineare Telegraphengleichung, weshalb wir gleich davon ausgehen. 
Das \textit{allgemeine} Integral, nennen wir es 1, besteht aus der \textit{allgemeinen} 
L\"{o}sung, 2, der dazugeh\"{o}rigen homogenen Differentialgleichung 
plus ein \textit{partikul\"{a}res} Integral, 3, der inhomogenen Gleichung. 
Unter allgemeines Integral der homogenen Gleichung verstehen 
wir dabei eine beliebige lineare Kombination zweier unabh\"{a}ngiger 
partikul\"{a}rer L\"{o}sungen, 2a + 2b. Das partikul\"{a}re Integral 
3 stellt diejenige L\"{o}sung von der inhomogenen Gleichung dar, 
die station\"{a}r ist und sich nach der St\"{o}rfunktion richtet. 
Das mathematische Verhalten eines bestimmten Systems wird aus 
1 = 2a + 2b + 3 durch Spezialisierung der Integrationskonstanten 
erhalten. Da nun 3 station\"{a}r ist, liefert 2 die ben\"{o}tigte 
Anpassung des Anfangszustands zum station\"{a}ren Schwingungszustand. 
2 liefert also s\"{a}mtliche m\"{o}gliche vor\"{u}bergehende Verhalten, 
weshalb wir es das \textit{allgemeine transiente Verhalten} nennen. 
Abgesehen von dessen Interpretation enth\"{a}lt ein Integral des 
Typs 2 nat\"{u}rlich, wenn die dazugeh\"{o}rigen Randbedingungen 
vorliegen, jede Spezialisierung der homogenen Gleichung von drei 
Argumenten, also auch fortschreitende ebene Wellenfronten\footnote{Ebene 
Wellen sind durch flache Wellenfronten kennzeichnet.}.

\textbf{Zur partikul\"{a}ren zeitabh\"{a}ngigen Transiente.} - Im Zusammenhang 
mit den eben besprochenen \textit{linearen mathematischen Transienten} 
m\"{o}chten wir er\"{o}r\-tern, warum raumzeitabh\"{a}ngige Integrale 
rein theoretisch an keine Raumvorstellung mehr kn\"{u}pfen k\"{o}nnen. 
Da wir uns jetzt ihrer Interpretation f\"{u}r elektrische Erscheinungen 
enthalten m\"{o}chten, kehren wir zum W\"{u}rfeln zur\"{u}ck. Nur bemerken 
wir im Vorbeigehen, dass mechanische Transienten im Allgemeinen 
ein sehr kompliziertes Aussehen haben, w\"{a}hrend es eine Menge 
vor\"{u}bergehender elektrischer Erscheinungen gibt, die sich mathematisch 
linear darstellen lassen. Das Verg\"{a}ngliche tut sich unserer 
Meinung nach schon in der Existenz nicht station\"{a}rer W\"{u}rfelflugbahnen 
kund, ohne deshalb auf Zerfall schlie{\ss}en zu lassen. Wir haben 
gesehen, dass sich das nichtstation\"{a}re Verhalten mathematisch 
auf die Unm\"{o}glichkeit einer t-Parametrisierung der r\"{a}umlichen 
Argumente bel\"{a}uft. Gibt es keine eindeutig bestimmbare Bahn, 
dann l\"{a}sst sich der Raum aus diesem Grund nicht abmessen. Das 
l\"{a}sst sich, weil die Bestimmungsst\"{u}cke v\"{o}llig willk\"{u}rlich 
sind, mit keiner Anzahl von \"{u}bersch\"{u}ssigen Parametern beseitigen.\\
\"{U}blich werden nun elektrische Schwingungen \textit{nach} Trennung 
der r\"{a}umlichen Argumente dahin gedeutet dass ein reibungsbedingtes 
Abklingen l\"{a}ngs der Ausbreitungsrichtung stattfindet. Aber 
die in Rede stehende ein bestimmtes System betreffende ,,Reibung`` 
muss sich mathematisch unabweisbar durch Spezialisierung des 
Integrals 2, will sagen der vollst\"{a}ndigen homogenen Gleichung, 
ermitteln lassen. Es folgt, dass keine explizite t-Abh\"{a}ngigkeit 
des allgemeinen Feldes
$ \textbf{F} = \textbf{F}(x, y, z, t)$
 Bahnen 
liefern kann, es bleibe nun dahingestellt, ob elektrisch die 
Niveaukurven station\"{a}re Laufbahnen -- d.h. mit Massenpunkten 
besetzte - oder lauter Flusslinien -- d.h. Geleise -- zu bedeuten 
h\"{a}tten.

Das ganze Problem mit den elektromagnetischen Feldgleichungen 
r\"{u}hrt daher, dass bei der \textit{allgemeinen} Integration linearer 
partieller Differentialgleichungen die genaue Funktionalabh\"{a}ngigkeit 
notwendig aus bleibt, und dass man leider je nach dem gew\"{a}hlten 
durchf\"{u}hrbaren Integrationsverfahren irgendeinen festen Funktionstyp 
zugrundelegt. Bis man umgekehrt, unter dem Vorwand, dass partikul\"{a}re 
Integrale prinzipiell stabil sind, das allgemeine Integral aus 
ihnen bildet, st\"{o}{\ss}t man nachtr\"{a}glich niemals auf \textit{partikul\"{a}re} 
Transienten. Der unumg\"{a}ngliche mathematische Grund daf\"{u}r 
beruht darauf, dass station\"{a}r Transienten nach Definition abgeklungen 
sind.

\textbf{Eine Handvoll L\"{o}sungen. --} Kehren wir zum formellen L\"{o}sungsverfahren 
der Wellengleichung als unendliche Summe von Termen zur\"{u}ck. 
Noch bevor wir mit unserem Vorhaben anfangen, ist es unbedingt 
wichtig die witzig anmutende Feststellung zu machen, dass weiterhin \textit{unwiederholbare 
fl\"{u}chtige Beobachtungen} aus der linearen Theorie abgestreift 
werden. Neu ist nur die Auffassung der gesetzm\"{a}{\ss}igen Erscheinungen.

Wenn man eine station\"{a}re Wellenfront f\"{o}rmlich nach Taylor 
entwickelt, meint man im Kleinen zu quadrieren. Die somit in 
einem infinitesimalen Bereich der reellen Achse gen\"{a}herte L\"{o}sung, 
stimmt aber im Limes entweder mit der gesuchten Funktion vollkommen 
\"{u}berein oder, wenn die gesuchte Funktion eben keine analytische 
Funktion ist, weicht von ihr \textit{wesentlich} ab. Auf die $\mathbb{C}$-Ebene 
\"{u}bertragen, stimmt n\"{a}mlich die kleinste Umgebung mit dem 
offenen Konvergenzkreis um den gegebenen Punkt \"{u}berein. Deshalb 
lassen sich partikul\"{a}re L\"{o}sungen praktisch als Potenzreihenentwicklungen 
nach einem Parameter niederschreiben \cite{kleinLin}.
 Die Bestimmung der Koeffizienten 
in der Reihe h\"{a}ngt offensichtlich von den Ableitungen im gew\"{a}hlten 
Punkt ab. Der topologische Zusammenhang der geometrischen Gebilde 
bestimmt aber diejenigen Integrale, die ineinander \"{u}bergehen, 
und es geht darum, jeweils einen einfach zu darstellenden Repr\"{a}sentanten 
der Reihe zu ermitteln. Im Raum k\"{o}nnen konform abbildende Funktionen 
als legendresche Polynome entwickelt werden. In $\mathbb{C}$ ist 
aber auch dieser Unterschied von geringer Bedeutung, da komplexe 
Zahlen genauso auf der argand-gau{\ss}schen Zahlenebene wie auf 
der riemannschen Zahlenkugel eingetragen werden k\"{o}nnen. Es 
erweist sich aus diesem Grund gleichg\"{u}ltig f\"{u}r die L\"{o}sungen, 
ob man $(x, y, z)$ oder $(r, \theta, \varphi)$ als Raumkoordinaten 
braucht.\\
Das Integral der \textit{homogenen Differentialgleichung} hei{\ss}t 
in $\mathbb{C}$ ganze Funktion, weil es keine Singularit\"{a}tsstellen 
im Endlichen besitzen darf. Zur mathematischen Existenz der als 
L\"{o}sung der inhomogenen Differentialgleichung gedachten Feldfunktion 
verlangt die Potentialtheorie, dass die im Endlichen gelegenen 
Pole mit endlichen Werten belegt werden. Ist nun im Fall von 
transversalen station\"{a}ren Wellen eine allgemeine L\"{o}sung nebst 
Pole und Nullstellen gegeben, dann kann man das vollst\"{a}ndige 
Integral, indem man die partikul\"{a}re L\"{o}sung als Summe von 
Partialbr\"{u}chen eindeutig\footnote{Zwei Funktionen mit denselben 
Polen unterscheiden sich h\"{o}chstens durch eine ganze Funktion.} 
nach Mittag-Leffler auswertet, erhalten. Dieser ist der Teil, 
welcher nach Abklingen der allgemeinen L\"{o}sung als station\"{a}rer 
Teil erhalten bleibt, und das nicht notwendig abbrechende gebrochene 
Polynom setzt sich also aus einem r\"{a}umlich abklingenden (dem 
hyperbolischen) und einem station\"{a}ren Teil zusammen. Die Darstellung 
der L\"{o}sung f\"{a}llt mit einer Laurent-Reihe mit einer endlichen 
Anzahl Summanden im Hauptteil, und mit Nebenteil zusammen. Wie 
jedes gebrochene Polynom kann diese Funktion, laut dem weierstra{\ss}schen 
Produktsatz als Produkt irreduzibler Faktoren zerlegt werden, 
wobei die Pol- und Nullstellen hervorgehoben werden. Wir haben 
damit bereits Bekanntes auf die komplexe Zahlenebene \"{u}bertragen.

\textbf{Wellenl\"{o}sungen als Abbildungen. -} Jetzt kommen wir zu einer 
notwendig erscheinenden Umdeutung der elektromagnetischen Felder 
im Vakuum, wenn als Definitionsbereich und Wertevorrat $\mathbb{C}$ 
gew\"{a}hlt wird.\\
Die maxwellschen Differentialgleichungen sind erster Ordnung. 
Um sie zu l\"{o}sen f\"{u}hrt man sich aber meistens auf Gleichungen 
zweiter Ordnung zur\"{u}ck. Anschlie{\ss}end \"{u}bertr\"{a}gt man die 
Gleichungen von differential auf algebraisch. Wenn man diese 
Weise Differentialgleichungen zu l\"{o}sen unabh\"{a}ngig von ihrer 
Deutung betrachtet, sieht man, dass mit den Bewegungsgleichungen 
algebraische Gleichungen zweiten Grades verkn\"{u}pft sind. Die 
dynamischen Gesetze sind von Haus aus so gebaut, wogegen man 
im elektromagnetischen Fall eigentlich Gleichungen erster Ordnung 
zu gen\"{u}gen sind.\\
Dasselbe gilt f\"{u}r die L\"{o}sungen. Ob irreduzible Polynome abgesehen 
von ihrer Vielfachheit nur \textit{lineare} oder auch \textit{quadratische} 
Faktoren enthalten, h\"{a}ngt vom zugrundegelegenen Zahlenbereich 
ab. Ist der Zahlenbereich $\mathbb R$, wie man zur L\"{o}sung des 
newtonschen Grundgesetzes verlangt, so kann man den Begriff der 
Bahnkr\"{u}mmung dazu benutzen, um die L\"{o}sungen mit quadratischen 
von denen mit rein linearen Faktoren zu unterscheiden. Weil aber 
dank des fundamentalen Theorems der Algebra in $\mathbb{C}$ lauter 
Linearfaktoren eingehen, trifft dieselbe Unterscheidung zwischen 
freien und kraftbedingten Zust\"{a}nden nicht mehr zu.

Angesichts der heute geltenden mathematischen Auffassung der 
fundamentalen L\"{o}sungen des freien elektromagnetischen Feldes 
als \textit{harmonischer} Wellen, scheint uns allerdings die \"{U}berlegung, 
dass es in $\mathbb{C}$ gleichwertig ist, ob man eine meromorphe 
Funktion auf der $\theta  = 0$ Achse in eine Potenzreihe oder 
nach elementaren Winkelfunktionen als Basisfunktionen entwickelt 
angebracht. Auf die einzelnen Glieder ihrer Reihenentwicklung 
kommt es schlie{\ss}lich bei der Deutung einer L\"{o}sung nicht an. 
Freilich werden mithin diejenige station\"{a}ren elektromagnetischen 
Felder verstanden, die gleichsam auf Fl\"{a}chen \textit{abbilden}. 
Wir nennen sie empfangenes Signal.

Wie ist die Deutung in $\mathbb{C}$ geometrisch zu vertreten? Wir 
greifen ein bisschen vor: die \"{u}bliche differentialgeometrische 
Deutung der \textit{passiven} optischen Distanzmessung setzt in \"{U}bereinkunft 
mit der geometrischen Optik voraus, dass Lichtstrahlen wie Stacheln 
eines Seeigels auf der K\"{o}rperoberfl\"{a}che des anvisierten Objekts 
sitzen. Wenn man einen Lichtb\"{u}ndel als Stachel fasst, betr\"{a}gt 
sein optischer Weg vom Stachelansatz zum Auge eine ganz bestimmte 
L\"{a}nge, wobei die Anforderung nicht mit der materiellen Beschaffenheit 
des zur Messung gebrauchten Lineals, sondern mit der Wahl einer 
Geometrie zusammenh\"{a}ngt \cite{helm}. 
Danach sehen zwei kugelf\"{o}rmige Sender wie zwei unterschiedliche 
Seeigel aus und es hat einen Sinn deren Lagen sowie die Kr\"{u}mmungsradien 
ihrer Skelette zur ersten N\"{a}herung aus der Lichtaussendung 
bemessen zu wollen. Laut der differentialgeometrischen Vorstellungen 
gilt ja die Kr\"{u}mmung als angeborene lokale Eigenschaft einer 
Fl\"{a}che. Deshalb sollte sie weder beeintr\"{a}chtigt werden, wenn 
man kleinere Gebiete in Betracht zieht, noch mit der Entfernung 
vom Beobachter abnehmen. Auf krumme ausstrahlende Fl\"{a}chen, 
die sich selber nicht \"{u}berlagern lassen, sollte aber das f\"{u}r 
ebene Wellen g\"{u}ltige Superpositionsprinzip keine Anwendung 
finden\footnote{Jener Behandlung nach treten Beugungseffekte zur zweiten 
N\"{a}herung, wegen der unterschiedlichen Kr\"{u}mmung der geometrisch 
gemeinten Wellenfronten ein. Sie werden auf einer kleinen Umgebung 
des Einfallstrahles erzeugt, und entstehen nicht infolge der 
Abblendung des ganzen empfangenen Signals als Modulation.}.\\
Die Abbildung s\"{a}mtlicher Fl\"{a}chen in derselben Brenn\textit{ebene} 
des Fernrohrobjektivs kommt, falls sie nicht gerade auf die Ebene 
abwickelbar sind, laut Differentialgeometrie einer Streckung 
der auf ihnen gezeichneten geradesten Linien gleich. Jede optische 
Anpeilung beruht jedoch im Prinzip darauf, dass man das Ziel 
scharf stellen kann, mit welchem Bed\"{u}rfnis ein Mal zu erblicken 
man sich im Endeffekt jeden weiteren Anspruch auf Messung im 
klassischen Sinn verdirbt\footnote{Soweit sich Interferenzmessungen 
aus der \"{U}berlagerung interferierender ebener ,,Wellen`` ergeben, 
geh\"{o}ren sie mit zu derselben Gruppe.}.

\subsection{
Geometrie und Modelle}

Wir haben bis jetzt zu betonen versucht, dass die Bewegungen 
keine Raumstruktur, geschweige denn einen Punktraum, untermauern. 
Diese Erkenntnis hatte bereits J. Maxwell dazu gef\"{u}hrt, Experimente 
unter Anwendung gezielter Analogien direkt an die Mathematik 
anzuschlie{\ss}en. Historisch war dann F. Klein in seiner ber\"{u}hmt 
gewordenen Erlanger Antrittsrede ein Vertreter der hier zur Diskussion 
stehenden ,,anschaulichen`` Rolle der Geometrie. Seine \"{U}berzeugung 
zur Geometrie bestand aus zwei Teilen.\\
{\textbullet}\tab 
Die im Erlanger Programm\footnote{,,Vergleichende Betrachtungen \"{u}ber 
neuere geometrische Forschungen``.} ausgedr\"{u}ckte relationale 
Auffassung der Geometrie.\\
{\textbullet}\tab 
Der st\"{a}ndig auf praktische Durchf\"{u}hrung geometrischer Aussagen 
angelegte Nachdruck. Nach unserem besten Wissen wurde dieses 
Verlangen auf keinen festen Namen getauft.

Das aus beiden Punkten bestehende Programm schiebt in das Wechselspiel 
des induktiv-deduktiven Verfahrens ein geometrisches Modell ein. 
Wir versuchen nun die von ihm andeutungsweise aufgestellte Beziehung: \textit{maxwellsche 
Gleichungen} $\Rightarrow$ \textit{Geometrie} $\Leftarrow$ \textit{optische 
Erscheinungen} zu erl\"{a}utern.

\textbf{Transformationen als Bewegungen. -} Klein stellte Bewegungsgruppen 
oder Umformungen als Axiome an die Spitze des Aufbaus jeder Geometrie 
und kam zum Ergebnis, dass die Geometrie des Raumes, nach Angabe 
der Grundbegriffe und Axiome, nichts mehr als logisches Verst\"{a}ndnis 
bietet. Das entspricht genau der um 1860 von den Mathematikern 
erlangten Auffassung, dass sich Beweise nicht auf Zeichnungen 
st\"{u}tzen d\"{u}rfen.\\
Im Gegensatz dazu fu{\ss}ten noch s\"{a}mtliche physikalische Untersuchungen 
des Raumes, es sei hier nur an das gro{\ss}artige helmholtzsche 
Unterfangen erinnert, auf dem Glauben von der beweisbaren \"{U}bereinstimmung 
einer Geometrie mit der ausmessbaren Struktur der Au{\ss}enwelt. 
Die Doktrin, dass die Geometrie ein Forschungsgebiet der angewandten 
Physik sei, geht auf Galilei zur\"{u}ck. Nicht dass letzterer gleichsam 
darauf bestanden h\"{a}tte. Weil er doch bekanntlich mit seinen 
hartn\"{a}ckigen Hinweisen darauf, dass die Naturereignisse f\"{u}r 
jedermann, der unbefangen daraus lernen will, zug\"{a}nglich seien, 
die Gewalt der Kirche eind\"{a}mmen wollte, wurde er dahin verstanden, 
dass er ihr ein verifizierbareres Dogma entgegenhielt. H\"{a}tte 
er \"{u}berhaupt nur f\"{u}r geometrische Anschaulichkeit der kirchlich 
gestatteten Naturerkl\"{a}rungen pl\"{a}diert, so h\"{a}tte er ohne 
Bann das Zeitliche gesegnet.\\
Nach Newton sch\"{a}tzte man seine vermeintliche Doktrin, d.h. 
die galileische Methode, so hoch, um den geometrischen Raum induktiv 
nach den Ma{\ss}ergebnissen im Laborzimmer zu gestalten. Dabei 
\"{u}bersah man was Einstein erkannte, n\"{a}mlich dass bis man keine 
fundamentalen Ma{\ss}st\"{a}be und Uhren aufst\"{o}bert, Messverfahren 
auf lauter Vereinbarungen beruhen. Folglich wurde auch das elektromagnetische 
Feld aufgezeichnet, indem man f\"{u}r allerlei Stellungen im Labor 
das Gleichgewicht auf Probek\"{o}rpern pr\"{u}fte\footnote{Der Nachweis 
des magnetischen Feldes anhand von Eisenfeilsp\"{a}nen beruht eben 
darauf, dass die Schnipsel auf dem Papier haften bleiben k\"{o}nnen.}, 
und die gefundenen rohen Werte direkt auf Papier brachte. Auf 
diese Weise wurden die S\"{a}tze von Coulomb und Biot-Savart formuliert, 
obwohl die elektromagnetischen Felder die vorhandene Raumstruktur 
eigentlich h\"{a}tten verzerren sollen.

 Im Anschluss daran hat 
sich dann Gau{\ss} auf das methodische Vorgehen Galilei berufen, 
um die optischen mit den herk\"{o}mmlichen Messmethoden zu vergleichen, 
und hat sie eventuell gleichwertig gefunden.
Der Anspruch Messungen numerisch auf ihre Richtigkeit hin zu 
pr\"{u}fen gr\"{u}ndet auf das Dogma, dass man Zahlen dingfest machen 
kann. Seitens der Mathematiker wird gerade das von Russell an 
bestritten.
Sobald man die naive \"{U}bereinstimmung der Natur mit der aufzustellenden 
Mathematik prinzipiell ablehnt, r\"{u}ckt die Frage, was aus einer 
m\"{o}glichst gut logisch begr\"{u}ndeten mathematischen Abhandlung 
eine naturwissenschaftliche Erkenntnis macht, recht in den Vordergrund. 
Vermutlich ist man mit dieser Frage immer vorsichtig umgegangen.

Nun nehmen wir den ersten Punkt der kleinschen Auffassung der 
Geometrie in Angriff. Zum deduktiven Verfahren meinte er etwa, 
dass auch ein physikalisch begabter Mathematiker den \textit{neuesten 
Stand} des physikalischen Verst\"{a}ndnisses nicht urteilen k\"{o}nne, 
was den Vertretern der axiomatischen Methode mitunter schlicht 
falsch erschien. Minkowski und Hilbert sind daf\"{u}r ausschlaggebend. 
Kleins Meinung wurde n\"{a}mlich zu einem Zeitpunkt vertreten, 
wo seine genannte Kollegen wohl behaupten konnten die Methoden 
der theoretischen Physik besser als die experimentell veranlagten 
Physiker zu meistern.

Der Standpunkt Kleins, dass sowohl induktive als auch deduktive 
Schl\"{u}sse nur innerhalb der Mathematik einen Sinn haben, war 
allerdings kein angeborener. Er war nach jahrzehntelanger Forschung 
geometrischer Modelle zum Ergebnis gekommen, dass sich vermutlich 
Naturerscheinungen jeder linearen Darstellung und angeblich auch 
jeder Logik\cite{rosinger}
entziehen, so dass man kaum von der Geod\"{a}sie annehmen kann, 
dass sie die Welt als solche darstelle. Er hat sich demzufolge 
in der Erw\"{a}gung, dass rein mathematische, aber ,,anschauliche`` 
Entwicklungen aus einem frisch erkorenen physikalischen Verst\"{a}ndnis 
brauchbarer sein w\"{u}rden gezwungen, nicht Hand an die damals 
im Bau befindlichen physikalischen Theorien zu legen. Er hat 
auch auf jegliche Anwendung der Mathematik auf die experimentellen 
Entdeckungen seiner Zeit verzichtet. Und doch hat er sein Modell 
f\"{u}r die bereits vorliegenden maxwellschen Gleichungen geliefert. 
Denn er selbst hatte den Gehalt der von A. Moebius zur Mathematisierung 
der Statik herangezogenen \textit{rein} geometrischen Modellierung, 
von stofflich auf relational verlagert. Dabei hatte er die Beziehung 
oder Relation mit der geometrischen Auffassung der Bewegung gleichgesetzt. 
Es leuchtet ein, dass das gegen\"{u}ber der bislang gewohnten kinematischen 
oder dynamischen Beschreibung der Kurven v\"{o}llig neu ist. Und 
gerade weil Klein die Geometrie sozusagen programmatisch als \textit{bildhaft 
zu darstellende Beziehung} erdacht hatte, konnte er es nicht unterlassen 
seine eigene Fassung der r\"{a}umlichen Transformationen mit der 
ebenfalls bereits formulierten speziellen Relativit\"{a}tstheorie 
zu vergleichen. Wir schieben die Besprechung seines Modells auf 
einen dazu gewidmeten Absatz.\\

\textbf{Pr\"{u}fbank der Geometrie. --} Nun kommen wir zum zweiten Punkt 
im kleinschen Programm\footnote{Elementarmathematik II S. 201 -- 202.}. 
Es betraf in erster Linie die bis vor Kurzem als schlicht unnat\"{u}rlich 
erachteten, und deswegen als nicht-existent liquidierten nicht\--eukli\-dischen 
Geometrien. Sie waren rein theoretisch erschlossen worden, jedoch 
fragte man sich urpl\"{o}tzlich, ob es \"{u}berhaupt m\"{o}glich w\"{a}re, 
einen euklidisch von einem nicht-euklidisch strukturierten Raum 
zu unterscheiden. Gau{\ss} hatte, wie erw\"{a}hnt zu diesem Zweck 
Messungen angestellt. Klein bestand dagegen bekanntlich auf praktische 
Ausf\"{u}hrungen der geometrischen Aussagen. Unnat\"{u}rlich hin, 
unnat\"{u}rlich her, immerhin hatte man die Mittel an die Hand, 
um sich jene Postulate bildlich zu vergegenw\"{a}rtigen, dachte 
er vielleicht. Der Bildhaftigkeit halber lie{\ss} er allerdings 
allerhand gipserne \textit{Fl\"{a}chen} anfertigen. Auf diese Weise 
merkte er recht bald, dass die Plastik, ganz unabh\"{a}ngig von 
der gemeinten Geometrie, deren Einbettung in die gewohnten metrischen 
Verh\"{a}ltnisse wiedergibt. Danach hielt er Ausschau auf neue 
Darstellungsmittel, und stie{\ss} auf die Perspektive. Nebenbei 
verwies er auf die r\"{a}umlich ausgedehnte Theaterdekoration, 
und beteuerte, dass sie denselben Postulaten wie die Geometrie 
der Lage gen\"{u}gt. Er meinte die erste als \textit{eine} m\"{o}gliche 
Verwirklichung der letzteren, die letzte aber als mathematische 
Behandlung \textit{a posteriori} (nach Einbezug m\"{o}glichst vieler 
Umst\"{a}nde) der B\"{u}hnendekoration. Kehren wir aber zu den nicht-euklidischen 
Geometrien zur\"{u}ck. An deren Modell-Veranschaulichung kn\"{u}pft 
auch die Frage, ob diejenigen Fl\"{a}chen, f\"{u}r die die euklidische 
Metrik nicht gilt, \textit{unbedingt krumm} auszusehen haben. Klein 
hat diese Frage schlie{\ss}lich mit ,,nein`` beantwortet. Um seine 
Aussage zu beweisen, hat er nichteuklidische Ebenen projektiv 
konstruiert und geometrisch begr\"{u}ndet.

\textbf{Projektive Ma{\ss}bestimmung.} -- Weil Klein sp\"{a}ter die \textit{Invarianz 
der Lichtgeschwindigkeit} genau nach derselben Vorschrift gedeutet 
hat, wollen wir hier auf seine Methode ausf\"{u}hrlicher eingehen. 
Nachdem Riemann das Kr\"{u}mmungsma{\ss} einer Mannigfaltigkeit von 
n Dimensionen differentialgeometrisch auf das Bogenelement gegr\"{u}ndet 
hatte, und gezeigt hatte, dass sich damit nichteuklidische Beziehungen 
in die Geometrie eingliedern lie{\ss}en, wurde das Problem Fl\"{a}chen, 
auf welchen die euklidische Geometrie nicht gilt, zu versinnbildlichen 
auf die Aufstellung zweier Hauptformen zur\"{u}ckgebracht (wenn 
man vom topologischen Zusammenhang absieht). Die von Riemann 
selber entdeckte elliptische Geometrie, f\"{u}r welche die Gerade 
nicht unendlich fortgesetzt werden kann, wurde auf die Kugel 
abgebildet. Was Gau{\ss} etwa als Kr\"{u}mmung verstand fiel hier 
positiv aus. Eine Fl\"{a}che mit konstanter negativer Kr\"{u}mmung\footnote{Gemeint 
ist, dass die W\"{o}lbung zwei Hauptnormalenschnitten mit in entgegengesetzten 
Richtungen weisenden Lotrechten besitzt.}, auf der das Parallelenaxiom 
hinf\"{a}llig wird, ist schwieriger zu veranschaulichen, weil man 
beim Aufbau immer auf Singularit\"{a}ten st\"{o}{\ss}t. Beltrami hat 
allerdings zeigen k\"{o}nnen, dass man sie \textit{st\"{u}ckweise} darstellt, 
wenn man eine Traktrix um ihre Achse drehen l\"{a}sst. Dieses zur 
anschaulichen Wiedergabe der hyperbolischen Geometrie g\"{u}ltige 
Fl\"{a}chengebilde wurde Pseudosph\"{a}re genannt.\\
Klein hat nun Fl\"{a}chen der beiden Arten auf der \textit{Halbkugel} 
gedeutet\footnote{Das \"{A}quator stellt nur dann einen Rand dar, wenn 
die Halbkugel im euklidischen Raum gedacht wird. Projektiv bildet 
aber die einseitige innere Halbfl\"{a}che die ganze hyperbolische 
Ebene ab. Deshalb wird das \"{A}quator als Ma{\ss}bestimmung gedeutet.}, 
und zwar die hyperbolische Fl\"{a}che auf der \textit{Kugelinnenseite}. 
Alsdann hat er beide Fl\"{a}chentypen in zwei Schritten \textit{konform} 
auf die Blattebene projiziert. Die Abbildung des \"{A}quators hat 
er als projektive Ma{\ss}bestimmung im Sinne von Cayley erkl\"{a}rt. 
Wir heben hervor, dass sich auf der Blattebene Geraden von Kreisb\"{o}gen 
projektiv nicht unterscheiden. Weiter k\"{u}rzt eine Projektion 
die Kugel um eine Ausdehnung, denn sie wird dann als Kugeloberfl\"{a}che 
verstanden. Zum Schluss werden s\"{a}mtliche euklidische bzw. nichteuklidische 
Strukturen auf dieselbe projektive Geometrie plus eine f\"{u}r 
jede Metrik charakteristische quadratische Form reduziert. Es 
ist hinterher spekuliert worden, dass sich auf Fl\"{a}chen von 
konstanter Kr\"{u}mmung kongruente Bewegungen durch die W\"{o}rtchen 
elliptische, euklidische oder hyperbolische kennzeichnen lassen. 
Das gilt ja nur bis sich die Bewegung aus der Struktur erschlie{\ss}en 
l\"{a}sst.

Was nun die Vermessung betrifft, verk\"{u}rzen sich die Ma{\ss}st\"{a}be 
l\"{a}ngs der Geraden dieser Ebenen genau wie es Lorentz verlangte. 
Um l\"{a}ngentreue Abbildungen handelt es sich also nicht. Was 
Klein hat beweisen k\"{o}nnen ist, dass projektive Abbildungen 
konforme Transformationen sind, und dass aus dem Bogenelement 
im Sinne Riemanns stereographisch eine quadratische Form wird
\cite{klein1928}. Zu diesem Zweck hat 
Klein aber die von Von Staudt gar abstrakt gefasste Geometrie 
herangezogen und sie pragmatisch ohne jede Metrik weiterentwickelt. 
Gelegentlich dieser Entwicklung hat er nicht nur die nichteuklidischen 
Geometrien, sondern auch die Geometrie der Lage sozusagen projektiv 
gedeutet.

\section{
Der Beitrag Felix Kleins zum Elektromagnetismus.}

Klein selber hat sich sowohl zu den ballschen Schrauben (als 
Bewegungen), als auch zur nicht-euklidischen Geometrie (als Raumstruktur) 
ge\"{a}u{\ss}ert. Er hat beide Male eine Quadrik in dieselbe synthetische 
projektive Geometrie, als die allgemeinste Geometrie, einf\"{u}hrt. 
In der speziellen Relativit\"{a}tstheorie ist er nun so verfahren, 
dass er Lorentzgruppe und elektromagnetische Gleichungen zugleich 
eingetragen hat. Wie leicht einzusehen liegt, wenn beide Quadriken 
identisch sind, nur eine Redundanz vor, weil dann die spezielle 
Relativit\"{a}tstheorie nach dem kleinschen Verfahren auf zweimalige 
Anwendung derselben Zutaten beruht. Wenn die Quadriken obendrein 
nicht einmal mathematisch \"{u}bereinstimmen, fragt es sich, ob 
es gemeint ist, dass sie sich ineinander \"{u}berf\"{u}hren lassen, 
oder ob es irgendwelche naturwissenschaftlichen Gr\"{u}nde gibt, 
um die eine davon zu bevorzugen. Das ist angeblich der Sinn der 
im letzten Brief von 1917 an Einstein gerichteten Frage. Aus 
dem Nachtrag von 1921 zur Arbeit von 1910 erhellt, dass Klein, 
trotzdem er keine eigene Theorie zu bieten hatte, mit dessen 
Antwort noch nicht einverstanden war. Ginge es nur darum, f\"{u}r 
die einsteinsche Theorie irgendeine feste Anzahl von Raumdimensionen 
zu statuieren, dann w\"{a}re nach so vielen Jahren und so vielen 
weiterf\"{u}hrenden Entwicklungen Klein auf diesem Gebiet nicht 
einmal mehr des historischen Erw\"{a}hnens w\"{u}rdig. Er scheint 
aber eine etwas kernhaltigere Fragestellung hinterlassen zu haben. 
N\"{a}mlich in etwa so: \textit{wenn die maxwellschen Gleichungen zur 
Beschreibung der elektrisch wichtigen Tatsachen reichen, kommen 
wir diesmal zwar nicht mit den uns jetzt gewohnten r\"{a}umlichen 
Vorstellungen aus, doch k\"{o}nnen wir hoffen, uns davon mit den 
uns zur Verf\"{u}gung stehenden Mitteln ein Modell zu verschaffen. 
Wenn aber die einsteinschen Erw\"{a}gungen notwendig dazu geh\"{o}ren, 
wird uns ein Hauptaspekt unserer Welt unzug\"{a}nglich bleiben}. 
Ein fr\"{u}hes Beispiel zum angef\"{u}hrten Ideenkreis sind die Versuche 
die M\"{o}glichkeit in ein vollst\"{a}ndig abgesperrtes Zimmer hineinzudringen 
und es wieder zu verlassen zu verdeutlichen. Diese Gabe scheint, 
trotzdem sie den elektrischen Erscheinungen anhaftet, uns nicht 
geg\"{o}nnt zu sein. Ferner ist es, wegen der Vertauschung von 
links und rechts, in unserer Welt unm\"{o}glich eine Spiegelung 
aus mechanischen Drehungen zusammenzubasteln. Trotzdem k\"{o}nnen 
Spiegel rechts und links, und obendrein oben und unten vertauschen.

\subsection{
Zu den geometrischen Grundlagen der Lorentzgruppe}

Die anf\"{a}ngliche Aussage Einsteins zur Kinematik bewegter K\"{o}rper 
ging nicht von einer vierdimensionalen Welt aus, sondern behandelte 
die Zeit als Messgr\"{o}{\ss}e auf derselben Grundlage wie die Raumlagen. 
Das bedeutet freilich, dass auch die Zeit zur dynamisch relevanten 
Ver\"{a}nderlichen wird, meint aber nicht, dass sie geometrisch 
den Raumvariablen gleichzusetzen sei. In diesem Sinn ist die 
vierdimensionale Welt erst von Minkowski aufgestellt worden.

\textbf{Das Verh\"{a}ltnis der Theorie zum Experiment.} -- Laut Einstein 
sind theoretische Ans\"{a}tze letzten Endes auf Messvorschriften 
zu st\"{u}tzen, weil Vermessungen die ganze Verkn\"{u}pfung zur Wirklichkeit 
ausmachen. Somit erwartet man, dass die Relativit\"{a}tstheorie 
genau mit getroffenen Messvorschriften \"{u}bereinstimmen soll. 
Aus welchem Grund auch immer, unterscheidet sich aber das Simultaneit\"{a}tsprinzip 
nicht nur von der absoluten Zeit Newtons, sondern auch von den 
zur Festlegung der Zeitzonen tats\"{a}chlich gefassten Bestimmungen.

\textbf{Das 1. Prinzip.} - Einstein hat zwei Prinzipien, denen alle 
physikalischen Gesetze unterstellt sein sollen aufgestellt.\\
Das 1. Prinzip verlangt, dass die physikalischen Gesetze f\"{u}r 
eine bestimmte Klasse von Beobachter analytisch unver\"{a}ndert 
lauten. Das hat Minkowski auf vier gleichberechtigte Argumente
 $x, y, z, it (c  = 1)$
 \"{u}bersetzt. Trotzdem Klein bestimmt nicht glaubte, 
dass $ict$  in der pseudo-euklidischen Geometrie eine wirkliche, 
unversehens der Au{\ss}enwelt anzuh\"{a}ngende Ausdehnung bedeuten 
sollte, enth\"{a}lt das von ihm f\"{u}r die Lorentzgruppe vorgeschlagene 
Modell 5 homogene projektive Punktkoordinaten
 $x_k , k = 1, 2, 3, 4, 5$, wie dasjenige von Kaluza und O. Klein. Nur stimmt bei 
F. Klein die ,,physikalische Dimension`` mit der Anzahl der Argumente 
der zugeh\"{o}rigen \textit{affinen} Mannigfaltigkeit, d.h.
 $x = x_1 /x_5$,
$ y = x_2 /x_5$,
$ z = x_3/x_5$
 und
$ ict = x_4/x_5$,
 \"{u}berein. 
Was gibt es aber physikalisches an einer Dimension? Wenn es ohnehin 
um Doppelverh\"{a}ltnisse geht, lassen sich Geometrien von 4 auf 
3 oder 2 Ausdehnungen \textit{projizieren}. Besteht man unbedingt 
auf 4 physikalisch relevante Gr\"{o}{\ss}en, so gibt es bereits im 
projektiven Raum von 3 Dimensionen vierdimensionale Gebilde. 
Im synthetisch-geometrischen Raum gibt es $\infty^4$ reelle 
Linien, so dass der uns gewohnte geometrische Raum bez\"{u}glich 
der Linien 4-dimensional ist. Dementsprechend hat Pl\"{u}cker eine 
beliebige Gerade im kartesischen Punktraum $(\xi, \eta, \zeta)$ 
durch die zwei Gleichungen: $\xi = x\zeta  + y$, $\eta  
= z\zeta + t$ analytisch ausgedr\"{u}ckt, weshalb dem Linienraum 
die rechtwinkligen inhomogenen Punktkoordinaten $x, y, z, t$ einer \textit{vierfach} 
ausgedehnten Mannigfaltigkeit zugeschrieben werden k\"{o}nnen.

Da sich die elektromagnetischen S\"{a}tze im Vakuum, wie wir demn. 
zeigen, synthetisch-geometrisch als Nullsystem darstellen lassen, 
dr\"{u}cken \textbf{E} und \textbf{H} Koordinaten von Liniengebilden aus, 
und die Invarianz des maxwellschen Gleichungssystem besagt, dass 
Schraubungen\footnote{Es handelt sich hier der projektiven Bewegungsgruppe.} 
ein Linienkomplex in sich \"{u}berf\"{u}hren.

\textbf{Das 2. Prinzip.} - Das 2. Prinzip schildert aufgrund eines 
Gedankenexperiments die Bedingung von der Konstanz der Lichtgeschwindigkeit 
beim \textit{Signal} analytisch als Invarianz von 
$x^2 + y^2 + z^2 - c^2t^2$. Nach der Weise, wie es Einstein eingef\"{u}hrt 
hat, liegt es nahe darunter die infinitesimale metrische\footnote{Klein 
f\"{u}hrt diese Auffassung der Ma{\ss}gr\"{o}{\ss}en auf Cayley zur\"{u}ck. 
Der Geometer M. Pasch war fest \"{u}berzeugt, dass Metrik 
und Infinitesimalrechnung, wegen der Definition von H\"{a}ufungspunkt, 
unvereinbar seien. Davon angeregt, hatte daraufhin Klein seine 
invariante Quadrik zur Festlegung der Metrik eingef\"{u}hrt. Doch 
kann man die ma{\ss}bestimmende Quadrik selber, ganz entgegen dem 
Laut des 2. Prinzips der Relativit\"{a}t, nicht metrisieren.} Bedingung 
f\"{u}r das Bogenelement nach Riemann zu verstehen.

Die lorentzsche Ma{\ss}bestimmung ist auf die kanonische Form\footnote{Nur 
nicht ausgeartete quadratische Gebilde ergeben dieselbe Figur 
wenn sie als Punkte (Gebilde zweiter Ordnung), oder als Tangentialebenen 
(Gebilde zweiter Klasse) erhalten werden. In diesem Fall benutzt 
Klein den Namen ,, Gebilde zweiten Grades`` f\"{u}r beide (Nicht-euklidische 
S. 85).}. L\"{a}gen ihr reelle homogene projektive Koordinaten 
zugrunde, entspr\"{a}che sie einer ovalen Fl\"{a}che in einer 3-dimensionalen 
Mannigfaltigkeit. Wenn man aber zur Aufrechterhaltung von metrischen 
Gr\"{o}{\ss}en verlangt, dass das unendlichferne Gebiet in sich selbst 
\"{u}bergef\"{u}hrt werden muss, handelt es sich eines in affinen 
Koordinaten ausgedr\"{u}ckten nullteiligen Hyperkegels, d.h. eines 
einmal ausgearteten Gebildes in einer vierdimensionalen Mannigfaltigkeit.

\textbf{Ma{\ss}bestimmung und Raum.} -- Nehmen wir an, der Raum lasse 
sich mittels der drei Punktkoordinaten \textit{x}, \textit{y} und \textit{z} 
darstellen. Dann werden Fl\"{a}chen im allgemeinen durch ebendiese 
Koordinaten ausgedr\"{u}ckt. Klein verbindet die Geometrie auf einer 
in diesem Raum eingebetteten \textit{Fl\"{a}che} mit der auf ihm geltenden 
Messvorschrift, indem er die ihr entsprechende Metrik an eine 
quadratische Gleichung kn\"{u}pft. Er setzt zwei reelle Wurzeln 
mit der Existenz zweier Parallelen, und also mit der Geometrie 
von Gau{\ss}, Bolyai, Lobatschefsky in Verbindung. Zwei imagin\"{a}re 
Wurzeln bezieht er auf eine riemannsche Geometrie (Metrik auf 
der Kugel), und er denkt sich den \"{U}bergangsfall einer doppeltz\"{a}hlenden 
reellen Wurzel mit unserer Raumanschauung \"{u}bereinstimmend.

Seit Gau{\ss} ist es bekannt, dass sich Fl\"{a}cheneigenschaften, 
darunter die metrischen Verh\"{a}ltnisse, auch mittels \textit{Fl\"{a}chenkoordinaten} 
$u, v$, angeben lassen. Es lassen sich nun \"{u}ber die sogenannten 
intrinsischen Koordinaten alle \textit{gew\"{o}lbten}, und trotzdem 
auf den euklidischen Raum abwickelbaren R\"{a}ume, durch eine Transformation 
des Quadrats des Bogenelements auf allgemeine inhomogene Koordinaten 
wie folgt bestimmen:
 $dx^2 + dy^2 + dz^2 = Q_1 dq_1^2 
+ Q_2 dq_2^2 + Q_3dq_3^2 + 2Q_23dq_2 dq_3 + 2Q_31dq_3dq_1  
+ 2Q_12dq_1 dq_2$, wobei
$ Q_i  = (\partial x/\partial q_i )^2 
+ (\partial y/\partial q_i )^2 + (\partial z/\partial q_i )^2$, 
und
$ Q_ij = (\partial x/\partial q_i )(\partial x/\partial q_j) 
+ (\partial y/\partial q_i )(\partial y/\partial q_j) 
+ (\partial z/\partial q_i )(\partial z/\partial q_j)$. 
Die Ausdr\"{u}cke der Bogenelemente f\"{u}r nichteuklidische R\"{a}ume\footnote{Parallelen 
sind von Haus aus eigentlich nur auf ebene Fl\"{a}chen definiert.} 
leisten genau dasselbe, bis diese R\"{a}ume ohne Dehnung \"{u}bereinander 
abgerollt werden k\"{o}nnen. Aber jene R\"{a}ume sind der Theorie 
nach auf eine Weise gew\"{o}lbt, die es verhindert, die oben angeschriebene 
Gleichung f\"{u}r die Transformation eines Bogen auf einen ,,flachen`` 
Raum metrisch anzuwenden. Intrinsische Koordinaten lassen nur \textit{zur 
ersten N\"{a}herung} nichteuklidische Ausdr\"{u}cke auf den euklidischen 
Raum beziehen, wobei diese N\"{a}herung nicht mit 
Differentialausdr\"{u}cken zu verwechseln ist.\\
Klein hat nun zwecks der Bestimmung von Bewegungsgruppen und 
dergleichen den Ausdruck des Bogenelements schon f\"{u}r die spezielle 
Relativit\"{a}t mathematisch als \textit{projektive} Ma{\ss}bestimmung 
gedeutet\footnote{Das besagt, dass er eigentlich die pseudoeuklidische 
Geometrie der speziellen Relativit\"{a}tstheorie noch vor der Diskussion 
zwischen Einstein und De Sitter \"{u}ber die massenbedingte Raumkr\"{u}mmung 
als Ma{\ss}bestimmung behandelt hat.}. Da die projektive Geometrie 
auf lineare Gebilde gr\"{u}ndet, wird das Bogenelement auf ihr 
nicht infinitesimal ausgedr\"{u}ckt. Statt dessen \"{u}bertragen 
sich die Eigenschaften des zur Einf\"{u}hrung einer Metrik n\"{o}tigen 
Bogenelements, da sich in unausdehnbaren Mannigfaltigkeiten von 
konstanter Kr\"{u}mmung Umgebungen unermesslich entfernter Stellen 
durch nichts von der Umgebung des Ursprungs unterscheiden, vom 
unendlichkleinen Bereich gleich auf den ganzen Raum. Das projektive 
quadratische Gebilde Cayleys liegt folglich, auch wenn es im 
Endlichen erscheint, in unermesslicher Entfernung. Es gibt einen 
weiteren Unterschied zwischen differentialer und projektiver 
Geometrie. Weil sich aus derselben Halbkugel sowohl eine elliptische 
als auch eine hyperbolische Ma{\ss}bestimmung auf dieselbe Ebene 
konform projizieren lassen, sind die drei Fl\"{a}chentypen, abgesehen 
von ihrer Ma{\ss}beziehung, projektiv ununterscheidbar. Das l\"{a}sst 
sich wohl aus dem Umstand erkl\"{a}ren, dass die projektive Gerade 
in sich geschlossen ist, und parallele Geraden projektiv zusammenlaufen. 
Wenn wir diese \"{U}berlegung auf den Raum erweitern, dann lassen 
sich aber projektiv die drei Geometriearten \textit{konform} aufeinender 
projizieren.\footnote{Die projektive Geometrie unterscheidet L\"{a}ngenmessungen 
nicht von Winkelmessungen.}.

\textbf{Zum projektiven Raum von vier Dimensionen.} - Ist der (synthetische) 
Raum vierfach ausgedehnt, dann gilt die projektive duale Entsprechung 
der Gebilde: Punkt \ensuremath{\Leftrightarrow} Ebene, Gerade \ensuremath{\Leftrightarrow} 
Gerade nicht mehr, und es tritt an seine Stelle: Punkt \ensuremath{\Leftrightarrow} 
Raum, Ebene \ensuremath{\Leftrightarrow} Gerade in Kraft. Bekanntlich f\"{u}hren 
auch die Koordinatensubstitutionen zu etwas unterschiedliche 
Bewegungen. In vier Dimensionen kommen n\"{a}mlich zentro-affin 
lauter Drehungen, darunter auch Drehungen um Ebenen, in Frage. 
In drei affinen Dimensionen sind dagegen Spiegelungen davon abzuscheiden. 
Das kommt freilich \textit{projektiv} auf dasselbe hinaus, wenn man 
bedenkt, dass ein ,,Kubus`` in einer vierfach ausgedehnten Welt 
auch Drehungen durchf\"{u}hrt, die, wenn sie auf den uns gewohnten 
Raum projiziert werden, Spiegelungen bewirken. Aber diese \"{U}berlegung 
erfordert eine projektive Auffassung der Bewegungen, welche die 
Mechaniker bisweilen noch nicht in Betracht gezogen zu haben 
scheinen. Um jenen Standpunkt mit einem Ausdruck aus Klein 1910 
wiederzugeben, scheinen immerhin \textit{alte} und \textit{neue}\footnote{D.h. 
die Elektrodynamik.} \textit{Mechanik} auf zwei experimentell voneinander 
unterscheidbare raumzeitliche Strukturen zu f\"{u}hren.

Wir m\"{o}chten diese Einsicht erg\"{a}nzen. Falls unsere Welt, aus 
welchem Anlass auch immer, eine \"{u}bersch\"{u}ssige Ausdehnung 
besitzen sollte, w\"{a}ren in ihr Drehungen um Ebenen tats\"{a}chlich 
durchf\"{u}hrbar. Diese Drehungen k\"{o}nnen wir mit den uns zur 
Verf\"{u}gung stehenden Mitteln experimentell nicht nachweisen. 
Wird aber die einsteinsche Forderung, was Ma{\ss}st\"{a}be und Uhren 
betrifft, gestrichen, dann ist das kleinsche Modell nicht mehr 
metrisch, sondern projektiv eingebettet. Wegen des hier erf\"{u}llten 
Dualit\"{a}tsprinzip werden aus unserer Sicht Punkte mit Ebenen 
vertauschbar, und somit werden Drehungen um Ebenen dualisiert. 
Sie werden mit unseren Mitteln eben als Spiegelung -- Projektion 
der \"{u}berz\"{a}hligen Ausdehnung in unsere Welt - nachgewiesen. 
Es erhellt zugleich, warum man mehr Aussichten die optischen 
Erscheinungen zu verstehen hat, wenn man beispielsweise Spiegelungen 
nicht unbedingt als mechanische Drehungen begreifen will\footnote{Schlie{\ss}lich 
d\"{u}rfen optische Messungen wohl nicht ohne irgendeine Eichung 
als ,,metrisch`` bezeichnet werden.}.

Aus den eben angegebenen Gr\"{u}nden geh\"{o}ren die von Einstein 
verlangte Signalinvariante (2. Gesetz) sowie die Kovarianz der 
elektromagnetischen Gleichungen (1. Gesetz) \textit{in der kleinschen 
Deutung} in demselben projektiven synthetischen Raum.

Entweder lassen sich die elektromagnetischen Gesetze\footnote{Beim 
Nullsystem sind die Quadriken einschalige Hyperboloide.} auf 
dasselbe Fundamentalgebilde bringen, oder ihre Quadriken k\"{o}nnen 
nicht durch \textit{reelle} Kollineationen ineinander \"{u}bergef\"{u}hrt 
werden, und bestimmen eventuell unterschiedliche Bewegungsgruppen. 
Die strenge Einteilung der quadratischen Formen nach dem Tr\"{a}gheitsindex 
gilt aber nur, bis man keine imagin\"{a}ren Substitutionen zul\"{a}sst, 
was wir wegen \textit{ict} und des Ausdrucks der Wellengleichung besonders 
hervorheben.

\subsection{
Zu den maxwellschen Gleichungen in der hertzschen Bezeichnung}\cite{kleinElem2}

Unter elektromagnetisches Gleichungssystem wird vielfach Verschiedenes 
gemeint\footnote{Maxwell selber k\"{o}nnte Integralgleichungen \"{u}ber 
topologisch nicht einfach zusammenh\"{a}ngende Gebiete gemeint 
haben. Er hatte, da er die faradaysche Induktion zu erkl\"{a}ren 
trachtete, zeitabh\"{a}ngige Erscheinungen im Sinn. Das kommt hier 
nicht mehr zur Diskussion. Allerdings hat er die vektorielle 
Potentialfunktion \textbf{A} eingef\"{u}hrt.}. Obwohl es also Maxwell 
anfangs aus den Versuchen Faradays abgeleitet hat, scheint die 
genaue Form, in der es den Beobachtungen entsprechen soll, nicht 
ganz fest zu sein. Die heute fasst universell als fundamentale 
Lehrs\"{a}tze des klassischen elektromagnetischen Feldes anerkannte 
Gleichungen sind immerhin Differentialgleichungen. Sie stellen 
Beziehungen zwischen 4 Feldver\"{a}nderlichen \textbf{E}, \textbf{D}, \textbf{H} 
und \textbf{B} dar, und den Gleichungen geht meistens ein Kapitel 
\"{u}ber Vektorrechnung voran. Der Deutlichkeit halber stellen 
wir sie deshalb f\"{u}r das Vakuum, unter Benutzung der Ausdr\"{u}cke \textbf{E} 
und \textbf{B} f\"{u}r die elektrische Feldgr\"{o}{\ss}e und die magnetische 
Induktion respektive, im gau{\ss}schen Ma{\ss}system wieder zusammen. 
Daneben tragen wir gleich deren Bezeichnung nach Hertz, wie wir 
sie aus Kleins Arbeit von 1910 entnommen haben, und wie sie \"{u}brigens 
auch in der grundlegenden einsteinschen Arbeit von 1905 stehen. 
Wir meinen nat\"{u}rlich nicht, dass die hertzschen Ausdr\"{u}cke 
(rechts), wie wir sie aus der kleinschen Arbeit abgeschrieben 
haben, die elektromagnetischen Gleichungen (links) in kartesischen 
Komponenten ausdr\"{u}cken.\\

\begin{longtable}{p{1.1in}p{1.3in}p{2in}}

% ROW 1
{\raggedright Coulombscher Satz. Keine freien elektrischen Ladungen} & 
{\raggedright $ \nabla\times\textbf{E} = 0$} & 
{\raggedright $ \partial X /\partial x + \partial Y /\partial y + \partial Z /\partial z = 0$} \\

&&\\

% ROW 2
{\raggedright Maxwell-amp\`{e}rescher Satz} & 
{\raggedright $ 1/c \partial \textbf{E}/\partial t = \nabla\times\textbf{B}$} & 
{\raggedright $ 1/c \partial X /\partial t = \partial M /\partial z-\partial N /\partial y$ \linebreak
$ 1/c \partial Y /\partial t = \partial N /\partial x-\partial L /\partial z$ \linebreak
$ 1/c \partial Z /\partial t = \partial L /\partial y-\partial M /\partial x$} \\

&&\\

% ROW 3
{\raggedright Faradayscher Satz} & 
{\raggedright $ 1/c \partial \textbf{B}/\partial t =-\nabla\times\textbf{E}$} & 
{\raggedright $ 1/c \partial L /\partial t = -[\partial Y /\partial z-\partial Z /\partial y]$ \linebreak
$ 1/c \partial M /\partial t = -[\partial Z /\partial x-\partial X /\partial z]$ \linebreak
$ 1/c \partial N /\partial t = -[\partial X /\partial y-\partial Y /\partial x]$} \\

&&\\

% ROW 4
{\raggedright Keine freien magnetischen Pole} & 
{\raggedright $\nabla  \cdot \textbf{B} = 0$} & 
{\raggedright $ \partial L /\partial x + \partial M /\partial y + \partial N /\partial z = 0$} \\

\end{longtable}

\textbf{Deutung der in den Grundgleichungen gemeinten Gr\"{o}{\ss}en.} 
-- Die elektromagnetischen Ver\"{a}nderlichen von heute und damals 
entsprechen auch ihrer Deutung nach einander kaum, obwohl die 
meisten Autoren nach wie vor zu einer energetischen Interpretation 
kommen. Heute schreibt man den \textbf{E}- und \textbf{B}- Felder dualistisch 
etwa eine Teilchennatur zu. Damals hat H. Hertz das ,,mechanische 
\"{A}quivalent`` der Elektrizit\"{a}t vor Augen geschwebt. Der Einheitlichkeit 
der \"{a}quivalenten mechanischen Wirkung wegen, hat er ohne weiteres 
unter \textbf{E} immer die elektrostatische Kraft zwischen Ladungen 
verstanden. Aber er hat die Wesensgleichheit, was das mechanische 
\"{A}quivalent betrifft, durchaus allgemeiner gefasst, da er auch \textbf{H} 
mit \textbf{E} f\"{u}r wesensgleich hielt, und insbesondere annahm, 
dass sich der \textit{lineare} Magnetismus vollst\"{a}ndig durch die 
elektrische Kraft aufheben lasse. Deshalb hat er 1884 elektrische 
und magnetische Kr\"{a}fte aus derselben Potentialfunktion \textbf{A} 
abgeleitet. Das sieht unter Zugrundelegung der maxwellschen Eichung 
${\rm div} \textbf{A} = 0$
 f\"{u}r das Strahlungsfeld ungef\"{a}hr folgenderma{\ss}en 
aus. Die Ausdrucksweise des coulombschen Satzes (rechts) lautet: 
${\rm div}\textbf{E} = - (1/c) {\rm div}(d\textbf{A}/dt) = - (1/c) d({\rm div}\textbf{A})/dt = 0$,
 sofern 
die r\"{a}umlichen Koordinaten nicht von der Zeit abh\"{a}ngen. Gibt 
es weiter nur geschlossene Str\"{o}me und keine magnetischen Ladungen, 
dann hei{\ss}t es in vektorieller Schreibweise
 ${\rm div}\textbf{J} = 0$
 und 
(wegen
$ \textbf{B} = \mu_0 \textbf{H}$,
$ \mu_0 = 1) {\rm div}\textbf{H} = 0$, 
was wiederum
$ \textbf{H} = {\rm rot}\textbf{A}$
 bedeutet. Der faradaysche Lehrsatz 
(rechts) folgt nun aus
$ d\textbf{H}/dt = d({\rm rot}\textbf{A})/dt = {\rm rot}(d\textbf{A}/dt) 
= - c {\rm rot}\textbf{E}$.
 Das amp\`{e}re-maxwellsche Gesetz 
$(1/c) \textbf{J} = (1/c) \partial \textbf{E}/\partial t = 
{\rm rot}\textbf{H}$,
 wobei
$ \textbf{J} = d\textbf{E}/dt$
\footnote{Erst beim Ringintegrieren 
tritt die Konstante $4\pi N\textbf{c}$ hinzu.},
 h\"{a}ngt dann direkt 
mit der Wellengleichung:
$ \Delta \textbf{A} - 1/c^2 d^2\textbf{A}/dt^2 = 0$
 zusammen\footnote{Man k\"{o}nnte mit Bezug zur Wellengleichung faradayschen 
und amp\`{e}re-maxwellschen Lehrsatz als Existenzbedingungen f\"{u}r \textbf{A} 
halten.}. Diese Vereinbarung durchschaut man nicht mehr, wenn 
man nachtr\"{a}glich \textbf{E} in $V/m$ und \textbf{B} in $Wb/m^2$ misst.

\textbf{Nullsystem und Spannungsfl\"{a}chen.} -- Nachdem A. M\"{o}bius 
begriffen hatte, dass die graphische Statik nicht ohne Weiteres 
auf den Raum zu erweitern ist, und au{\ss}erdem auf Zeichenkunst\footnote{Man 
k\"{o}nnte, trotzdem es aus Eschers Zeichnungen wohl hervorzugehen 
scheint, manche seiner Geb\"{a}ude tats\"{a}chlich nicht zum ,,stehen`` 
veranlassen.} fu{\ss}t, hat er das Nullsystem in Betracht gezogen. 
Somit hat er die Statiklehre, ob sie mit den Naturwissenschaften 
irgend zu tun hat oder nicht, mathematisch begr\"{u}ndet, und bildlich 
als geometrisches Modell untermauert. ,,Nullsystem`` ist deswegen 
der in Bezug auf die Anwendungen der Statik gepr\"{a}gte Beiname 
des linearen Komplexes, einer Geometrie, welche keinen Anspruch 
auf \"{A}hnlichkeit mit der von uns bewohnten Welt hat. Von Haus 
aus ist das Nullsystem eine besondere projektive Raum\textit{verwandtschaft}, 
die sich rein synthetisch behandeln l\"{a}sst. Laut M\"{o}bius ist 
der projektive geometrische Raum starr, weshalb ein gegebenes 
Kr\"{a}ftesystem im Prinzip f\"{u}r jeden Raumpunkt berechnet werden 
kann, sobald es f\"{u}r eine gen\"{u}gende Anzahl von Punktlagen 
bekannt ist. Als projektive Verwandtschaft gen\"{u}gt das Nullsystem, 
weil L\"{a}ngen- und Winkelma{\ss} \"{u}bereinstimmen, dem \textit{Dualit\"{a}tsprinzip} Punkt \ensuremath{\Leftrightarrow} 
Ebene (wobei dual die Gerade sich selbst entspricht), und dem \textit{Invarianzprinzip 
des Doppelverh\"{a}ltnisses} zwischen je 4 mit einer Geraden inzidierenden 
Punkten/Ebenen.

Es werden beim Nullsystem Quadriken mit der Eigenschaft ausgezeichnet, 
dass jede ihrer Ber\"{u}hrungsebenen den ihr durch die Verwandtschaft 
entsprechenden Punkt enth\"{a}lt, und umgekehrt. In der Statiklehre 
wird diese Besonderheit, ohne weiter auf die dualen Beziehungen 
R\"{u}cksicht zu nehmen, sowohl zur Befriedigung der Fundamentalgleichungen 
als auch zur Ermittlung der Belastung von Fachwerken benutzt. \\
Als Klein 1904 \"{u}ber \textit{reziproke Diagramme} in den maxwellschen 
Fachwerken zu sprechen kommt, notiert er, dass der Statik zuliebe 
die Achsen des reziproken Diagramms in dem cremonaschen Kr\"{a}fteplan 
um einen rechten Winkel im Uhrzeigersinn gegen\"{u}ber denen des 
maxwellschen Diagramms gedreht sind. Infolge der Drehung sieht 
die Raumverwandtschaft zwischen direkter und reziproker Figur 
anders, und insbesondere sind die Fundamentalgebilde mit der 
Besonderheit, dass ihre Punkte auf der jeweiligen dualen Ebene 
liegen auch unterschiedlich. Die r\"{a}umlichen Gebilde, f\"{u}r welche 
diese Beziehung in der Statik gilt, sind nun die ausgezeichneten 
Quadriken des Nullsystems. Sie sehen, infolge einer Drehung, 
anders als die maxwellschen Spannungen aus.\\
Weil durch Angabe einer invarianten \textit{Quadrik} duale und reziproke 
Zuordnung vollkommen bestimmt werden, kann die Verwandtschaft 
stets auf den ganzen Raum erstreckt werden. Deshalb ersetzt die 
Auszeichnung auch eines einzigen Grundgebildes, unter Ausfall 
des Punkt-Ebene-Inzidenz Merkmales f\"{u}r die dann nicht auf dem 
Gebilde liegenden Punkte, geometrisch das statische Feld. Der 
Unterschied zwischen metrischer und projektiver Auffassung besteht 
darin, dass das metrische, jedoch nicht das projektive Gebilde 
normalisiert werden kann. Projektiv hat man immer mit unendlich 
vielen Grundgebilden zu rechnen.

\begin{figure}
  \includegraphics[scale=0.25]{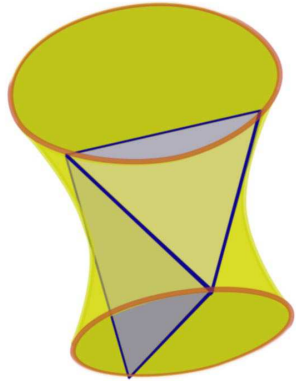}
  \caption{Einschaliges Hyperboloid. Die Tetraederkanten, welche Leitlinien bilden, geh\"{o}ren verschiedenen Gewinden an.}
  \label{fig_1}
\end{figure}

\begin{figure}
  \includegraphics[scale=0.25]{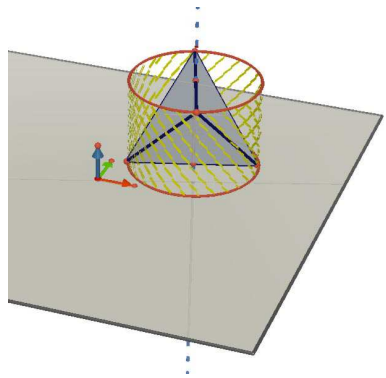}
  \caption{Kreiszylinder. Eine der Achse koaxiale H\"{u}llfl\"{a}che. Sind die Punkte auf den Kreisumf\"{a}ngen Nullpunkte, dann ist das Tetraeder aus 4 Nullebenen herausgeschnitten. Die Nullebenen schneiden einander zu je 2 l\"{a}ngs der Kreisdurchschnitte. Die Tetraederkanten, welche keine Kreisdurchschnitte sind, geh\"{o}ren 2 verschiedenen Regelschargewinden an.}
  \label{fig_2}
\end{figure}

\textbf{Synthetisch-geometrische Gestalt des Nullsystems.} -- Das Nullsystem 
ist haupts\"{a}chlich eine r\"{a}umliche lineare Gruppierung von 
dreifach unendlich vielen Geraden, die aus Kreisregelschargebilden 
besteht. Weil es sich \textit{nicht} auf die Blattebene einer Zeichnung
\cite{smith}\footnote{R. Descartes nennt 
auf S. 302 seines Buches die ebene Geometrie ,,gew\"{o}hnliche 
Geometrie``.} abbilden l\"{a}sst, stellt es sozusagen eine Erweiterung 
der statischen Graphik auf r\"{a}umliche Verh\"{a}ltnisse dar, wobei 
die Graphik gleichsam ausf\"{a}llt. Obwohl es keine Figur im euklidischen 
Sinn darstellt, haben wir es teilweise\footnote{Das volle Gebilde 
ist ein projektives Torus. Geometrisch als riemannsches Gebilde 
aufgefasst, hie{\ss}e es \textit{elliptisches Gebilde}.} auf ein St\"{u}ck 
Papier gezeichnet. Wir meinen die in Abb. \ref{fig_1} (s. auch Abb. \ref{fig_2}) dargestellte Projektion 
eines als Regelschar abgebildeten einschaligen Hyperboloids.

\begin{figure}
  \includegraphics[scale=0.25]{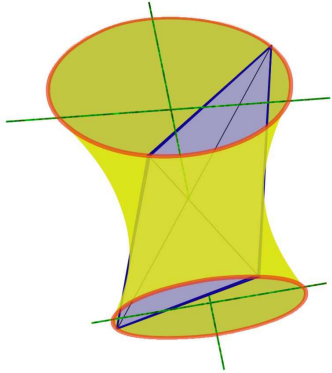}
  \caption{Einschaliges Hyperboloid. Einander zugeordnete Polaren.}
  \label{fig_3}
\end{figure}

Die Leitlinien sind lauter paarweise einander zugeordnete Polaren 
in Bezug auf die Zentralachse, die wir in Abb. \ref{fig_2} hervorgehoben haben. Wir 
haben, auf sie st\"{u}tzend, auch ein Bezugstetraeder, gebildet. Diejenige Ebene, in 
der sich die zwei fett durchgezogenen Tetraederkanten kreuzen, 
ist eine Tangentialebene. Ist sie eine \textit{Nullebene}, so kommt 
der mit ihr vereinigte duale Punkt, der \textit{Nullpunkt}, auf ihr 
zu liegen. Er ist Mittelpunkt s\"{a}mtlicher Nullachsen auf seiner 
Nullebene, also insbesondere ist er Nullpunkt einer der durch ihn hindurchgezeichneten 
Regelscharlinien. Wenn man Schraubentransformationen zul\"{a}sst, 
kommen auf dieser H\"{u}llfl\"{a}che s\"{a}mtliche Nullpunkte auf die Regelschar mit derselben 
Neigung zu liegen. Das Tetraeder scheint hier, wie das Hyperboloid, 
ein festes Volumen zu besitzen, aber es sieht halt bei dieser 
Projektion so aus. Es ist dar\"{u}ber hinaus gar nicht gemeint, 
dass eine projektive Schraubung, d.h. eine lineare \"{U}bertragung \textit{des 
ganzen Raumes} in sich, gerade dieses Tetraeder in sich \"{u}berf\"{u}hrt. 
Aus Abb. \ref{fig_3} erhellt, warum Tetraeder projektiv reine Illusionen 
oder, um uns mit M\"{o}bius Worten auszudr\"{u}cken, reine au{\ss}ergew\"{o}hnliche 
Polyeder sind. W\"{a}re nicht das Dualit\"{a}tsprinzip, wobei dann 
jeder Punkt sozusagen auf die ihm entsprechende Ebene projiziert 
wird, so k\"{o}nnten die beiden Regelscharen derart gestellt werden, 
dass das einschalige Hyperboloid ausartet und mithin den projektiven 
Raum st\"{u}rzt.

\begin{figure}
  \includegraphics[scale=0.25]{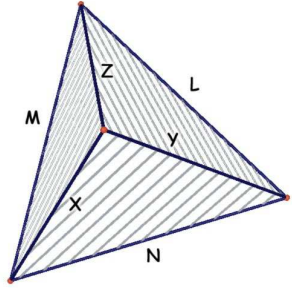}
  \caption{ X:Y:Z:L:M:N als Linienkoordinaten im speziellen Fall.}
  \label{fig_4}
\end{figure}

Ungeachtet der Schwierigkeit, das Nullsystem auf einem Blatt 
Papier darzustellen, kann es analytisch charakterisiert werden. 
F\"{u}hrt man homogene Punktkoordinaten $x, y, z, t$ f\"{u}r eine dreidimensionale 
Mannigfaltigkeit ein, dann l\"{a}sst es sich auf das fundamentale 
Tetraeder analytisch beziehen. Es seien dazu zwei beliebig auf 
einer gegebenen Gerade des Systems liegende Punkte
$ P \rightarrow (x, y, z, t)$
 und
 $P' \rightarrow (x', y', z', t')$ angenommen. 
Dann lassen sich die ebenfalls homogenen Linienkoordinaten bez\"{u}glich 
dem Tetraeder folgenderma{\ss}en ausdr\"{u}cken:
$ X = xt' - tx'$,
$ Y = ty' - yt'$,
$ Z = zt' - tz'$,
$ L = yz' - zy'$,
$ M = xz' - zx'$
 und 
$N = xy' - yx'$. Wir haben unten in Abb. \ref{fig_4} die Koordinaten eines 
weiteren Tetraeders projiziert. Die Koordinaten
 $X: Y: Z: L: M: N$
 befriedigen in diesem Fall die spezielle Gleichung
$ XL + YM + ZN = 0$ identisch, und bestimmen somit eine beliebige Linie 
des Nullsystems durch 4 Linienkoordinaten nach Gra{\ss}mann. Wir 
haben aber die Linienkoordinaten aus derselben Gerade $PP'$ berechnet. 
Wenn wir statt dessen verlangen, dass eine bestimmte Linie 
$X'_0:Y'_0: Z'_0: L'_0: M'_0: N'_0$
 geschnitten werde, lautet die 
Gleichung 
$X'_0L + Y'_0M + Z'_0N + L'_0X + M'_0Y + N'_0Z =0$.
 Die allgemeine bilineare Gleichung
$ XL' + YM' +ZN' + LX' + MY' + NZ' = 0$
 bestimmt schlie{\ss}lich s\"{a}mtliche konjugierte 
Leitlinien.

\textbf{Ableitung des maxwellschen Satzes von Gleichungen aus dem 
Nullsystem.} -- Wenn man die eingangs gegebenen formellen Ans\"{a}tze 
des Elektromagnetismus und der Statik miteinander vergleicht, 
so scheint die erste Theorie nicht sowohl auf dynamische als 
auf kinetische Grunds\"{a}tze gegr\"{u}ndet. Das bedeutet, dass man 
bez\"{u}glich der Gleichungen die Vereinbarung
$ \textbf{E} \Leftrightarrow (X, Y, Z)$
 und
$ \textbf{H} \Leftrightarrow (L, M, N)$
 trifft (wir bemerken 
dabei gleich, dass mitunter auch die andere Zuordnung getroffen 
worden ist). Ferner lauten die Linienkoordinaten:
$ E_x : E_y : E_z : H_x : H_y : H_z$,
 und man kann die f\"{u}r das Nullsystem 
typische bilineare Gleichung auf elektromagnetische Koordinaten 
laut:
$ E_xH'_x + E_yH'_y + E_zH'_z + H_xE'_x + H_yE'_y + H_zE'_z = 0$
 ausdr\"{u}cken. Deswegen d\"{u}rfte die Einf\"{u}hrung 
infinitesimaler Transformationen, etwa $E_xds = - dV$ usw., und 
$V = T dx/ds$ usw., wobei $T$ die nach den Koordinatenachsen zerlegte 
Spannung bedeutet, insbesondere wenn man die G\"{u}ltigkeit des 
Doppeltverh\"{a}ltnisses 
$d\textbf{E}/\textbf{E} = d\textbf{H}/\textbf{H}$
 im Betrag 
voraussetzt, kaum von Belang sein. \textbf{E} und \textbf{H} modellieren 
gleichwohl nach statischem Ermessen keine sich ausbreitenden 
L\"{o}sungen, sondern bestenfalls Momentbilder von sehr langsam 
ver\"{a}nderlichen mechanisch wirkenden Gr\"{o}{\ss}en dar. Hingegen 
beruht die Deutung, im markanten Unterschied zur Dynamik, wo \textit{beliebige, 
jedoch f\"{u}r einen partikul\"{a}ren Massenpunkt kennzeichnende} Anfangsbedingungen 
aufgelistet zu werden brauchen, auf das Entsprechen der 6 Koordinaten 
mit sechs f\"{u}r die Unbeweglichkeit notwendigen Bindungen. Das 
sind bestimmt keine physikalischen Bedingungen f\"{u}r das Einhalten 
des Gleichgewichts, wie es Ausdehnung und Undurchdringlichkeit 
der (starren) K\"{o}rper es sind. Deshalb sind diese Bindungen 
stets mit Schraubungen, will sagen, Raumtransformationen vertr\"{a}glich.\\
Zumal wir keine Deutung der Gleichungen nach statischen Ansichten 
erstreben, sondern uns ausschlie{\ss}lich f\"{u}r die Zul\"{a}ssigkeit 
des kleinschen Gedankengangs interessieren, geht es uns vorz\"{u}glich 
darum den Gleichungen einen geometrischen Gehalt innerhalb des 
Nullsystems zu verleihen. Sollte der Satz von Gleichungen relational 
im Sinne F. Kleins erstellt werden, dann w\"{a}re dass Nullsystem 
sowieso als rein geometrische Raumverwandtschaft aufzufassen\footnote{Je 
nach der gew\"{a}hlten Invariante m\"{o}gen die Linien irgendeine 
andere quadratische Regelschar bezeichnen. W\"{a}hrend nun das 
maxwellsche Modell auf Schraubensymmetrie gr\"{u}ndet, ist Kleins 
Einf\"{u}hrung der Imagin\"{a}ren Geraden f\"{u}r die Darstellung der 
Bewegungen wichtig, weil sie die Mittel an die Hand gibt, die 
Quadriken ineinander zu \"{u}berf\"{u}hren.}. Dazu w\"{a}hlen wir ohne 
irgendwelche Begr\"{u}ndung die am Anfang dieses Kapitels ausgestellte 
Differentialform.\\
Die Gleichungen lesen wir folglich als:\\

\begin{longtable}{p{1.1in}p{1.5in}p{2in}}

% ROW 1
{\raggedright Coulombscher Satz, keine freien Ladungen} & 
{\raggedright $ \lambda \textbf{E}_1 + \mu \textbf{E}_2 + \nu \textbf{E}_3 = 0$} & 
{\raggedright $ \partial X /\partial x + \partial Y /\partial y + \partial Z /\partial z = 0$} \\
&&\\
% ROW 2
{\raggedright Maxwell-amp\`{e}rescher Satz} & 
{\raggedright $ \rho \textbf{E}_1 = \mu'\textbf{H}_2 -  \nu'\textbf{H}_3$ \linebreak
$ \sigma \textbf{E}_2 = \nu'\textbf{H}_3 -  \lambda'\textbf{H}_1$\linebreak
$ \tau \textbf{E}_3 = \lambda'\textbf{H}_1 -  \mu'\textbf{H}_2$ } &
{\raggedright  $ 1/c \partial X /\partial t = \partial M /\partial z - \partial N /\partial y$\linebreak
$ 1/c \partial Y /\partial t = \partial N /\partial x - \partial L /\partial z$ \linebreak
$ 1/c \partial Z /\partial t = \partial L /\partial y - \partial M /\partial x$} \\

&&\\

% ROW 3
{\raggedright Faradayscher Satz} & 
{\raggedright $ \rho' \textbf{H}_1 = -[\mu\textbf{E}_2 - \nu\textbf{E}_3]$
$ \sigma' \textbf{H}_2 = -[\nu\textbf{E}_3 - \lambda\textbf{E}_1]$\linebreak
$ \tau' \textbf{H}_3 = -[\lambda\textbf{E}_1 - \mu\textbf{E}_2]$}
\footnote{Das Minuszeichen deutet an, dass das Dreieck im Vergleich 
zum vorangehenden Fall mit umgekehrten Sinn um die Achse dreht. 
In der Abbildung scheint das Dreieck dabei umzuklappen. Aber 
der Satz von Desargues ist projektiv.} & 
{\raggedright $ 1/c \partial L /\partial t = -[\partial Y /\partial z - \partial Z /\partial y]$\linebreak
$ 1/c \partial M /\partial t = -[\partial Z /\partial x - \partial X /\partial z]$\linebreak
$ 1/c \partial N /\partial t = -[\partial X /\partial y - \partial Y /\partial x]$} \\

&&\\

% ROW 4
{\raggedright Keine freien magnetischen Monopole} & 
{\raggedright $ \lambda' \textbf{H}_1 + \mu' \textbf{H}_2 + \nu' \textbf{H}_3 = 0$} & 
{\raggedright $ \partial L /\partial x + \partial M /\partial y + \partial N /\partial z = 0$} \\

\end{longtable}

\begin{figure}
  \includegraphics[scale=0.25]{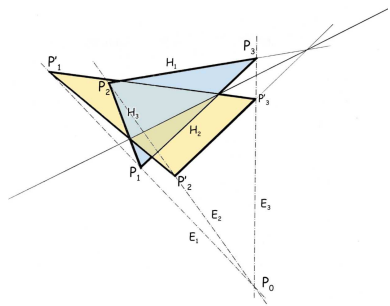}
  \caption{Zum Theorem von Desargues.}
  \label{fig_5}
\end{figure}

Nach diesen Glossen seien
 $P_i  = (x_i , y_i , z_i ), i = 0, 1,2, 3$
 vier Punkte mit den inhomogenen Koordinaten $x, y, z$. Es 
werden die Bedingungen niedergeschrieben, unter welchen eine 
durch sie hindurchgehende Gerade mit 
$\rho \textbf{E}_i  = (x_i -x_0,y_i -y_0, z_i -z_0) = (E_ix, E_iy, E_iz)$
 gleichgerichtet ist. 
Durch $P_1$  geht die durch das Verh\"{a}ltnis 
$(x - x_1)/E_1x = (y - y_1 )/E_1y = (z - z_1 )/E_1z$
 bestimmte Gerade $g_1$ , was 
die drei folgenden Gleichungen ergibt:
$ E_1x(y - y_1 ) = E_1y(x - x_1 )$,
$ E_1x(z - z_1 ) = E_1z(x - x_1 )$,
$ E_1y(z - z_1 ) = E_1z(y - y_1 )$.
 Ganz auf dieselbe Weise lassen sich zu 
$\rho \textbf{E}_2$  
und $\rho \textbf{E}_3$ gleichgerichtete Geraden $g_2$  und $g_3$ durch 
die Punkte $P_2$  und $P_3$ respektive anlegen. Da wir die
$ \rho \textbf{E}_i$  
s\"{a}mtlich durch $P_0$ gew\"{a}hlt haben, kann man nat\"{u}rlich einen 
B\"{u}ndel durch $P_0$ definieren. Dr\"{u}ckt man den B\"{u}ndel als
 $\lambda \textbf{E}_1 + \mu\textbf{E}_2  + \nu\textbf{E}_3 = 0$
 aus, so bestimmt
 $P_0+ \lambda\textbf{E}_1  = 0$
 au{\ss}er $P_1$  noch weitere Punkte $P_1'$, 
$P_1''$, u.s.w., so dass, wenn wir
 $P_3 - P_2  = \textbf{H}_1$ , 
$P_1  - P_3 = \textbf{H}_2$,
$ P_2  - P_1  = \textbf{H}_3$
 nennen, sich die Strahlen
$ \mu'\textbf{H}_2- \nu'\textbf{H}_3 = 0$
 (l\"{a}ngs der
$ \rho \textbf{E}_1$  Geraden),
$ \nu'\textbf{H}_3 - \lambda'\textbf{H}_1  = 0$
 und
 $\lambda'\textbf{H}_1 - \mu'\textbf{H}_2  = 0$
 (l\"{a}ngs der beiden anderen Geraden) 
weiterhin in $P_0$ treffen. Wir haben das in Abb. \ref{fig_5} zu wiedergeben 
versucht.
In Abbildung \ref{fig_5} ist $\textbf{E}_i$  als Geradenb\"{u}ndel durch einen 
Punkt, $\textbf{H}_i$  aber als ebenes Dreiseit zu verstehen.
 Wenn diese Ausdr\"{u}cke nicht verschwinden, werden
 $P'_1 P'_2  P'_3$,
$ P''_1  P''_2  P''_3$ 
usw. keine Dreiecke. Nach demselben Theorem von Desargues bestimmt 
man dual, dass sich die gleichnamigen Schenkel zweier Dreiseiten
 $\textbf{H}_1  \textbf{H}_2  \textbf{H}_3$, 
und $\textbf{H}'_1  \textbf{H}'_2  \textbf{H}'_3$ auf einer 
Achse kreuzen. Diese Punkte sind die B\"{u}schelmittelpunkte von
$ \nu\textbf{E}_3 - \mu\textbf{E}_2  = 0$,
 $\lambda\textbf{E}_1  - \nu\textbf{E}_3 = 0$
 und
$ \mu\textbf{E}_2  - \lambda\textbf{E}_1  = 0$.
 Eine Drehung 
um eine Achse findet aber im Nullsystem im Allgemeinen nicht 
statt.\\
Da der Satz von Desargues nach den oben geschriebenen Gleichungen 
f\"{u}r alleinstehende Tripel $\textbf{E}_i$  und $\textbf{H}_i$  nicht zur Geltung 
kommt, zeigt die am Anfang stehende \"{U}bersicht einen Satz Gleichungen, 
deren Verwandtschaft weder Fluchtpunkt noch -Achse zul\"{a}sst. 
Das Nullsystem besitzt diese Eigenschaft.

\begin{figure}
  \includegraphics[scale=0.25]{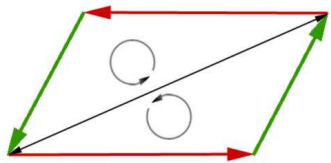}
  \caption{Kr\"{a}fteparallelogramm nach M\"{o}bius.}
  \label{fig_6}
\end{figure}

\textbf{Fundamentalgleichungen der Statik.} -- M\"{o}bius hat das Nullsystem 
auf den Fall angewendet, dass zwei zueinander windschiefe Kr\"{a}fte 
auf einen (geometrischen) K\"{o}rper wirken, und die Gleichgewichtsstellungen 
untersucht. Um dabei polare und axiale Vektoren nicht miteinander 
addieren zu m\"{u}ssen, hat er bereits in der Ebene aus dem Kr\"{a}fteparallelogramm 
herausgelesen, dass sich im Gleichgewicht zwei einander entgegengerichtete 
Paare bilden, die sich nach Abb. \ref{fig_6} stets das Gleichgewicht halten. 
Mit dieser Festsetzung schlie{\ss}en in der Ebene Kraftvektoren 
immer orientierte Fl\"{a}chen ein. Dieser Begriff l\"{a}sst sich 
nun auf das allgemeinste System von zueinander windschiefen Kr\"{a}ften 
im Raum erweitern. Wenn man die Darstellung auf den Raum \"{u}bertr\"{a}gt, 
k\"{o}nnen zwar Fl\"{a}chen ein Volumen fassen, dem eine \textit{relative} 
Zahl als Inhalt zukommt. Jedoch wird dann dieselbe Kante aus 
zwei angrenzenden Fl\"{a}chen jeweils in umgekehrter Richtung durchgelaufen. 
M\"{o}bius hat obendrein gelegentlich darauf aufmerksam gemacht, 
dass man Fl\"{a}chen mit sich kreuzenden Kanten sogar \textit{keinen 
festen Wert} des Fl\"{a}cheninhalts mehr verleihen kann. Dieses 
Problem tritt f\"{u}r K\"{o}rper mit einseitigen Fl\"{a}chen ein, da 
sie lauter au{\ss}erordentliche Polyeder sind. Beil\"{a}ufig wollen 
wir noch erw\"{a}hnen, dass man darum historisch das projektive 
Linienelement als Raumelement angenommen hat, weil es selbstdual 
ist.

Die Fundamentalgleichungen der Statik lauten, wenn sie weder 
s\"{a}mtlich von einem Punkt ausgehen (Resultante), noch sich auf 
eine einzige Ebene reduzieren (Paar): ${\bf R} = 0$ und $\textbf{M} = 0$. 
M\"{o}bius hat darauf hingewiesen, dass es m\"{o}glich ist zwei windschiefe 
Kr\"{a}fte auf eine einzige Weise auf eine Resultante {\bf R} und 
ein gleichgerichtetes Kr\"{a}ftepaar \textbf{M} zu reduzieren. Das entspricht 
etwa der Transformation von Abbildung \ref{fig_1} zu Abbildung \ref{fig_3} auf der 
Walze. War die Nullachse anfangs f\"{u}r das statische Gleichgewicht 
kennzeichnend, so werden durch die Verwandtschaft s\"{a}mtliche 
Leitlinien auf der gezeichneten H\"{u}llfl\"{a}che Nullachsen.

\textbf{Geometrische Symmetrien und Prinzipien der Dynamik.} -- M\"{o}bius 
selber hat sich, obwohl er auch den Bewegungen eine geometrische 
Grundlegung zu verleihen gedachte, nicht besonders damit befasst. 
Einen Ansatz hat er trotzdem hinterlassen: seine Lehrs\"{a}tze 
zum Gleichgewicht von an elastischen K\"{o}rpern angreifenden Kr\"{a}ftesystemen 
stehen, da er eine zur ersten Ordnung kleine Beweglichkeit gegen\"{u}ber 
der nullten Ordnung heranzieht, der Potentialtheorie bemerkenswert 
nahe.

Wie schon angedeutet war auch Pl\"{u}cker der Meinung, dass sich 
die Geometrie auf die Bewegung anwenden lie{\ss}e. Pl\"{u}ckers Vorstellung 
\"{u}ber die Zuordnung von Kraft und \textit{Kinetik} beruhte doch offensichtlich 
nicht auf Ursache und Wirkung, wie bei Newton, sondern eher auf 
geometrische Symmetrieeigenschaften der Figuren. So wurden, weil 
beides mal eine Gerade in sich \"{u}berf\"{u}hrt wurde, Kr\"{a}fte 
mit elementaren\footnote{Das Wort ,,elementar`` bezieht sich auf die 
M\"{o}glichkeit jede beobachtete Bewegung aus einfachen Urtypen 
zusammenzustellen. Klein hat elementar mit infinitesimal ersetzt. 
Das mag aber nicht stimmen, weil die hier in Betracht gezogenen 
Symmetrien mit der Infinitesimalrechnung nichts zu tun haben. 
Es ist auch so, dass infinitesimale Drehungen hart von Verschiebungen 
zu unterscheiden sind.} Rotationen koordiniert. Dabei dr\"{u}ckte 
die Gerade einmal die Wirkungslinie der Kraft, das andere Mal 
die Rotationsachse aus. Die Aussage, dass es diese invariante 
Gerade auch f\"{u}r Kr\"{a}ftepaar und Verschiebung gibt, kann allerdings 
nicht strikt innerhalb der euklidischen Geometrie bewiesen werden\footnote{Das 
geht nur durch adjungieren des Unendliches.}. Der Inbegriff aller 
Kr\"{a}fte und Paare, bzw. aller Verschiebungen und Drehungen, 
wurde trotzdem \textit{Dyname} genannt und zur Gesamtheit der Raumlinien 
bezogen. Der grundlegende Ansatz zum geometrischen Begriff der 
Bewegung hat erst Klein geliefert. Er beruht v\"{o}llig auf den 
Erlanger Programm.

Die \"{u}bliche, \textit{metrisch verstandene} Statiklehre, aus der 
die Bewegungslehre abgeleitet wird, geht jedenfalls weiter als 
das Nullsystem, und deutet {\bf R} als Kraftresultante l\"{a}ngs 
der Schraubenachse und \textbf{M} als ihr zugeordnetes Paar in einer 
der Achse senkrechten Ebene\footnote{Man dachte zeitweilig, dass sich 
die Schraubung als allgemeinste Bewegung aus den Elementarbewegungen 
Verschiebung und Drehung zusammensetze. Sp\"{a}ter wurde aber die 
Verschiebung als Resultierende zweier entgegengesetzten Drehungen 
erkl\"{a}rt. R\"{a}umliche Drehungen sind jedoch keine einfachen 
Drehungen, weshalb man endlich Elementarbewegungen auf infinitesimale 
Bewegungen zur\"{u}ckf\"{u}hrte.}.

Der springende Punkt ist, dass sich dieser Unterschied nicht 
aufrechterhalten l\"{a}sst, sobald es einen Anlass dazu g\"{a}be, 
also im Fall \textit{station\"{a}rer} Bewegungen. Geometrisch fu{\ss}t 
n\"{a}mlich die Bewegung auf Symmetrieeigenschaften der Figuren. 
Nun werden s\"{a}mtliche Nullgeraden auf den H\"{u}llfl\"{a}chen des 
Nullsystems als geod\"{a}tische Linien erkl\"{a}rt. Sie bilden auf 
doppelkegelartigen Regelscharen und geraden Kreiszylindern Geraden. 
Schiefwinklige Mantelgeraden werden mitunter auch ber\"{u}cksichtigt, 
und werden, obwohl sie geometrisch lauter gerade Strecken auf 
rektifizierbaren Fl\"{a}chen sind, als \textit{Windungen} gedeutet. 
Die H\"{u}lle bleibt bei Schraubenbewegungen\footnote{Weil hier ein 
System von Regelscharen in betracht kommt, ist es angebracht 
daran zu erinnern, dass projektiv die Linien von einer Regelschar 
zur anderen \"{u}berspringen k\"{o}nnen.} der schiefwinkligen Mantelgeraden 
aus Symmetriegr\"{u}nden invariant, wobei es je nach der Schraubung 
eine auf die andere als fest gedachte Regelschar sich als ganzes 
verschiebt. Geometrisch gelten auch reine Drehungen noch als 
Symmetrietransformationen. Die Statiklehre deutet im Nachhinein 
diese Art der Raum\"{u}bertragungen als Erhaltungss\"{a}tze. Sie 
k\"{o}nnte genauso gut auf diese Interpretation verzichten.

\textbf{Zum absoluten Raum.} -- Schraubentransformationen sind, wie 
wir bis jetzt gezeigt haben, die ,,Symmetrietransformationen 
der Statik``. Maxwells elastomechanische Spannungen des \"{A}thers 
weisen freilich eine andere Invariante auf. Einstein hat schlie{\ss}lich 
den Satz der Invarianz der Lichtgeschwindigkeit aufgestellt. 
Man k\"{o}nnte nun behaupten, dass Koordinatentransformationen 
zur Wiedergabe der Relativit\"{a}t des Bewegungszustandes begrifflich 
anschaulicher seien. Stabile statische Spannungsverteilungen 
sind tats\"{a}chlich, ohne einen ziemlichen Aufwand M\"{u}he an K\"{o}rpern 
kaum\footnote{Ob und inwiefern Aush\"{o}hlungen die elektromagnetische 
Wirkung auf die Materie an Ort und Stelle zu messen gestatten, 
m\"{o}chten wir jetzt nicht besprechen. Jedenfalls werden heute 
zunehmend solche Signalmessungen vorgezogen, die gleich einen 
ganzen von Au{\ss}en bestrahlten K\"{o}rper durch ihre Verteilung 
kennzeichnen. Auch typisch mechanische, z.B photoelastische Messungen 
werden an besonders dazu hergestellten durchsichtigen Probek\"{o}rpern 
vorgenommen.}, und im \"{A}ther schon gar nicht messbar. Was aber 
die synthetisch-geometrischen Modelle betrifft, weisen sie s\"{a}mtlich 
eine Invariante Quadrik auf. Die St\"{a}rke des geometrischen Modells 
besteht eben darin, dass die entsprechenden Theorien, ohne Besonderheiten 
der mathematischen Hilfsmittel oder t\"{u}ckische Erscheinungen 
bei der Deutung beachten zu m\"{u}ssen, untereinander vergleichbar 
werden.

\textbf{Zur Ausschaltung der Fernwirkung.} - Nun versuchen wir zu 
er\"{o}rtern, was mit der Erkl\"{a}rung verbunden sein k\"{o}nnte, \textit{es 
g\"{a}be keine von fern wirkenden Kr\"{a}fte}.\\
Auf der Ebene darf man das Moment immer durch eine \"{a}quivalente 
Kraft im Unendlichen ersetzen, was damit gleichkommt, dass man 
geometrisch entweder \textbf{H} auf \textbf{E} oder umgekehrt reduziert. 
Das will man im Elektromagnetismus von Haus aus nicht haben, 
weil man damit nicht \"{u}ber die elektro- und magnetostatischen 
Aussagen ginge, w\"{a}hrend doch Induktion und Maxwell- Amp\`{e}resche 
Gleichung \textbf{E} und \textbf{H} auf nicht-triviale Weise zueinander 
beziehen\footnote{Wegen des projektiven Ausfalls des Unterschiedes 
zwischen Drehungen und Verschiebungen, bedeutet das nicht, dass 
sich die eine Gr\"{o}{\ss}e als Drehgr\"{o}{\ss}e, die andere aber als 
Verschiebung deuten lie{\ss}e.}. F\"{u}r das Nullsystem als r\"{a}umliche 
Geometrie soll also das gegenseitige Reduzieren irgendwie nicht 
mehr zutreffen.

Der klaffende Widerspruch entsteht, weil die Kr\"{a}fte zur Reduktion 
eines Kr\"{a}ftesystems der Parallelogrammregel nach l\"{a}ngs ihrer 
Richtung bis zu einem gemeinsamen, eventuell unendlich weit entfernten 
Angriffspunkt verschoben werden m\"{u}ssen. Will man nun keine \textit{unendlichferne 
unendlichkleine Kraft} zulassen, dann st\"{u}rzt man im Raum beil\"{a}ufig 
den Parallelismusbegriff. Gibt es n\"{a}mlich keinen Punkt im Unendlichen 
aus dem sich untereinander parallele Ebenen projizieren lassen, 
dann liegen die parallelen Kr\"{a}fte entweder bereits auf einer 
gemeinsamen Ebene des Raumes, oder sie sind auf Regelscharen 
verteilt, gelten als zueinander windschief und lassen sich nimmer 
auf eine Resultante reduzieren. Das nichtentartete Nullsystem 
stellt den zweiten Fall dar.

Schlie{\ss}t man umgekehrt nach Maxwell im Voraus jede Fernwirkung 
aus, dann ist man dazu berechtigt den Unterschied zwischen den 
im Endlichen und Unendlichen gelegenen Punkten fallen zu lassen.

\textbf{Zur Ausschaltung der Laufbahnen.} -- Die lineare geometrische 
Bewegungsbeziehung, die aus dem Nullsystem hervorgeht weist keine 
Punktbahnen auf. Sollte man unter Beibehaltung des Modells Punkte 
als Elemente einbeziehen, w\"{a}ren unbedingt Ebenen als Elemente 
mitzunehmen. Aus diesem Grund passt keine projektive Geometrie, 
wie sie gew\"{o}hnlich verstanden wird, zur Darstellung von Ladungen 
und konvektiven Str\"{o}men.

Trotz dass die Raumverwandtschaften wie ,,Drehungen um Hauptachsen`` 
behandelt werden (Eigenwerteprobleme), wobei keine oder zwei 
Strahlen projektiv unbewegt bleiben k\"{o}nnen\footnote{In entarteten 
F\"{a}llen kommt auch ein einziger Strahl als Invariante vor.}, 
kommt ihnen ferner keine Drehfrequenz zu. Die Linienelemente 
selber bedeuten \textit{unteilbare} Strahlen/Achsen, was an die \"{a}ltere 
und links liegen gelassene alternative zur Infinitesimalrechnung 
ankn\"{u}pft (J. Bell), und nicht mit derjenigen Bahnparametrisierung 
zu verwechseln ist, deren zweite Ableitung nach t zum Kraftgesetz 
f\"{u}hrt.

Es bleibt zu zeigen, dass die projektive Geometrie \"{u}berhaupt 
keine Bestimmung einer punktierten Bahn erlaubt. Wir erinnern 
daran, dass auf Linienkoordinaten keine Raumpunkte bezeichnet 
werden, obwohl die Transformationen des Nullsystems Punkte in 
Punkte oder Ebenen (da sie nicht selbstdual sind) \"{u}berf\"{u}hren. 
Projektiv kann \textit{ein} Punkt nur als B\"{u}ndelmittelpunkt erkl\"{a}rt 
werden. Im Fall, dass der Punkt unermesslich fern zu liegen kommt, 
kann er dann mit \textit{einer einzigen Richtung} verbunden werden, 
aber, wegen des Dualit\"{a}tsprinzips, trotzdem mit keinem \textit{einzelnen} 
Strahl.

\textbf{Geometrisch zu fassende Unterschiede zwischen Mechanik und 
Elektromagnetismus. --} Klein meinte wohl, jenseits aller physikalischen 
Vorstellungen, der Welt eine einzige beobachtbare Struktur beimessen 
zu sollen, weil er also schlie{\ss}t: ,,Infolgedessen k\"{o}nnen 
wir bei den meisten Beobachtungen der Physik und im besonderen 
der Astronomie die G\"{u}ltigkeit der klassischen Mechanik voraussetzen, 
ohne merkliche Fehler zu begehen.``\footnote{Letzte Zeile in den Vorlesungen 
\"{u}ber nicht-Euklidische Geometrie, f\"{u}r den Druck neu bearbeitet 
von W. Rosemann, im 1928.} Wir glauben dagegen nicht im Geringsten 
die Einheitlichkeit der Welt auf unsere Erfahrung st\"{u}tzen zu 
k\"{o}nnen.\\
Selbstredend sind projektive Darstellungen keineswegs zur Auffassung 
der elektrischen Erscheinungen erfunden worden. Es ist ferner 
nicht anzunehmen, dass die dargestellten Gr\"{o}{\ss}en je nach der 
gew\"{a}hlten Mathematik automatisch irgendwie zu einer bestimmten 
Gruppe von Erscheinungen passten. So sind \textbf{E}, \textbf{H} nicht darum 
physikalisch Felder, weil sie sich einem Nullsystem oder gar 
einer Wellengleichung f\"{u}gen. Darum mochte die Frage, inwiefern 
Maxwell als lineare elektromagnetische Erscheinungen andere Erscheinungen 
als Newton meinte, eine interessante sein. Der gew\"{a}hlten mathematischen 
Ausdrucksform wegen, m\"{o}chten wir immerhin dreierlei unterschiedlich 
beurteilen.\\
{\textbullet}\tab 
Erfahrungsgem\"{a}{\ss} k\"{o}nnen K\"{o}rper niemals \"{u}berlagert werden. 
Newton ging davon aus, dass sich K\"{o}rper nicht ohne ein schreckliches 
Zusammenprallen durchdringen. Die Anfangsbedingungen in der Bahndarstellung 
dienen dazu, dieser M\"{o}glichkeit in den gesetzlich erlaubten 
Bahnen vorzubeugen. Die mathematischen Punkte untergehen n\"{a}mlich 
selber keine Sto{\ss}prozesse. Im Gegenteil k\"{o}nnen wir so viele 
Punkte \"{u}berlagern, wie wir wollen. Auch betreffs der Beobachtung 
ist Newtons Auffassung nicht unbedingt selbstverst\"{a}ndlich, 
denn wir sehen ja wie sich Mond und Sonne w\"{a}hrend einer Sonnenfinsternis 
\"{u}berlagern. Der fortschreitenden Wellenl\"{o}sungen kann man 
wie der Punkte so viele \"{u}berlagern, wie man Lust dazu hat, 
und der Unterschied zur mechanischen Vorstellung liegt nur darin, 
dass Rundfunksendungen im Modulierungs- Entmodulierungsprozess 
gerade davon \textit{Anwendung machen}.\\
{\textbullet}\tab Erfahrungsgem\"{a}{\ss} ist die Lage von K\"{o}rpern 
(bzw. von Kraftfeldern) immer messend festzustellen, was man 
\"{u}blich auf den Begriff der Lokalisierbarkeit zur\"{u}ckf\"{u}hrt. 
Die ebenen Wellen oder Wellenpakete sind in der raumzeitlichen 
Darstellung nicht lokalisierbar. Aber darum handelt es sich eben 
nicht. Indem Newton die Massenpunkte mittels Anfangsbedingungen 
charakterisiert, nimmt er diese als K\"{o}rpereigenschaften an. 
Mathematisch ist es allerdings nicht leichter, einen Punkt auf 
einer Gerade als eine Wellenl\"{o}sung im Raum zu lokalisieren. \textit{Der 
Gesichtssinnerfahrung nach} l\"{a}sst dazu die optische Lokalisierung 
all derjenigen K\"{o}rper, die selber oder abgebildet dicht am 
Brennpunkt eines optischen Apparates zu liegen kommen zu w\"{u}nschen 
\"{u}brig.\\
{\textbullet}\tab Als er die Ableitungen einf\"{u}hrte, hat sich Newton 
von der Vorstellung der Bewegung als von einer stetigen \"{U}berf\"{u}hrung 
der Punkte leiten lassen. Es sieht so aus, als ob die Erscheinungen 
diese Ansicht best\"{a}tigten. Auch die Ansicht, dass sich die 
unmittelbare Nachbarschaft eines Punktes w\"{a}hrend der Bewegung 
im Sinn des lokalen Zusammenhangs beliebig wenig vom Punkte entfernt, 
schein best\"{a}tigt. Beide fallen bei Beobachtungen am Spiegel 
aus. Kein Bild passiert die Brennebene einer Linse oder eines 
Spiegels stetig, oder bleibt sich dabei \"{a}hnlich, obwohl es 
als Ganzes abgebildet wird. Trotzdem lassen sich die eben genannten 
Beobachtungen mathematisch durch stetige Funktionen darstellen.

\textbf{Zum kleinschen Modell.} - Die von Klein betrachtete Geometrie 
kann nun, wenn man den r\"{a}umlichen Sehsinn ausnutzt\footnote{Weil 
sich die r\"{a}umlichen Gegenst\"{a}nde auf der zwar gekr\"{u}mmten, 
doch flachen Netzhaut abbilden, und wir sie obendrein unwillk\"{u}rlich 
scharf stellen, entspricht die r\"{a}umliche Ausdehnung der optischen 
Bilder einer verwickelten Erfahrung.}, auch rein visuell angeschaut 
werden. Es handelt sich nat\"{u}rlich nur von \textit{einer anderen} 
plastischen Deutung. Mit optischen Mitteln bildet man aber \textit{r\"{a}umliche} 
Beziehungen, ohne auf Fl\"{a}cheneinbettungen im Euklidischen Raum 
angewiesen zu werden, direkt projektiv ab. Anschaulich sind die 
Bilder freilich nicht, weil man einer ziemlichen Ein\"{u}bung bedarf, 
um sie \"{u}berhaupt als r\"{a}umlich zu entschl\"{u}sseln. Die Losl\"{o}sung 
von den mechanischen Vorstellungen in der Modellierung von elektromagnetischen 
Signaltheorien setzt allerdings die Erfassung solcher Beziehungen 
voraus.

Wird die beim ersten Blick nicht ganz anschauliche, doch recht 
bildhafte optische Ausf\"{u}hrung der Geometrie der Lage nach Klein 
gedeutet, dann muss man beispielsweise gewahr sein, dass sich 
optisch keine Raumstruktur, sondern bestenfalls das \textit{empfangene} 
Signal abbildet. Die die Abbildung erzeugende Transformation 
soll versuchsweise als Bewegung im Mittelpunkt der Modellierung 
verlegt werden, und das Wort ,,Abbildung`` nur dazu benutzt werden, 
um die Plastik der geometrischen Darstellung hervorzuheben. Mit 
optischen Bildern beliebiger K\"{o}rper d\"{u}rfen also die mathematischen 
Beziehungen des Elektromagnetismus wie gesagt modelliert, jedoch 
nicht identifiziert werden, weil sie im Modell Transformationen 
sind. Zudem sollen einfache geometrische Gebilde wie Punkte, 
Ebenen und Strahlen ausgemacht und physikalisch gedeutet werden.\\
Wenn man mittels physikalischer Vorstellungen zur Konstruktion 
des r\"{a}umlichen Modells gelangt, m\"{o}gen unterschiedliche Merkmale 
die Wahl bestimmen. So w\"{u}rden wir jetzt rein optisch, und unbeachtet 
der mathematischen Schwierigkeiten das Element mit dem ,,Bild 
des sehr weit entfernten Fremdstrahlers`` gleichsetzen, und dementsprechend 
eher die Punkt/Richtung und die ihr duale Ebene in betracht nehmen. 
Sollte die Theorie unseren jetzigen Ansichten \"{u}ber optische 
Signale entsprechen, dann l\"{a}gen ihr dar\"{u}ber hinaus, wie bereits 
von Klein verlangt, allgemeinere Transformationen der invarianten 
Quadrik zugrunde.

Zum Schluss m\"{o}chten wir einige Umst\"{a}nde erw\"{a}hnen, die zur 
Tilgung des kleinschen Modells beigetragen haben k\"{o}nnten.\\
1.\tab 
Eben hatte man noch geglaubt, Raum und Ausdehnung w\"{a}ren identisch. 
Man kann nicht erwarten, dass Klein selber die Metrik in der 
pl\"{o}tzlich relational verstandenen Geometrie etwa klar und deutlich 
als reiner Ballast empfinden sollte.\\
2.\tab 
Die (flache) Perspektive war in der Malerei kanonisiert, so dass 
oft einzig die mit der optischen Projektion verbundene Graphik 
als projektive Geometrie galt. Umgekehrt wurden Parallaxeeffekte 
aus den optischen Erscheinungen standardm\"{a}{\ss}ig abgestreift.\\
3.\tab Bis fasst zu Kleins Zeiten war der geometrische Raum mit dem 
analytischen Punktraum von drei Dimensionen gleichgeschaltet 
worden.\\
4.\tab Um die Jahrhundertwende zum 1900 wurden viele verschiedene 
Systeme zur Darstellung der Drehungen vorgeschlagen. Bestimmend 
f\"{u}r die Anwendungen wurde dann der bequeme Anschluss an die 
Analysis. Deshalb wurde die Vektoralgebra popul\"{a}r.\\
5.\tab Die auf den Linienelement gebaute Geometrie ist angeboren r\"{a}umlich, 
was leicht zu erkennen ist: sind keine \textit{Punkte} als Elemente 
gegeben, dann besteht auch nicht die M\"{o}glichkeit sich durch 
Projizieren aus einem Punkt auf die Geometrie in der Ebene zu 
reduzieren.\\
6.\tab 
Die sogenannte geometrische Optik benutzt Lichtstrahlen zur graphischen 
Auswertung der flachen Abbildungen. Entsprechend wurde die Gerade 
als Raumelement anschaulich sofort mit einem Lichtb\"{u}ndel identifiziert. 
Das hatte keinen Anhaltspunkt.

Das hat dem Bestehen des kleinschen Modells nichts geholfen.

\subsection{
Schlussbemerkungen}

Wir fassen diese Auseinandersetzung mit dem kleinschen Modell 
des Elektromagnetismus noch auf eine andere Weise zusammen.\\

\begin{longtable}{llll}

% ROW 1
\multicolumn{1}{p{2.205in}}
{\raggedright
Geometrisierung. \linebreak
} & 
\multicolumn{1}{p{2.205in}}
{\raggedright
Modellierung. \linebreak
} \\

% ROW 2
\multicolumn{1}{p{2.205in}}
{\raggedright
Sie gilt nur zur \"{u}bersichtlichen \"{U}berpr\"{u}fung der Widerspruchslosigkeit 
der Gleichungen. Dazu braucht man mathematische Ausdr\"{u}cke eindeutig 
durch Gebilde darzustellen. \linebreak
Relativit\"{a}tstheorie und statistische Mechanik nehmen aber hin, 
dass sich die geometrische Nachpr\"{u}fung der logischen Koh\"{a}renz 
nicht auf beobachtbare Raumeigenschaften st\"{u}tzen kann. \linebreak
} & 
\multicolumn{1}{p{2.295in}}
{\raggedright
Sie kn\"{u}pft auf eine zugleich synthetisch geometrisch und mathematisch 
geltende Relation. Die Liniengebilde im Linienraum k\"{o}nnen so 
verteilt werden, dass ihre Bewegung ein System von Quadriken 
in sich \"{u}berf\"{u}hrt. Dann bilden die Linien Regelscharen, und 
die Bewegungen sind (hyperbolische) Schraubungen. Diese werden 
proiektiv auf lineare Substitutionen in der komplexen Zahlenebene 
und somit auf die Funktionentheorie bezogen\footnote{Zur hyperbolischen Bewegung muss bemerkt werden, dass die 
projektive Ebene einseitig, die zweischalige Hyperboloiden und 
die euklidische Ebene dagegen zweiseitig sind.}. \linebreak
} \\

% ROW 3
\multicolumn{1}{p{2.205in}}
{\raggedright
Die Relevanz einer einschl\"{a}gig gestalteten Mathematik f\"{u}r 
das Verst\"{a}ndnis der unvermittelt beobachteten Tatsachen wird 
untersch\"{a}tzt. Denn man sagt, dass die Wirklichkeit durchaus 
nicht den Beobachtungen zu entsprechen braucht. \linebreak
} & 
\multicolumn{3}{p{2.295in}}
{\raggedright
Die projektiven \textit{Raum}beziehungen k\"{o}nnen optisch veranschaulicht 
werden. Wenn man die Spiegelung nach der Strahlenoptik rein graphisch 
fasst, kommt man zu hyperbolischen Bewegungen. \linebreak
} \\

\end{longtable}

Wenn man dem logischen Ausbau der Geometrie Verwandtschaften 
zugrundelegt, kommen im Nullsystem dieselbe hyperbolischen Bewegungen 
vor, wie in der Theorie der automorphen Funktionen. Ein verallgemeinerter 
Zusammenhang, mit Einbeziehung imagin\"{a}rer Transformationen, 
zwischen synthetischer Geometrie und Funktionentheorie hat F. 
Klein ausgebaut. Das Einbeziehen imagin\"{a}rer Transformationen 
ist auch notwendig, wenn man unbesorgt von den maxwellschen Gleichungen 
zur Wellengleichung \"{u}berzuwechseln w\"{u}nscht. Das entspricht 
einer Erweiterung im Sinne Reye der von M\"{o}bius f\"{u}r die Statik 
aufgestellten Geometrie. Diesen Vorschlag hat wiederum schon 
Klein an Einstein aus denselben Grund heraus gemacht. Das w\"{u}rde 
das geometrische Modell zu einer elektromagnetischen Theorie 
des empfangenes Signals liefern. Nat\"{u}rlich begr\"{u}ndet eine 
Versinnbildlichung der Geometrie f\"{u}r sich noch nicht die elektromagnetische 
Theorie als Signaltheorie. Hier waren wir aber vorerst an einen 
Vorschlag zum Vordringen interessiert. Sp\"{a}ter, wenn man so 
viele Arten von empfangenen Signalen wie m\"{o}glich unterzubringen 
trachtet, werden mutma{\ss}lich weitere Erw\"{a}gungen mehr in den 
Vordergrund treten.

\end{document}